\documentclass[11pt,tightenlines,aps,prd,floatfix,onecolumn,superscriptaddress,nofootinbib,longbibliography]{revtex4-2}
\usepackage[utf8]{inputenc}

\usepackage{amsmath, amsthm, amssymb,slashed,url,bbold,xcolor,upgreek,xspace, silence}
\usepackage{graphicx}
\usepackage{adjustbox}
\usepackage{tabularx}
\usepackage[justification=raggedright, singlelinecheck=false]{caption}
\usepackage[singlelinecheck=false]{subcaption}
\usepackage{epstopdf}
\usepackage{slashed}
\usepackage{physics}
\usepackage{siunitx}
\ExplSyntaxOn
\msg_redirect_name:nnn { siunitx } { physics-pkg } { none } %the conflict is between the command \qty definition in the packages physics and siunitx, the warning happens by design when you load both packages, so i silenced it (we never use \qty anyway)
\ExplSyntaxOff
\usepackage{hhline}
\usepackage{appendix}

\usepackage[pdfpagelabels, pdfencoding=auto, psdextra]{hyperref}
\hypersetup{%
 pdfsubject=Paper,
 pdfkeywords={nuclear physics} 
 unicode = true,
 breaklinks = true,
 colorlinks = true,
 linkcolor = blue,
 menucolor = blue,
 citecolor = blue,
 urlcolor = blue
}

\newcommand{\reflectance}[0]{\mathbb{R}}
\newcommand{\transmittance}[0]{\mathbb{T}}

\newcommand{\mean}[1]{\left\langle#1\right\rangle}

\newcommand{\pd}[0]{\mathcal{P}_V}
\newcommand{\Psingle}[0]{\mathbb{P}}

\newcommand{\change}[1]{#1}

\begin{document}

\preprint{AIP/123-QED}

\title{Dark Matter Haloscope with a Disordered Dielectric Absorber}

\author{Stewart Koppell}
\thanks{Both authors contributed equally to this work.\\  skoppel2@jh.edu, obittencourt@perimeterinstitute.ca}
\affiliation{Research Laboratory of Electronics, Massachusetts Institute of Technology, Cambridge, MA 02139, USA}

\author{Otavio D. A. R. Bittencourt}
\thanks{Both authors contributed equally to this work.\\  skoppel2@jh.edu, obittencourt@perimeterinstitute.ca}
\affiliation{Perimeter Institute for Theoretical Physics, Waterloo, Ontario N2L 2Y5, Canada}
\affiliation{Department of Physics and Astronomy, York University, Toronto, ON M3J 1P3, Canada}

\author{Dip Joti Paul}
\affiliation{Research Laboratory of Electronics, Massachusetts Institute of Technology, Cambridge, MA 02139, USA}

\author{\\Junwu Huang}
\affiliation{Perimeter Institute for Theoretical Physics, Waterloo, Ontario N2L 2Y5, Canada}

\author{Masha Baryakhtar}
\affiliation{Department of Physics, University of Washington, Seattle, WA 98195-1560, USA}

\author{Karl K. Berggren}
\affiliation{Research Laboratory of Electronics, Massachusetts Institute of Technology, Cambridge, MA 02139, USA}

\begin{abstract} 
Light dark matter candidates such as  axions and dark photons generically couple to electromagnetism, yielding dark-matter-to-photon conversion as a key search strategy.  In addition to resonant conversion in cavities and circuits, light dark matter bosons efficiently convert to photons on material interfaces, with a broadband power proportional to the total area of these interfaces. In this work, we make use of interface conversion to develop a new experimental dark matter detector design: the disordered dielectric detector. We show that a volume filled with dielectric powder is an efficient, robust, and broadband target for axion-to-photon or dark-photon-to-photon conversion. 
We perform semi-analytical and numerical studies in small-volume 2D and 3D disordered systems to compute the conversion power as a function of dark matter mass. 
We also discuss the power gathered onto a sensitive photodetector in terms of the bulk properties of the disordered material, making it possible to  characterize the predicted dark-matter-to-photon conversion rate across a wide range of wavelengths. Finally, we propose DPHaSE: the Dielectric Powder Haloscope SNSPD Experiment which is composed of a disordered dielectric target, a veto system, and a photon collection chamber to maximize the coupling between the powder target and a low noise superconducting nanowire single photon detector (SNSPD). \change{With ambitious but realistic improvements to sensor area and detection efficiency at low energy, the projected reach in the 10 meV-eV mass range is sensitive to QCD axion-photon couplings and exceeds current constraints on dark photon dark matter by up to 5 orders of magnitude.}  %The projected reach,  in the 10 meV-eV mass range, is sensitive to QCD axion-photon couplings and exceeds current constraints on dark photon dark matter by up to 5 orders of magnitude.
\end{abstract}

\maketitle

\tableofcontents

\section{Introduction}
\label{sec:introduction}

Clear evidence exists for the presence of dark matter (DM) in the Universe through its gravitational interaction~\cite{Rubin:1970zza,Spergel:2003cb, Aghanim:2018eyx}; its microphysical nature, however, remains a mystery. A large variety of particles have been proposed as DM candidates, motivating a broad range of experimental programs to search for them in the laboratory. In this paper, we propose a novel experimental scheme to search for light bosonic DM in the meV to eV mass range based on disordered dielectric materials.

Light bosons, such as the QCD axion---proposed to solve the strong-CP problem of the Standard Model (SM)~\cite{Peccei+1977,Peccei:1977ur,Weinberg1978,Wilczek1978}---, axion-like particles, and dark photons~\cite{Holdom:1985ag,Okun:1982xi}, are a class of particularly compelling DM candidates. Theoretically, axion-like particles  arise in the paradigm of String Axiverse~\cite{Arvanitaki:2009fg} as a low energy consequence of the existence and the complexity of extra dimensions and topology in string theory. 
In the early universe, axions are produced through the misalignment mechanism~\cite{Abbott:1982af,Dine:1982ah,Preskill:1982cy}, or as remnants of axion string and domain wall evolution~\cite{Sikivie:1982qv,Vilenkin:1982ks,Vilenkin:1984ib,Davis:1986xc}. While for the QCD axion typical early universe production mechanisms generally favor axions with masses in the range of $\mu {\rm eV}$ to $\rm{meV}$~\cite{Borsanyi:2016ksw,Dine:2017swf,Buschmann:2021sdq,Benabou:2024msj},   the QCD axion mass can be at or above the meV scale both in the misalignment~\cite{Arvanitaki:2019rax,Cyncynates:2023esj} and string and domain wall decay~\cite{Gorghetto:2020qws,Saikawa:2024bta,Kim:2024wku,Benabou:2024msj} scenarios.  Dark photons can be produced gravitationally during inflation~\cite{Graham:2015rva}, as well as in the late universe through conversion or decay processes~\cite{Agrawal:2018vin,Co:2018lka,Cyncynates:2023zwj,Cyncynates:2024yxm}. Recent studies of the consistency of dark photons as DM candidates motivate small kinetic mixings and masses in the meV to eV range, so as to avoid vortex formation in the early universe that can deplete dark photon DM before recombination~\cite{East:2022rsi,Cyncynates:2023zwj,Cyncynates:2024yxm}.

The  coupling of light bosons to the SM photon has been a key target for laboratory searches of DM~\cite{Adams:2022pbo,Antypas:2022asj,Baryakhtar:2025jwh}.  In particular,  the axion-photon coupling parametrized by $g_{a\gamma \gamma}$~\cite{Kim:1979if,Shifman:1979if,Dine:1981rt,Zhitnitsky:1980tq} and the dark photon kinetic mixing $\varepsilon$~\cite{Holdom:1985ag, Okun:1982xi} appear in generic models. The laboratory-based search for axion DM has made the most progress to date with boson-to-photon conversion in resonant microwave cavities, in particular with the Axion Dark Matter experiment (ADMX) recently reaching the sensitivity to more weakly-coupled DFSZ axion at $\mu {\rm eV}$ masses~\cite{ADMX:2021nhd}. Other resonant cavity experiments searching in the $\mu$eV range include HAYSTAC~\cite{HAYSTAC:2018rwy,HAYSTAC:2020kwv}, CAPP and CAPP-12TB~\cite{CAPP:2020utb,Yi:2022fmn,CAPP:2024dtx}, ORGAN~\cite{McAllister:2017lkb,Quiskamp:2022pks} and QUAX-$a\gamma$~\cite{Alesini:2020vny}, as well as the SQuAD~\cite{Dixit:2020ymh} demonstrator for the dark photon. 

At higher masses, the conversion in a resonant cavity does not yield enough power to be detectable with current technology, and two main alternative approaches have been developed. The first approach builds upon the idea of coherently summing over signals from a volume spanning multiple wavelengths using resonant detectors. This resonant approach includes proposals for dielectric haloscopes---MADMAX~\cite{Caldwell:2016dcw,Millar:2016cjp, MADMAX:2019pub,MADMAX:2024jnp,Garcia:2024xzc}, LAMPOST~\cite{Baryakhtar:2018doz,Chiles:2021gxk} and Orpheus~\cite{orpheus,Cervantes:2022epl,Cervantes:2022yzp}---as well as plasma haloscopes~\cite{ALPHA:2022rxj} which are currently being developed to search for axions and dark photons across the $\sim 100$ $\mu {\rm eV}$-${\rm eV}$ mass range. Prototypes have demonstrated reach beyond previous constraints for both dark photon~\cite{Chiles:2021gxk,Cervantes:2022epl,Cervantes:2022yzp,MADMAX:2024jnp} and axion~\cite{Garcia:2024xzc} DM. The second approach builds upon the dish antenna idea~\cite{Jaeckel:2013sqa}, in which broadband axion-to-photon conversion can be viewed instead as occurring on material interfaces~\cite{Jaeckel:2013eha}. Prototypes including FUNK~\cite{PhysRevD.102.042001} and BREAD~\cite{BREAD:2021tpx,Hoshino:2025fiz} have been designed to search for axions over a wide range of masses, with the former placing constraints on dark photons and the latter on axions, albeit still within astrophysical bounds.

Despite this concerted experimental effort, the most stringent constraints  in the meV to eV mass range to date largely arise not from laboratory searches but from astrophysical observations. For the axion, the strongest constraints arise from the evolution of the Sun and Globular Cluster stars~\cite{Vinyoles:2015aba,Ayala:2014pea,Dolan:2022kul}, as well as searches for axions emitted from the Sun with the CAST experiment~\cite{alvarezmeconrades}. Recent searches with JWST data have also yielded constraints on dark photons and axions in the eV mass range~\cite{Janish:2023kvi,Roy:2023omw,Pinetti:2025owq,Saha:2025any}. Similarly, for the dark photon, the strongest constraints are set by solar cooling and searches for solar dark-photon radiation in underground dark-matter detectors~\cite{PhysRevD.102.115022,XENON:2021qze}, with the exception of the LAMPOST prototype in the $0.7$-$0.8$~eV range~\cite{Chiles:2021gxk}. Combined with the theoretical interest in this mass range and the rapid progress of single photon detector technology at visible and infrared (IR) frequencies~\cite{Taylor2023,Luskin2023}, new experimental approaches in the optical to far-IR range warrant development.

The choice between a broadband and resonant  conversion target approach depends largely on the experimental noise and on the efficiency of coupling the signal to a detector~\cite{Chaudhuri:2018rqn, Lasenby:2019hfz}. At fixed target volume, the rate of dark-matter-to-photon conversion integrated over a range of frequencies, and hence the scanning speed of the DM parameter space, is proportional to the total surface area of the interface and is independent of resonant enhancement in the absence of noise~\cite{Baryakhtar:2018doz}.  Increasing the quality factor of the setup increases the dark-matter-to-photon conversion rate at the resonance frequency at the expense of reducing the bandwidth, which can be beneficial in the presence of noise.  In the dielectric haloscope, the total surface area and the quality factor both grow with the number of dielectric layers~\cite{Baryakhtar:2018doz, Caldwell:2016dcw}. Theoretically, a resonant approach based on half-wave stacks is the most efficient at converting DM to photons at high frequencies~\cite{Baryakhtar:2018doz, Lasenby:2019hfz}. Moreover, due to 1-dimensional periodicity of the target, the resulting photons are collimated and can be focused down to a small area and low noise photon counter. Experimentally, however, half-wave stacks can be difficult to manufacture, scale up to large volume, and characterize robustly~\cite{Chiles:2021gxk}.

The challenge of fabricating half-wave stacks, combined with the understanding that scanning speed is proportional to the total surface area of interfaces rather than their precise arrangement,  motivates novel broadband detector designs. Concurrently, significant improvements have been made in fabrication of large-area low-dark-count superconducting nanowire single photon detectors (SNSPDs). For example, the LAMPOST experiment demonstrated an SNSPD with area \SI{0.2}{\mm\squared} and a dark count rate of less than 1 per day~\cite{Chiles:2021gxk}. Since then, micrometer-wide superconducting wires fabricated using optical lithography have made cm-scale detectors feasible~\cite{korneeva2018optical}, while the best far-infrared sensitivity for nanowires has been extended to $\SI{29}{\mu m}$ \cite{Taylor2023}. These advancements reduced the necessity of being able to focus the light emitted from the conversion target, leading to the possibility of more complex meta-material structures.

Maximizing the total surface area available for conversion at fixed volume while giving up resonant conversion and signal focusing leads us to a novel DM to photon conversion target which relies on randomized metamaterials in 2- or 3- dimensions. Such a  target can consist of dielectric spheres (``3D powder'') or a set of aligned dielectric fibers (``2D powder''). In order to achieve a smooth, broadband conversion power as a function of DM mass in a densely-packed powder, it is necessary to combine structures with a variety of sizes to suppress interference and resonance effects.\footnote{Reference~\cite{Bloch:2024qqo} recently considered the role of disorder in another context, as applied to axion-nucleon interactions and randomly oriented nuclear spins in a medium.} The smoothing effect of this disorder in higher-dimensional metamaterials stands in contrast to its effect in 1-dimensional structures, such as a dielectric stack; in 1D, introducing a large amount of disorder (such as variable spacing between dielectric layers) leads to unpredictable, rapidly varying power as a function of DM mass~\cite{Baryakhtar:2018doz}.

In this paper we lay an initial theoretical groundwork for DM conversion by disordered dielectric systems. We build upon existing literature on the properties of light scattering in disordered media and apply this to  design and calibrate a DM search. High purity dielectric powders (of, for example, alumina or NaCl) with high transparency and low loss in the mid- and far-infrared, are commercially available and inexpensive. We outline an experimental proposal to take advantage of these ideas and demonstrate that competitive sensitivity can be achieved in an inexpensive and  scalable setup.

The paper is divided as follows: Section \ref{sec:Section II} reviews the main techniques for laboratory search of heavy axions and dark photons. In Section \ref{sec:Section III}, we study DM conversion in disordered media and present our main theoretical results and numerical simulations. In Section \ref{sec:transport}, we discuss the optical properties of the powder which govern the transport of signal photons to the detector. In Section~\ref{sec:exp_setup}, we describe the proposed experimental setup and  we compare our sensitivity with current experiments in Section~\ref{sec:reach_plot}. We conclude in Section \ref{sec:conclusion}. Appendix~\ref{app:symbols} summarizes the symbols used in the text.
 Throughout the main text, we adopt natural units, with $\hbar = c = 1$, whereas SI units are temporarily restored in Apps.~\ref{app:computation} and~\ref{app:collector}, where details of the computations are presented. 

%%%%%%%%%%%%%%%%%%%%%%%%%%%%%%%%%%%%%%%%%%%%%%%%%%%%%%%%%%%%%%%%%%%

\section{A Review of Light-Boson-to-Photon Conversion}\label{sec:Section II}

In this Section we provide a brief review of DM to photon conversion followed by a summary of the main ideas behind some of the proposed laboratory experiments. We discuss both resonant and broadband conversion targets, highlighting that the scanning speed of such experiments scales with the total dielectric interface surface area. This motivates the use of disordered systems to exploit this scaling and enhance conversion efficiency.

The interactions of the axion with photons are parametrized by  the coupling $g_{a\gamma \gamma}$~\cite{Kim:1979if,Shifman:1979if,Dine:1981rt,Zhitnitsky:1980tq} and enter as an additional term in the Lagrangian,
\begin{equation} \label{eq:Laxion}
		\mathcal{L}  \supset \frac{1}{2} (\partial_\mu a)^2 -  \frac{1}{2} m_a^2 a^2
-\frac{1}{4} g_{a \gamma \gamma} a F_{\mu \nu} \tilde{F}^{\mu\nu},
\end{equation}
where $a$ is the axion field, $m_a$ the axion mass, $F^{\mu\nu}$ is the photon field strength and $\tilde{F}_{\mu\nu}$ its dual.  For the dark photon, the leading interaction is specified by the kinetic mixing $\varepsilon$~\cite{Holdom:1985ag, Okun:1982xi},
\begin{equation}
	\mathcal{L} \supset -\frac{1}{4} F_{\mu\nu} F^{\mu\nu}
	- \frac{1}{4} F'_{\mu\nu} F'^{\mu\nu}  
	- \frac{1}{2} \varepsilon F_{\mu\nu} F'^{\mu\nu}  + \frac{1}{2}m_{A'}^2 {A_{\mu}'}^2,
    \label{eq:Ldarkphoton}
\end{equation}
where $A'_{\mu}$ is the dark photon field with mass $m_{A'}$ and field strength $ F'^{\mu\nu}  $.

The resulting equations of motion yield  modified Maxwell equations with an addition DM `current' term,
\begin{equation}
    \partial_\mu F^{\mu\nu} = j^{\nu} + j_{\rm DM}^{\nu}, 
\end{equation} 
where $j_{\rm DM}^{\nu} \equiv g_{a\gamma\gamma} (\partial_{\mu}a )\tilde{F}^{\mu\nu}$ for the axion, and $j_{\rm DM}^{\nu}\equiv \varepsilon m_{A'}^2 {A'}^{\nu}$ for the dark photon. The DM fields have small velocity $v\ll 1$, and therefore, negligible momentum.
Keeping only the leading terms in the DM velocity expansion, and (in the axion scenario) assuming that the applied magnetic field $\vec{B}_{\rm ext}$ is much larger than any other electromagnetic fields, the effect of the DM can be treated as an additional current density, and the Ampere–Maxwell law is modified to~\cite{PhysRevLett.58.1799,Graham:2015rva}
\begin{equation}\label{eq:mod_maxwell}
\begin{split}
&\nabla \times \vec{H} - \partial_t\vec{D} = \vec{j} + g_{a\gamma\gamma} \partial_t a \vec{B}_{\rm ext}. \\
&\nabla \times \vec{H} - \partial_t\vec{D} = \vec{j} + \varepsilon m_{A^\prime}^2 \vec{{A^\prime}}, \\
\end{split}
\end{equation}
where $\vec{D}$ and $\vec{H}$ are the electric displacement field and magnetic field strength, respectively, and $\vec{j}$ is the current of free electrons, zero in our setup with dielectric materials. The massive DM fields oscillate with a frequency that is equal to the DM mass in the same $v\rightarrow 0 $ limit,
\begin{align}\label{eq:fourier}
a(t)=a_0 \cos (m_a t), \quad
\vec{A}^\prime(t)=\vec{A}_0^\prime \cos (m_{A^\prime} t)~,
\end{align}
which can source oscillations of the electromagnetic fields, i.e., photons, at the same frequency. The amplitudes of the axion and dark photon field are
$a_0$ and $\vec{A_0^\prime}$, respectively. 

\begin{figure}   
    \includegraphics[width=.9\textwidth]{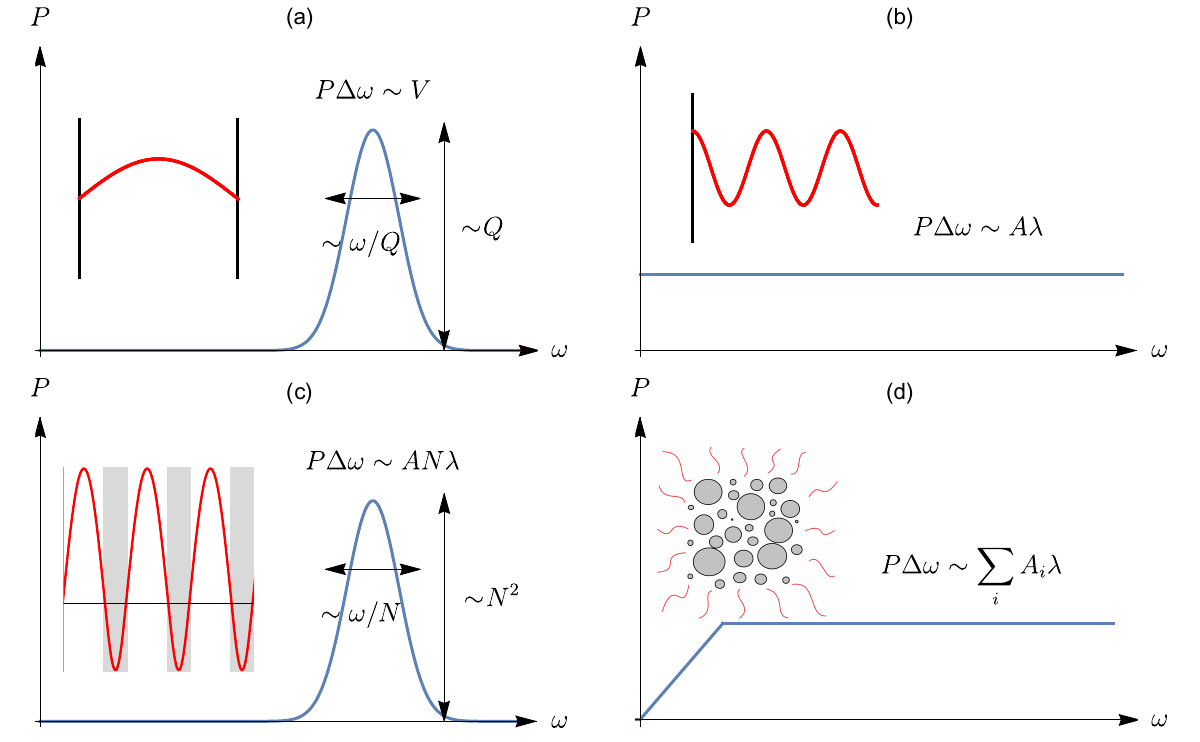}
    \caption{Structures used for DM-photon conversion and a schematic representation of the resulting photon power P as a function of converted photon frequency $\omega$ (wavelength $\lambda$). (a)~Resonant cavities with quality factor $Q$ have peak converted power  proportional to the quality factor over an $\omega/Q$ bandwidth around the resonant frequency;  the scanning speed, $P\Delta \omega$, is a function of the volume in the zero background limit. (b) Multilayer haloscopes with $N$ layer pairs of area $A$ have an effective volume of $\sim AN\lambda$ and a quality factor $\sim N$. (c) Dish antennae have a broadband response and an effective volume of $A\lambda$. (d) Disordered systems (proposed here) can be used to achieve a broadband response. The power scales with the total dielectric interface area $ {\sum_i A_i}$.}
    \label{fig:sketch}
\end{figure}

The massive axions and dark photons cannot convert to massless photons in vacuum due to inability to conserve both energy and momentum~\cite{Baryakhtar:2018doz}. However, in structures that break translational symmetry, the conversion is allowed. Figure~\ref{fig:sketch} provides a schematic overview of these structures along with their respective scanning speed. The first such example is the microwave cavity haloscope, where the light bosons resonantly convert to a microwave cavity mode, with (axion) conversion power given by \cite{PhysRevD.32.2988}
\begin{equation}\label{eq:signalADMX}
     P_{\rm peak}= \mathcal{G} g_{a\gamma\gamma}^2 V B_{\rm ext}^2 \rho_{a} \frac{Q}{m_{a}}.
\end{equation}
This conversion power is proportional to the volume of the cavity $V$ and the quality factor $Q$, over a bandwidth of $\Delta \omega \sim m_a/Q$ around the resonant frequency $\omega_{\rm res} = m_a \simeq V^{-1/3}$. $\mathcal{G}$ is a geometrical factor of $\mathcal{O}(1)$. 

The sensitivity to DM can then be computed by comparing this signal power to the detector and environmental noise. Dark matter haloscope experiments like ADMX, in general, consist of a DM conversion target, which converts the light bosons into photons, and a sensitive photon detector which reads out the result either as excess electromagnetic power or single photons. A sketch of the experimental geometry, frequency profile of the signal,  and scanning speed of cavity experiments is shown in Fig.~\ref{fig:sketch}(a). In absence of background, the scanning speed of the DM parameter space is given by the signal power integrated over frequency, or more schematically, the peak signal power multiplied by the bandwidth, labeled as $P\Delta\omega$. In the case of resonant cavities, in the limit of zero backgrounds, the scanning speed\footnote{For a more detailed expression of scanning speed in the presence of noise see~\cite{Chaudhuri:2018rqn}, Section IV.D. In the presence of noise, scanning speed increases with higher quality factor.}
\begin{equation}
    P\Delta\omega \approx P_{\rm peak} \frac{Q}{m_a} = \mathcal{G} g_{a\gamma\gamma}^2 V B_{\rm ext}^2 \rho_{a},
\end{equation}
is independent of the quality factor.

Another example of an effective translation symmetry breaking structure is the dish antenna~\cite{Horns:2012jf}, proposed as conversion targets that extend the sensitivity to light boson DM to higher frequencies, with a broadband signal power of $P_{\rm sig} \simeq g_{a\gamma\gamma}^2  B_{\rm ext}^2 \rho_{a} A/m_a^2$. Comparing with Eq.~\eqref{eq:signalADMX}, such a scaling corresponds to an effective volume of $V_{\rm eff} \sim A\lambda$, where $\lambda$ is the Compton wavelength of the axion and the converted photon and $A$ is the surface area. As shown in Fig.~\ref{fig:sketch}(b), the scanning speed is proportional to the surface area $A$. Consequently, in \cite{Horns:2012jf}, it is shown that the lack of the resonant enhancement can be compensated by a large surface area. The BREAD mirror design \cite{BREAD:2021tpx}  fits efficiently into a magnet bore geometry and at the same time the emitted signal photons can be focused onto a small, sensitive photon detector. In this approach, it is natural to think of the scaling in terms of the surface area, rather than the effective volume.

In addition to experiments involving DM-to-photon conversion on metallic surfaces, this process can also occur at dielectric-dielectric and dielectric-vacuum interfaces~\cite{Caldwell:2016dcw}. 
Here, it is important to remember that $\vec{D}=\epsilon \vec{E}$ in dielectric materials, where $\epsilon$ is the (relative) dielectric permittivity of the medium (not to be confused with the kinetic mixing of the dark photon $\varepsilon$).
It is convenient to work with an oscillating, spatially uniform electric field $\vec{E}_0$ induced by the DM current density. For DM axions and dark photons, respectively, $\vec{E}_0$ takes the form
\begin{equation}\label{eq:E_0}
 \vec{E}_0 \equiv  g_{a\gamma\gamma}a_0 \vec{B}_{\rm ext}, \quad
 \vec{E}_0 \equiv \varepsilon \vec{E}_0^{\prime} = \varepsilon m_{A^\prime} \vec{A_0^\prime},
\end{equation}
where $\vec{E}_0^{\prime}$ is the amplitude of the dark electric field. This DM induced electric field $\vec{E}_0$ enters the wave equation as a source $\vec{E}_0/\epsilon$ in a dielectric medium (see App.~\ref{app:computation} for more details). At dielectric-dielectric and dielectric-vacuum interfaces, the differences in $\epsilon$ result in differences of this source across the material interface, which lead to DM to photon conversion at the interface~\cite{Caldwell:2016dcw,Baryakhtar:2018doz}.

Periodic layers of dielectric materials as used in the MADMAX~\cite{Caldwell:2016dcw}  and LAMPOST~\cite{Baryakhtar:2018doz} experiments provide an ideal structure for converting DM to photons. These structures can produce a collimated signal that can be focused on a small detector area  while enhancing the peak signal power: 
\begin{equation}\label{eq:signalLAMPOST}
       P_{\rm peak}= \mathcal{G}(n_1,n_2)g_{a\gamma\gamma}^2 B_{\rm ext}^2 \rho_{a} \frac{N^2 A}{m_a^2}, 
\end{equation}
where $N$ is the number of periods of dielectric material. $\mathcal{G}(n_1,n_2) $ is a factor that depends on the refractive indices $n_i$ of the dielectric material and on geometry, maximized by a half-wave stack~\cite{Baryakhtar:2018doz}. The $N^2$ scaling can be understood as a combination of the increased effective volume $V_{\rm eff}\sim A N/ m_a$  (relative to a dish antenna) and a quality factor $Q\sim N$ (bandwidth of $\Delta \omega \sim m_a/N$), in accordance with Eq.~\eqref{eq:signalADMX}. A sketch of the periodic dielectric layers, the corresponding spatial and frequency profile of the signal, and scanning speed of the experimental setup is shown in Fig.~\ref{fig:sketch}(c). 
Theoretically, as some of the authors described in more detail in \cite{Baryakhtar:2018doz} (see also \cite{Millar:2016cjp}), in the absence of noise, the scanning rate of these experiments is also independent of the quality factor,
\begin{equation}
    P \Delta \omega \approx P_{\rm peak} \frac{m_a}{N}\propto g_{a\gamma\gamma}^2 B_{\rm ext}^2 \rho_{a} \frac{N A}{m_a},
\end{equation}
also enhanced by a factor of $N$ compared to the dish antenna setup.

Periodic dielectric structures are commonly available, relatively inexpensive, and clearly provide a promising enhancement in reach to axion and dark photon DM at high masses. Dielectric haloscopes also benefit from a collimated signal which allows focusing of the converted photons onto a small sensitive detector, as for the dish antenna. Challenges include the accurate fabrication and calibration of narrow band periodic dielectric structures over a broad range of light boson masses in a cryogenic environment~\cite{Chiles:2021gxk,Egge:2020hyo,Egge:2023cos,MADMAX:2024pil}. These challenges motivated some of the authors to explore the possibility of broadening the bandwidth at the expense of a reduction of the peak power by utilizing 1D structures with variable spacing across the structure. A chirped stack~\cite{Baryakhtar:2018doz} is one promising way of increasing the bandwidth of a single dielectric structure by up to $30\%$, and could be integrated into a BREAD-like reflective focusing setup~\cite{Baryakhtar_talk,Fan:2024mhm}. In parallel, Ref.~\cite{Baryakhtar:2018doz} found  that due to significant interference effects, dielectric stacks with randomly varying---rather than constant or smoothly varying---layer spacing result in a signal power which is strongly peaked at unknown frequencies and therefore cannot be used as a robust DM conversion target. 

The exploration of variable dielectric spacing scenarios such as the chirped stack and random variations in the system motivates us to understand the scanning speed of the the dielectric haloscope as a sum over emission of the dielectric interfaces,
\begin{equation}\label{eq:signalinterfaces}
    P \Delta \omega \simeq \mathcal{G}(n_1,n_2)g_{a\gamma\gamma}^2 B_{\rm ext}^2 \frac{\rho_{a}}{{m_a}} {\sum_i A_i},
\end{equation}
where the sum runs over all dielectric interfaces and $\sum_i A_i = N A$ for multilayered dielectric haloscopes.  Through a minor change of perspective, this approach leads us to find a new design to search for light bosonic DM. We will see that Eq.~\eqref{eq:signalinterfaces} is a general result beyond planar dielectric haloscopes with periodicity in one dimension (1D haloscopes) and can be generalized to two and three dimensional structures while reducing the detrimental interference in 1D haloscopes.

We will now describe a novel design: the disordered Dielectric Powder Haloscope SNSPD Experiment (DPHaSE). We demonstrate that for a fixed volume, an ensemble of dielectric powder with a range of radii can be used to maximize the area of interfaces, and thus the scanning speed, while maintaining a broadband dark-matter conversion power at short wavelengths by minimizing  interference  between interfaces. This scaling is symbolically shown in Fig.~\ref{fig:sketch}(d) as a function of frequency. The theoretical discussion can be applied to a variety of experimental setups. The only requirement on the property of this dielectric powder is a refractive index that is sufficiently different from unity, and can be made from any common, low loss dielectric material. While we focus on a collection of dielectric spheres (``3D powder''), a similar experiment could be designed with conducting spheres, or aligned dielectric fibers (``2D powder'').

%%%%%%%%%%%%%%%%%%%%%%%%%%%%%%%%%%%%%%%%%%%%%%%%%%%%%%%%%%%%%%%%%%%

\section{Randomized Dielectric Systems as Light Boson conversion Targets}\label{sec:Section III}

In this Section, we lay the theoretical foundations for the conversion of DM to photons in disordered systems. While the main experimental proposal presented in this work is based on 3D powder, the 2D system is tractable with semianalytic approaches, so we discuss both in detail to gain intuition about disordered systems.
\change{Before we delve into the details, it is worth noting that light propagation in random media has been studied extensively in the literature with a variety of methodologies (see \cite{RevModPhys.71.313} for a review). Among these, the inclusion of photon emission at material interfaces 
due to DM-photon conversion is most natural in the formalism we present, based on multiple scattering theory \cite{RevModPhys.23.287} and the T-matrix formalism \cite{PhysRevD.3.825}.
In this formalism, as demonstrated in effectively 1D analysis in \cite{Millar:2016cjp,Baryakhtar:2018doz}, interface DM-photon conversion can be easily implemented as a source term in Eq.~\eqref{eq:recursive}. Such treatment, as we will demonstrate in the following,
can be extended to 2D and 3D systems. Note also that diffraction and near-field effects in 3D systems (open axion haloscopes) have been studied using finite element methods and Fourier optics \cite{Knirck2019AFL}.}

We begin by solving the modified Maxwell's equations, Eq.~\eqref{eq:mod_maxwell}, in the presence of an isolated dielectric fiber in Sec.~\ref{sec:iso2D}, and a sphere in Sec.~\ref{sec:iso3D}. Similar to a planar dielectric interface \cite{Millar:2016cjp}, photons are emitted perpendicular to the surface of the fiber or sphere. The converted power is obtained by integrating the Poynting vector over the full angular domain, and the scanning speed is found to be proportional to the total area of the dielectric-vacuum interface in the limit of high frequencies. Next, we study the conversion by multiple objects, beginning in Sec.~\ref{sec:disorder2D} with a bundle of dielectric fibers (2D powder), where the solution is tractable semi-analytically using the transfer-matrix formalism \cite{Millar:2016cjp,Maystre,Tayeb:97,Felbacq:94,PhysRevB.58.9587,yasumoto2004electromagnetic,yasumoto2018electromagnetic}. In Sec.~\ref{sec:disorder3D}, we extend the analysis to three dimensions using numerical simulations as in \cite{Jeong:2023bqb}. The simulation volumes are limited by our computational power, so in Sec.~\ref{sec:largenumber} we discuss the behavior of larger volumes of powder and summarize the results in Sec.~\ref{sec:summary}.

\subsection{Isolated Dielectric Fiber}\label{sec:iso2D}

We begin by analyzing DM-photon conversion in the presence of a single dielectric fiber. Axion DM can convert into photons in a strong external magnetic field $\vec{B}_{\rm ext}$ applied parallel to the fiber, while dark photon DM can convert to photons in an analogous manner without the applied magnetic field. We take the fiber to have length $L$ and radius $R$, and to be composed of a non-magnetic dielectric material of index of refraction $n$. 
If $L\gg R, \lambda$, where $\lambda$ is the wavelength of the converted photon, the fiber can be considered infinite along the $z$-axis, making the system effectively two-dimensional. 

\begin{figure}\
\newdimen\imageheight
        \settoheight{\imageheight}{% <==========================================
  \includegraphics[width=0.45\textwidth]{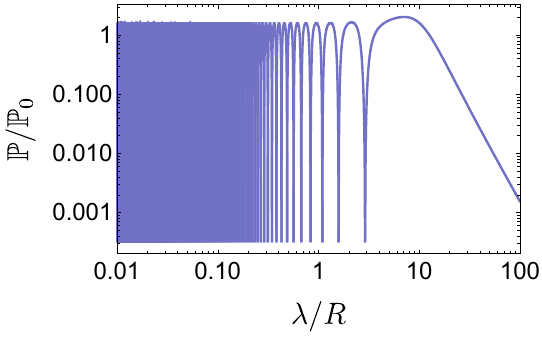}%
}
    \centering
    \begin{subfigure}{0.45\linewidth}\hspace{1cm}(a)
        \centering
        \includegraphics[width=\linewidth]{Figures/power_1_cylinder.pdf} 
    \end{subfigure}
    \begin{subfigure}{0.45\linewidth}\hspace{1cm}(b)
        \centering
        \includegraphics[width=\linewidth, height=\imageheight]{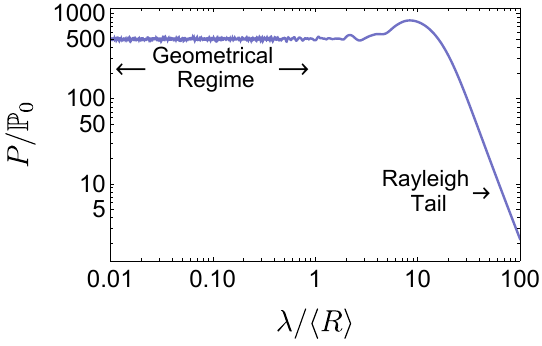} 
    \end{subfigure}
    
    \caption{(a) Power converted by a single dielectric fiber given by Eq.~\eqref{eq:power_single_fiber} normalized by the frequency averaged power $\Psingle_0$ in the geometrical regime in 2D [Eq.~\eqref{eq:P_0}] as a function of the converted photon wavelength in units of the fiber radius $R$. (b) The incoherent sum of the power converted by $500$ fibers also normalized by $\Psingle_0$ as a function of the converted photon wavelength in units of the average fiber radius $\langle R \rangle$. The radii are chosen from a truncated Gaussian distribution of average radius $\langle R \rangle$ and standard deviation $\sigma_R=0.5\langle R \rangle$. The fibers are taken to be made out of alumina, with an index of refraction $n=1.77$.} 
    \label{fig:power_cylinder_no_int}
\end{figure}

We can solve Eqs.~\eqref{eq:mod_maxwell} and~\eqref{eq:fourier} to find the photon wave from DM conversion propagating away from the fiber in the radial direction. The only non-zero component of the electric field is along the $z$-axis, and the magnetic field is transverse to the direction of propagation.
The full computation is presented in App.~\ref{app:DMone} and the resulting Poynting vector is given by Eq.~\eqref{eq:app_poyn_cyl}:
\begin{equation}\label{eq:poynting_single}
    \begin{split}
        \vec{S}(r) &= \frac{E_0^2}{2} \left(1-\frac{1}{n^2}\right)^2  \\ &\times \frac{|J_0^\prime (k_1 R)|^2}{\Big| H_0^{(1)} (k_0 R) J_0^\prime(k_1 R)  - (1/n) J_0(k_1 R) H_0^{(1)\prime} (k_0 R) \Big|^2}\text{Re}\left[ i  H_0^{(1)} ( k_0 r) \left(H_0^{(1)\prime} ( k_0 r)\right)^* \right] \hat{r}~.
    \end{split}
\end{equation}
Here, $\text{Re}[\cdot]$ denotes the real part, $J_n(x)$ and $H_n^{(1)}(x)$ are the Bessel function and Hankel function of the first kind, and $k_0=2\pi/\lambda$ and $k_1=n k_0$ are the converted photon momenta in the vacuum and the dielectric, respectively. The induced field $\vec{E}_0$ was defined in Eq.~\eqref{eq:E_0}. Here, it is parallel to the fiber, such that $\vec{E}_0=g_{a\gamma\gamma}a_0\vec{B}_{\rm ext}$ for axions and $\vec{E}_0=\varepsilon (\vec{E}^\prime)_{\parallel}$, for dark photons, with $(\vec{E}^\prime)_{\parallel}$  the component of the dark photon electric field parallel to the fiber. The radial distance in cylindrical coordinates is given by $r \geq R$ with the fiber centered at the origin. Primes denote taking a derivative with respect to the argument of the function. The power flowing away from the fiber is found by integrating Eq.~\eqref{eq:poynting_single} over a cylindrical surface encompassing the fiber. Since the Poynting vector has no angular- or $z$-dependence, the integration is straightforward and we obtain the power per unit length at a distance $r$ from the fiber,
\begin{equation}\label{eq:power_single_fiber}
    \frac{\Psingle(r)}{L}=2\pi r \left|\vec{S}(r) \right| ~.
\end{equation}
Here we use $\Psingle$ to refer to the power converted by a single object, and reserve $P$ for the power converted by multiple objects as discussed in the following subsections. The power as a function of wavelength in Eq.~\eqref{eq:power_single_fiber} for radii $r\gg R$ is shown in Fig.~\ref{fig:power_cylinder_no_int}(a).

The Poynting vector of the photon converted by DM, given in Eq.~\eqref{eq:poynting_single}, has a behavior similar to the usual scattering of incoming light by a dielectric fiber. To make this more explicit, we take two interesting limits of Eq.~\eqref{eq:poynting_single}: when the wavelength is much smaller or larger than the radius of the fiber, 
\begin{equation}\label{eq:cyn_approx}
\begin{split}
    \frac{\Psingle_G}{L} & \approx \frac{E_0^2}{2} \left(1-\frac{1}{n^2}\right)^2 \frac{2\pi R}{1- \frac{1}{n^2} - \frac{2}{n^2}\frac{1}{\sin(\frac{4 n \pi R}{\lambda})-1}} \hspace{0.74 cm} \text{if} ~~ \lambda \ll 2\pi R~, \\
    \frac{\Psingle_R}{L} & \approx \frac{E_0^2}{2} n^4\left(1-\frac{1}{n^2}\right)^2 \frac{\pi^2}{4} R^4 \left(\frac{2\pi}{\lambda}\right)^3 \hspace{1.93 cm} \text{if} ~~ \lambda \gg 2\pi R~,
\end{split}
\end{equation}
where the subscripts $G$ and $R$ refer to the geometrical and Rayleigh scattering regimes of light, respectively. For large wavelengths, the power has a $\lambda^{-3}$ scaling, analogous to Rayleigh scattering of light in 2D. This scaling arises because the electric field induced by a spatially homogeneous DM field is itself homogeneous [see Eq.~\eqref{eq:app_hom+part} for the electric field solution], resembling a plane wave with a very long wavelength.
At smaller wavelengths, we have a highly oscillatory function, similar to the geometrical optics regime where the diffraction through the scatterer leads to multiple maxima and minima. Notice, however, that the exact form of the oscillations is not the same; the DM conversion power oscillates as the Bessel function in the numerator of Eq.~\eqref{eq:poynting_single}, while the scattering of light oscillates as the fiber T-matrix given in Eq.~\eqref{eq:app_T}.

Taking the average over frequency in the geometrical regime, denoted by $\langle \cdot \rangle_\omega$, gives
\begin{equation}\label{eq:average_frequency}
    \frac{\langle \Psingle_G \rangle_\omega}{L} = \frac{E_0^2}{2} \left(1-\frac{1}{n}\right)\left(1-\frac{1}{n^2}\right) 2\pi R ~.
\end{equation}
Notice that the scanning speed as defined in Sec.~\ref{sec:Section II} is just $\langle \Psingle_G \rangle_\omega$ multiplied by the bandwidth $\Delta \omega$ and scales with the surface area of the fiber $\langle \Psingle_G\rangle_\omega \Delta \omega \propto 2\pi RL$, recovering the expected scaling presented in Eq.~\eqref{eq:signalinterfaces}. This behavior is the same as $\langle \Psingle_G \rangle_\omega \Delta \omega$ found in~\cite{Baryakhtar:2018doz} for effectively 1D dielectric structures. However, unlike in effective 1D systems, where the oscillations in frequency cannot be easily smoothed out in realistic situations, these oscillations in frequency can now be smoothed out by averaging over many dielectric fibers of varying radii. For example, assuming that the fiber radii are chosen from a Gaussian distribution of mean $\langle R \rangle$, the average can be computed by expanding the first line of Eq.~\eqref{eq:cyn_approx} in a Fourier series, multiplying by the Gaussian distribution and integrating. After some manipulation, it can be shown that
\begin{equation}\label{eq:average_radius}
    \frac{\langle \Psingle_G \rangle_R}{L} \approx \frac{E_0^2}{2} \left(1-\frac{1}{n}\right)\left(1-\frac{1}{n^2}\right) 2\pi \langle R \rangle + \mathcal{O}(\lambda/\langle R\rangle) ~,
\end{equation}
where $\langle \cdot \rangle_R$ denotes averaging over radii. In the limit of $\lambda/2\pi \langle R \rangle \rightarrow 0$, we recover the same result as the average over frequency, Eq.~\eqref{eq:average_frequency}, if the fiber has a radius $R=\langle R \rangle$. This is analogous to the ``area law" derived in \cite{Millar:2016cjp}. We take this average to be the reference power $\Psingle_0$, used in Fig.~\ref{fig:power_cylinder_no_int},~\footnote{A truncated Gaussian distribution is used in the numerical studies to ensure non-negative powder radii. This truncation introduces errors of up to $3\%$ in the results shown in the figures, which we disregard in our analysis.}
\begin{equation}\label{eq:P_0}
    \frac{\Psingle_0}{L}\equiv\frac{E_0^2}{2} \left(1-\frac{1}{n}\right)\left(1-\frac{1}{n^2}\right) 2\pi \langle R \rangle~.
\end{equation}
Note that the assumption of a Gaussian distribution of radii can be alleviated. We expect that the average power (over radii) to be equal to the reference power in Eq.~\eqref{eq:P_0} as long as the distribution is smooth on scales of order $\lambda$. We will consider non-Gaussian distributions in Sec.~\ref{sec:disorder3D}.

If we incoherently sum over the power converted by $N$ fibers with radii chosen from a Gaussian distribution, for large enough $N$, the resulting power is simply $N$ times the average power. For the geometrical regime this is $N$ times Eq.~\eqref{eq:average_radius}, while the sum in the Rayleigh regime is straightforward to compute from Eq.~\eqref{eq:cyn_approx}. The incoherent sum of the power converted by multiple fibers, in the geometrical and Rayleigh regimes are, respectively,
\begin{equation}\label{eq:power_2D}
\begin{split}
    &\frac{\langle P_G \rangle}{L} = N\frac{\langle \Psingle_G \rangle}{L} = N\frac{E_0^2}{2} \left(1-\frac{1}{n}\right)\left(1-\frac{1}{n^2}\right) 2\pi \langle R \rangle,\\
    &\frac{\langle P_R \rangle}{L} = N\frac{\langle \Psingle_R \rangle}{L} = N\frac{E_0^2}{2}  n^4\left( 1-\frac{1}{n^2} \right)^2 \frac{\pi^2}{4} \langle R^4 \rangle \left( \frac{2\pi}{\lambda} \right)^3~,
\end{split}
\end{equation} 
where we drop the subscript $R$ here and all averages are understood to be over the radius unless specified explicitly.\footnote{Note that the power scales as $n^4$ in the limit of $n\rightarrow \infty$ in the Rayleigh regime. This unphysical divergence arises from an inconsistent order of limits: we expanded Eq.~\eqref{eq:poynting_single} for small $n R/\lambda$, yet later took $n\rightarrow \infty$. However, when the $n\rightarrow \infty$ limit of Eq.~\eqref{eq:poynting_single} is taken before the $R/\lambda \rightarrow 0$ limit, we get the standard result of ${\langle \Psingle_R \rangle}/{L} \propto 1/\left|\log[R/\lambda]\right|^2$ in 2D, which is not divergent in the $n\rightarrow \infty$ limit \cite{jackson}.}
 This analytical formula for the converted power agrees with the numerical result of incoherently summing over many dielectric fibers shown in Fig.~\ref{fig:power_cylinder_no_int}(b). 
As we expected, the average converted power in the geometrical regime is broadband after averaging over $N= 500 $ dielectric fibers, and is proportional to the average total surface area of the interfaces: $\langle P_G \rangle \propto N 2\pi \langle R \rangle L$.
This incoherent sum offers a good first approximation to photon converted by multiple fibers, which we will further discuss in Sec.~\ref{sec:disorder2D}. In particular, in the geometrical regime, the broadband power of converted photons will continue to scale with the average total dielectric surface area when interference effects are taken into account. However, before studying the interference effects, we will first consider DM-photon conversion by a 3D isolated dielectric sphere.

\subsection{Isolated Dielectric Sphere}\label{sec:iso3D}

\begin{figure}
\newdimen\imageheight
        \settoheight{\imageheight}{% <==========================================
  \includegraphics[width=0.45\textwidth]{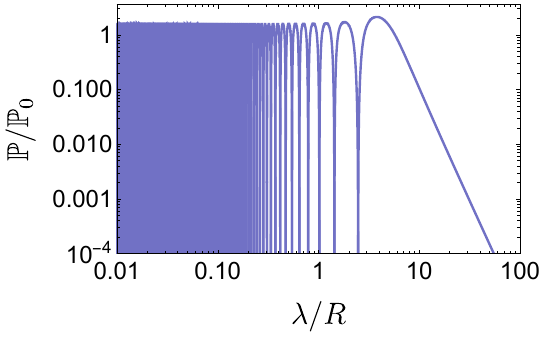}%
}
    \centering
    \begin{subfigure}{0.45\linewidth}\hspace{1cm}(a)
        \centering
        \includegraphics[width=\linewidth]{Figures/power_1_sphere.pdf} 
    \end{subfigure}
    \begin{subfigure}{0.45\linewidth}\hspace{1cm}(b)
        \centering\vspace{-0.001cm}
        \includegraphics[width=\linewidth, height=\imageheight]{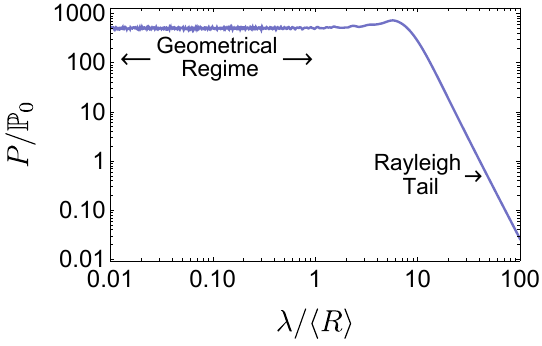} 
    \end{subfigure}
    
    \caption{(a) Power converted by a single dielectric sphere normalized by the frequency average in the geometrical regime in 3D [Eq.~\eqref{eq:P0_3D}, with $ \Psingle_0 \propto \langle R^2\rangle = R^2$ for a single object of radius R],  as a function of the converted photon wavelength in units of the sphere radius $R$.  (b) Normalized power converted by the incoherent sum of $500$ spheres with radii chosen from a truncated Gaussian distribution with radius average $\langle R\rangle$ and standard deviation $\sigma_R = 0.5 \langle R \rangle$,  as a function of the converted photon wavelength in units of $\langle R\rangle$ . In this case  $ \Psingle_0 \propto \langle R^2\rangle = R^2+\sigma_R^2$. The spheres are made of alumina with an index of refraction $n=1.77$.}
    \label{fig:power_sphere_no_int}
\end{figure}

Consider a non-magnetic dielectric sphere of radius $R$ and index of refraction $n$ in vacuum, in the background of a DM field. In the axion case, a strong external magnetic field $B_{\rm ext}$ is also applied. The same conclusions drawn in Sec.~\ref{sec:iso2D} apply to this 3D scenario. As detailed in App.~\ref{app:computation}, we solve the modified Maxwell's equations, Eq.~\eqref{eq:mod_maxwell}, to find the Poynting vector of the photons converted by a dielectric sphere, which is given by Eq.~\eqref{eq:app_poyn_sphere}, 
\begin{equation}\label{eq:poyn_single_sphere}
    \begin{split}
        \vec{S}(r) = &\frac{E_0^2}{2} \left(1-\frac{1}{n^2}\right)^2 \sin^2\theta\\
        \times &\Bigg| \frac{n k_1 R j_1(k_1 R)}{h^{(1)}_1(k_0 R) [x j_1(x)]^\prime|_{x=k_1 R} - n^2 j_1(k_1 R) [x h^{(1)}_1(x)]^\prime|_{x=k_0 R}} \Bigg|^2 \text{Re}\left[ i \frac{h_1^{(1)*}( k_0 r) }{k_0 r} [x h_1^{(1)}(x)]^{\prime}\Big|_{x=k_0r} \right] \hat{r},
    \end{split}
\end{equation}
The spherical Bessel function and spherical Hankel function of the first kind are denoted by $j_n(x)$ and $h_n^{(1)}(x)$, respectively. The radial coordinate in spherical coordinates is $r\geq R$, and $\theta$ is the angle between the radial direction $\hat{r}$, and the direction of the applied external magnetic field in the axion case, or the direction of the dark photon electric field in the dark photon case. The wave vectors in the vacuum and inside the sphere are, respectively, $k_0=2\pi/\lambda$ and $k_1=nk_0$ and $E_0$ is defined in Eq.~\eqref{eq:E_0}. The power can be found by integrating Eq.~\eqref{eq:poyn_single_sphere} over the full angular domain and is shown in Fig.~\ref{fig:power_sphere_no_int}(a).

We once again expand the power in the geometrical and Rayleigh regimes, keeping the same notation as before, 
\begin{equation}\label{eq:sphere_approx}
\begin{split}
    \Psingle_G & = \frac{E_0^2}{2} \left(1-\frac{1}{n^2}\right)^2  \frac{8\pi}{3} R^2 \frac{1}{1+\frac{1}{n^2} \tan^2(\frac{2\pi n R}{\lambda})} \hspace{0.75 cm} \text{if} ~~ \lambda \ll 2\pi R~, \\
    \Psingle_R & = \frac{E_0^2}{2} \left( \frac{n^2 - 1}{n^2 + 2 } \right)^2  \frac{8\pi}{3} R^6 \left( \frac{2 \pi}{\lambda} \right)^4  \hspace{2.5 cm} \text{if} ~~ \lambda \gg 2\pi R ~.
\end{split}
\end{equation}
For large wavelengths, we recover the usual $1/\lambda^4$ Rayleigh scaling in 3D, while for short wavelengths, the oscillations of the spherical Bessel function dominate. Taking the average of $\Psingle_G$ over frequency or over radius, we find,
\begin{equation}
\begin{split}
    &\langle \Psingle_G\rangle_\omega = \frac{E_0^2}{2} \left(1-\frac{1}{n}\right)\left(1-\frac{1}{n^2}\right)  \frac{8\pi}{3} R^2,\\
    &\langle \Psingle_G \rangle_R = \frac{E_0^2}{2} \left(1-\frac{1}{n}\right)\left(1-\frac{1}{n^2}\right)  \frac{8\pi}{3} \langle R^2 \rangle + \mathcal{O}(\lambda/\langle R\rangle)~.
\end{split}
\end{equation}
Similar to Sec.~\ref{sec:iso2D}, both averages agree in the $\lambda/2\pi \langle R \rangle \rightarrow 0$ limit,
suggesting that a broadband DM converted photon power that saturate the scaling of Eq.~\eqref{eq:signalinterfaces} can also be achieved by averaging over many dielectric spheres of varying radii, that is, $\langle \Psingle_G \rangle_R \propto 4 \pi \langle R^2 \rangle$, the average surface area of the dielectric powder. 

We take this average to be the reference power for Fig.~\ref{fig:power_sphere_no_int}, for a given distribution of radii,
\begin{equation}\label{eq:P0_3D}
    \Psingle_0= \frac{E_0^2}{2} \left(1-\frac{1}{n}\right)\left(1-\frac{1}{n^2}\right)  \frac{8\pi}{3} \langle R^2 \rangle~.
\end{equation}
In the limit of large $N$, the incoherent sum of the power converted by $N$ spheres is given by $N$ times the average in Eq.~\eqref{eq:P0_3D} (here we again drop the subscript $R$),
\begin{equation}\label{eq:power_3D}
\begin{split}
    &\langle P_G \rangle= N\langle \Psingle_G\rangle = N \frac{E_0^2}{2} \left(1-\frac{1}{n}\right) \left(1-\frac{1}{n^2}\right) \frac{8\pi}{3}  \langle R^2 \rangle ~, \\
    &\langle P_R \rangle = N\langle \Psingle_R\rangle = N \frac{E_0^2}{2} \left( \frac{n^2-1}{n^2+2} \right)^2 \frac{8\pi}{3} \langle R^6 \rangle  \left( \frac{2\pi}{\lambda}\right)^4~.
\end{split}
\end{equation} 
This analytical formula for the converted power agrees with the numerical result of incoherently summing over many dielectric spheres shown in Fig.~\ref{fig:power_sphere_no_int}(b). 
As we expected, the average converted power in the geometrical regime is broadband after averaging over $N= 500 $ dielectric spheres, and is proportional to the average total surface area of the interfaces $\langle P_G \rangle \propto N 4\pi\langle R^2 \rangle$.
In the following, we address the problem of whether these simple scalings of the converted power
in the $\lambda \gg 2\pi R$ and $\lambda \ll 2\pi R$ limits remain true when interference effects are taken into account,  first for dielectric medium consisting of fibers (2D powder) semi-analytically, before solving for dielectric medium consisting of spheres (3D powder) fully numerically.

\subsection{Disordered Dielectric System in 2D}\label{sec:disorder2D}

In this subsection, we study the DM-photon conversion inside an effectively 2D disordered dielectric system.  This analysis aims to demonstrate that the key features found when incoherently summing the power of many fibers in Sec.~\ref{sec:iso2D} remain qualitatively the same when interference is taken into account, with some minor differences. These features are:
\begin{itemize}
    \item At short wavelengths ($\lambda \ll 2\pi \langle R\rangle $) we enter the geometrical optics regime. The photons propagate diffusely through the powder and the oscillations due to interference average out. The conversion power is proportional to the total surface area and independent of frequency.
    \item At long wavelengths ($\lambda \gg 2\pi \langle R\rangle$) we enter the Rayleigh regime. The power scales as $\lambda^{-3}$ (notice we are in 2D). However, the introduction of a new length scale, namely the distance between each fiber, changes the onset of the Rayleigh tail.
    \item The converted power can still be compared to the scattering of incident light. This is important since we can characterize the transport of light in this medium by the same experimental measurements used to characterize the optical response of the powder. At wavelengths comparable to the fiber radius ($\lambda \simeq 2\pi \langle R\rangle$), interference effects can lead to phenomena such as resonances, bandgaps and localization, which we wish to avoid.
\end{itemize}

With these considerations in mind, we now present the problem at hand. Consider a bundle of $N$ parallel non-magnetic dielectric fibers, aligned and taken to be infinite along the $z$-axis.  Each fiber has a distinct radius, $R_l$, and index of refraction, $n_l$, and is located at an arbitrary position $(r_l,\phi_l)$ in cylindrical coordinates. The fibers are assumed to be in vacuum;  an external magnetic field $B_{\rm ext}$ is applied along the $z$-axis in the axion case. As discussed in Sec.~\ref{sec:iso2D}, each fiber converts the background DM field into photons, which then travel through the disordered media until they exit the bulk. Our goal is to solve Eq. \eqref{eq:mod_maxwell} and compute the resulting converted power by the dielectric powder. We base our method on the multiple scattering formalism, the scattering of light by randomly arranged parallel dielectric fibers has already been studied in this framework \cite{Felbacq:94,Maystre,Tayeb:97,PhysRevB.58.9587,yasumoto2004electromagnetic,yasumoto2018electromagnetic}. The distinction here is that in the presence of DM, the fibers themselves act as both the sources and scatterers for the photons, rather than just scatterers being illuminated by an incoming photon wave.

The electric field of the converted photon will point in the same direction as the external magnetic field \cite{Millar:2016cjp}, which we take to be the $z$-axis, while its other components are zero. Further, each fiber is illuminated by the photons converted by the surrounding fibers, which are subsequently scattered \cite{Felbacq:94}.
The scattered electric field also points along the $z$-axis. Therefore, the total electric field emitted from a fiber, consisting of both the converted and scattered contributions, is directed along the $z$-axis. Since it must propagate away from the fibers, it can be expanded in Hankel functions. Therefore, the total emitted field can be written as  
\begin{equation}\label{eq:totalfield}
    \vec{E}_{\rm emitted} = \sum_l \vec{E}_{\rm emitted}^l = \sum_l \sum_m D_{l,m} H_m (k_0 r_l) e^{im\phi_l} \hat{z},
\end{equation}
where $l$ runs over all fibers in the configuration and $m$ over the harmonics of the Hankel function. $\vec{E}_{\rm emitted}^l$ is the electric field emitted by the $l$-th fiber and $k_0$ is the wave vector in vacuum. The coefficients $D_{l,m}$ contain information about the signal photon, and need to be determined.

This determination can be achieved by noting that the incoming wave on a given fiber is the sum of the scattered wave by every other fiber, which is given by the same coefficients $D_{l,m}$ times a coupling matrix, $G_{m,n,j,l}$, that propagates the wave from one fiber to the other \cite{Maystre}. The relationship of the incoming and scattered wave is given by the known T-matrix of the fiber, allowing us to find a recursive relation that can be solved numerically, 
\begin{equation}\label{eq:recursive}
    \textbf{D}_j = \textbf{T}_j \cdot \left( \sum_{l\neq j} \textbf{G}_{j,l} \textbf{D}_l \right) + \textbf{S}_j ~.
\end{equation}
The term in parentheses represents the incoming wave, which is the emitted wave by the $l$-th fiber transformed into an incoming wave in the $j$-th fiber by the matrix $\textbf{G}_{j,l}$. This incoming wave is then scattered by the $j$-th fiber, as encoded in the $\textbf{T}_j$ matrix, and added to the converted DM field, given by the $\textbf{S}_j$ matrix.
The bold letters are matrices in harmonic space (e.g. $\textbf{D}_j$ denotes $D_{m,j}$ while $\textbf{G}_{j,l}$ denotes $G_{m,n,j,l}$). The indices $j,l$ run over all the fibers. $\textbf{T}_j$ is the usual T-matrix for the $j$-th fiber, $\textbf{G}_{j,l}$ encodes the geometry of the configuration and depends only on the positions of the $j$-th and $l$-th fibers \cite{Maystre}. See App.~\ref{app:computation} for explicit expressions and a detailed derivation. Finally, $\textbf{S}_j$ is the term sourced by DM conversion and is given by [see Eq.~\eqref{eq:poynting_single}],
\begin{equation}
    S_{m,j} = \delta_{m,0} E_0 \left(1-\frac{1}{n_j^2}\right) \frac{J_0^\prime (k_j R_j)}{H_0^{(1)} (k_0 R_j)  J_0^\prime(k_j R_j)- (1/n_j) J_0(k_j R_j) H_0^{(1)\prime} (k_0 R_j)}~,
\end{equation}
with $k_0$ and $k_j$ being the wave vectors in vacuum and inside the $j$-th fiber, respectively. Notice that the wave converted by a single fiber only contains the $m=0$ harmonic, while waves of different $m$ will mix due to the presence of other fibers, the first term of Eq.~\eqref{eq:recursive}. Physically, the fiber emits photons in the radial direction.
These photons, however, are not generally incident on the other fibers in the radial direction. Therefore, the angular distribution changes and higher $m$ modes get populated when the wave scatters with the other fibers.

To numerically solve Eq.~\eqref{eq:recursive} we begin by constructing the geometry. We specify the positions, radii, and refractive indices of each fiber. We take all the fibers to have the same index of refraction, $n$. The system is built using a Python script that randomly selects the radii of the fibers from a truncated Gaussian distribution (i.e., a Gaussian distribution with strictly positive values) with mean value $\langle R \rangle$ and standard deviation $\sigma_R$. The fibers are then randomly placed inside a box with cross-sectional area $A_\text{box}=V_{\rm box}/L$. Due to the limited computational resources, this volume is much smaller than the proposed experimental setup. Therefore, the results presented in this subsection are better interpreted as the power converted by a small volume element of dielectric powder in the experimental apparatus.
A discussion of the extrapolation to very large volume will be presented in Sec.~\ref{sec:largenumber}.
Once we have the geometry, we build the required matrices and solve for $D_{j,m}$. Then, we can substitute $D_{j,m}$ into Eq.~\eqref{eq:totalfield} to find the electric field at all spatial locations, while the magnetic field can be found with Faraday's law, $\vec{B}_t = i \hat{z} \times \nabla_{t} \vec{E}_z$, where $\nabla_{t}$ denotes the transverse derivative (see definitions in App.~\ref{app:computation}). The Poynting vector is then computed and integrated over the whole angular direction,  from $0$ to $2\pi$. Since we integrate over the whole angular domain, the power is independent of the distance from the powder. The code used is an adaptation of existing literature (see, for example, Chapter 3 of \cite{10.1088/978-1-6817-4301-1}).

Apart from the index of refraction, the box volume and the radius distribution, another variable of our problem is the filling factor, $f = N \pi \langle R^2\rangle/A_{\rm box}$, with $N$ being the number of fibers. Despite the more complicated theoretical framework, we still expect the power to be given by Eq.~\eqref{eq:power_2D}, as was discussed at the end of Sec.~\ref{sec:iso2D}. Here the averages are understood to be over the radius, in terms of the new variables we have,
\begin{equation}\label{eq:power_2D_final}
    \begin{split}
        &\langle P_G \rangle = \frac{E_0^2}{2} \left( 1-\frac{1}{n} \right)\left( 1-\frac{1}{n^2} \right) f V_{\rm box} \frac{2\langle R \rangle}{\langle R^2 \rangle},\\
        &\langle P_R \rangle =G(f) \frac{E_0^2}{2} n^4 \left( 1-\frac{1}{n^2} \right)^2 f V_{\rm box} \frac{\pi}{4}\frac{\langle R^4 \rangle}{\langle R^2 \rangle}  \left( \frac{2\pi}{\lambda} \right)^3.
    \end{split}
\end{equation}
Note that $fV_{\rm box}$ gives us the total volume occupied by the fibers, while the filling factor dependent geometrical factor $G(f)$ will be defined and discussed later when we describe the Rayleigh regime.

\begin{figure}[th!]
    \centering
    
    \begin{subfigure}{0.6\linewidth}
        \centering
        \hspace{-3cm}\includegraphics[width=\linewidth]{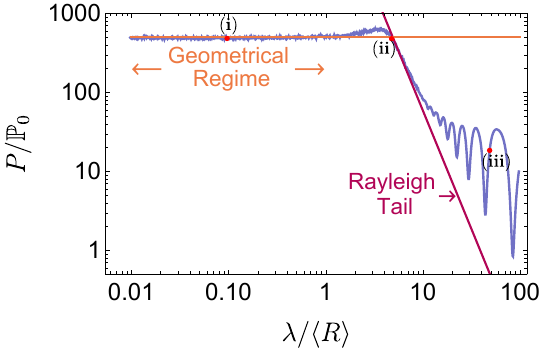}
    \end{subfigure}
    
    \begin{subfigure}{\linewidth}
        \centering
        % trim={<left> <lower> <right> <upper>}
	 \includegraphics[ clip, width=\textwidth]{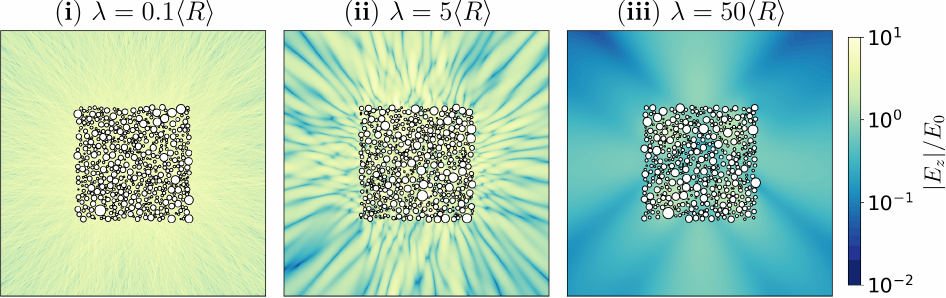}
    \end{subfigure}

    \caption{(a) Power converted by a benchmark configuration of aligned dielectric fibers, normalized by $\Psingle_0$, as a function of converted photon wavelength normalized by the average fiber radius $\langle R \rangle$ as in Fig.~\ref{fig:power_cylinder_no_int}. The fiber radii are chosen from a truncated Gaussian distribution with average radius $\langle R \rangle$ and standard deviation $\sigma_R = 0.5 \langle R \rangle$. All fibers are made of alumina with index of refraction $n=1.77$ and are randomly distributed in a box of cross-sectional area $20\sqrt{10}~\langle R\rangle ^2$ and filling factor $f=0.5$. The geometrical and Rayleigh regime are easily identifiable. The electromagnetic field amplitude $|E_z|$ normalized to the reference field amplitude as induced by the DM $E_0$ [Eq.~\eqref{eq:E_0}] is shown at three example wavelengths, labeled (b-d) in panel (a). (b) In the geometrical regime, the photons are completely randomized by the time they leave the bulk, leading to fast-varying maxima and minima that form a speckle pattern. This is the reason why the power is seen as uniform for a sufficiently large detector. (c) For $\lambda\simeq  2\pi  \langle R\rangle$, localization and bandgap effects are most prevalent, see Sec.~\ref{sec:largenumber} for a more detailed discussion. (d) Wavelengths close to the size of the simulation begin to excite the modes of the box, which leads to the oscillations in the Rayleigh tail as visible in (a).  These are not physically relevant as a realistic experiment will have an extent much larger than the individual particle sizes. }
    \label{fig:electric_field}
\end{figure}

To demonstrate that this prediction is justified, the converted power of a benchmark configuration is solved numerically, and compared to Eq.~\eqref{eq:power_2D_final} in Fig.~\ref{fig:electric_field}, which also shows the electric field of the configuration at different wavelengths. The physical picture is as follows: as light is converted within the bulk of the powder, it must travel through a disordered medium before escaping. If $\lambda \ll 2\pi R$, the direction of propagation and the phase of the photons are randomized by the time they exit the bulk. This regime can be modeled by photon diffusion, see Sec.~\ref{sec:transport} for further details. This randomization leads to an electric field with an angular distribution exhibiting multiple maxima and minima, forming a speckle pattern \cite{Goodman:76}, as shown in Fig.~\ref{fig:electric_field}(b). Consequently, for a sufficiently large detector positioned outside the configuration, the power it measures averages out the electric field fluctuations to a constant value. By increasing the detector size to cover all the angles, the variation of the oscillations decreases, and we recover a power similar to Fig.~\ref{fig:power_cylinder_no_int}(b). To capture this effect the simulation is run for 100 configurations and only the configuration-averaged results are presented in Fig.~\ref{fig:electric_field}(a). This approach mimics the behavior of increasing the total number of particles or increasing the detector area.

\begin{figure}[th]
\newdimen\imageheight
        \settoheight{\imageheight}{% <==========================================
  \includegraphics[width=0.49\textwidth]{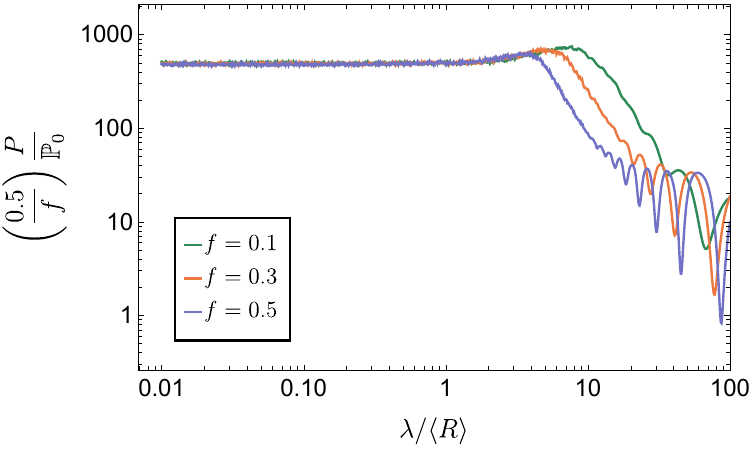}%
}
    \begin{subfigure}[b]{0.49\textwidth}\hspace{1.3cm}(a)
        \centering
        \par\hspace{1cm}
        \includegraphics[width=\textwidth]{Figures/changfill.pdf}
        \label{fig:sub2}
    \end{subfigure}
    % Second row
    \vspace{0.3cm}
    \begin{subfigure}[b]{0.49\textwidth}\hspace{1.3cm}(b)
        \centering
        \par\hspace{1cm}
        \includegraphics[width=\textwidth, height=\imageheight]{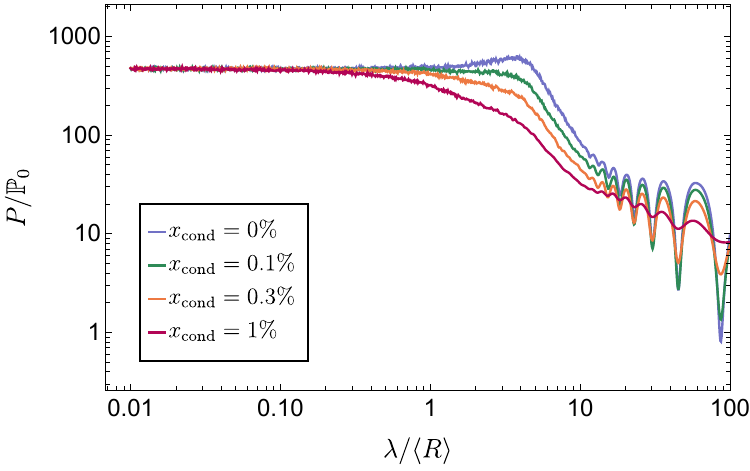}
        \label{fig:sub3}
    \end{subfigure}
    \hfill
    \begin{subfigure}[b]{0.5\textwidth}\hspace{1.35cm}(c)
        \centering
        \par\hspace{1cm}
        \includegraphics[width=\textwidth]{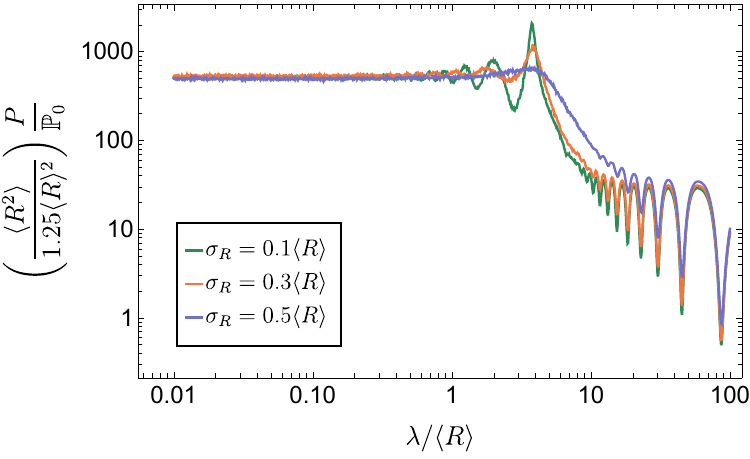}
        \label{fig:sub4}
    \end{subfigure}
    
    \caption{ We illustrate the dependence of the power converted by fibers in a box as in Fig.~\ref{fig:electric_field}(a) on  (a) the filling fraction $f$, (b) dielectric losses and (c) variance in radii. (a) As the filling factor $f$ increases the Rayleigh tail begins at a shorter wavelength, as characterized by the geometrical factor $G(f)$. The power at short wavelengths is proportional to $\mathbb{P}_0 f$, Eq.~\eqref{eq:power_2D_final}. (b) Same configuration as in (a) but at a constant filling factor $f=0.5$. We include dielectric losses as the imaginary part of the index of refraction, see Eq.~\eqref{eq:complex_index}, with $x_{\rm cond}$ the conductivity in units of $1/\langle R\rangle$. Larger losses lead to suppression of power in the intermediate wavelength regime. (c) Same as (a) at a constant filling factor of $f=0.5$ and varying  standard deviation of the distribution of radii $\sigma_R$. Smaller variations produce sharper features and interference at intermediate wavelengths. Note that the cross-sectional area of a cylinder is proportional to $\langle R^2\rangle$. Since the filling factor $f$ and mean radius $\langle R \rangle$ are held fixed, the three curves in panel (c) correspond to systems with different total numbers of cylinders $N\propto 1/\langle R^2\rangle$. Therefore, we normalized the ratio $ P/\mathbb{P}_0$ by the 
    factor $\langle R^2\rangle/1.25\langle R\rangle^2$ so that the curves coincide at short wavelengths, and that the curve for $\sigma_R = 0.5\langle R \rangle$  matches the results in Fig.~\ref{fig:electric_field}.}
    \label{fig:power_int}
\end{figure}

The Rayleigh regime differs from the incoherent sum in Eq.~\eqref{eq:power_2D_final}. By arranging the fibers on a random grid, we introduce a new length scale into the system, namely, the size of the gaps between the fibers. If the filling factor is increased while keeping $R$ and $\sigma_R$ constant, these gaps eventually become smaller than the size of an average fiber. More importantly, the gaps with irregular shapes can have features on length scales much smaller than their sizes. In this limit, the system behaves, instead of dielectric fibers in vacuum, more like vacuum gaps in a dielectric medium. Since the Rayleigh suppression is determined by the smallest length scale, the Rayleigh tail begins at a shorter wavelength compared to the configurations with a lower filling factor, see Fig.~\ref{fig:power_int}(a). This offset has to be measured experimentally by shining light in the configuration. We include this effect in a filling-factor-dependent geometrical factor $G(f)$.

Something remains to be said about the transition between the geometric regime and the Rayleigh tail. For small standard deviation $\sigma_R$, there is a clear peak at $\lambda \sim 2\pi \langle R \rangle$ due to the resonance of the fibers. As we increase $\sigma_R$ this peak is smoothed out since the fibers can have a wider variety of radii, suppressing the resonance. This is shown in Fig.~\ref{fig:power_int}(c). Whereas our results seem to indicate that the DM-conversion power is smooth across the region where $\lambda \sim 2\pi \langle R \rangle$, we acknowledge that the behavior of a system with a much larger number of fibers, as will be the case with our experimental apparatus, can be quite different due to the collective effects of Anderson localization and photonic bandgaps \cite{FroufePrez2017BandGF,scheffold2022transport}. These effects, though unpredictable, can be characterized by transmission measurements, since the wavelengths at which these effects occur in a specific realization of powder do not depend on whether the photon comes from DM conversion or an external light source. As the wavelength becomes comparable to the size of the box, we begin to excite the modes of the box, seen as the oscillations at large $\lambda$ in Fig.~\ref{fig:electric_field}(a). This is a physical effect that should occur in both the simulations and the experiment, however, for a realistic experimental setup, with large numbers of dielectric fibers, these oscillations start at parametrically longer wavelengths than the mass range of interest.

The last effect that needs to be accounted for is dielectric loss. This can be easily implemented in our formalism by generalizing the index of refraction to have an imaginary part, the most general form is 
\begin{equation}\label{eq:complex_index}
\underline{n}=n + i\kappa + ix_{\rm cond} \frac{\lambda}{2\pi\langle R\rangle},
\end{equation}
where $\kappa$ is the frequency-independent dielectric loss, while the second term is the loss due to conductivity with $x_{\rm cond}$ being a dimensionless conductivity in units of $1/\langle R\rangle$.  For concreteness, we choose to focus on the limit where conductivity-induced losses dominate; this will likely apply for the semi-conducting materials we plan to utilize.
As shown in Fig.~\ref{fig:power_int}(b), the signal power around $\lambda \gtrsim \langle R\rangle$ is significantly reduced when $x_{\rm cond} \gtrsim 0.3 \%$. An $\mathcal{O}(1)$ suppression of the signal power at $\lambda \lesssim  \langle R\rangle$ for $x_{\rm cond} \simeq 1 \%$ confirms the expectation that the signal photon 
diffuses through the dielectric medium, propagating as a random walk in the geometric regime.
This can be seen by noting that, in a dielectric medium, the average displacement of the traveling photon is proportional to the square root of the number of encounters with a dielectric particle, $\sim N_{\rm en}^{1/2} \langle R\rangle$. To exit the box, this displacement must be larger than the size of the region containing the dielectric particles, which, in 2D, grows with the number of dielectric particles as $N^{1/2}$. This suggests that the number of encounters with the dielectric powder is proportional to the number of particles $N_{\rm en} \sim N$ in 2D. As a result, the distance the photon travels in powder is $N_{\rm en} \langle R \rangle \approx N \langle R\rangle$, and correspondingly, the dielectric loss is $\log \left[P (x_{\rm cond}) /P (x_{\rm cond} =0)\right] \simeq - N x_{\rm cond}$. Whereas this result is obtained with a simplified scenario in 2D, it does highlight the importance of choosing low loss materials, as the size of the experiment (i.e., number of dielectric powder particles) is limited by dielectric loss. In an actual experimental setup, it is neither realistic nor optimal to surround a box of powder with photon detectors. With a photon collection chamber, the effective powder volume can be maximized for a fixed amount of dielectric loss, a more detailed explanation of the relationship between dielectric loss and the effective experimental volume can be found in Sec.~\ref{sec:transport} and~\ref{sec:exp_setup}. 

Further, the result shown in Fig.~\ref{fig:power_int}(b) suggests that the dielectric loss decreases as the wavelength decreases. The dielectric loss $\log \left[P (x_{\rm cond}) /P (x_{\rm cond} =0)\right]$ depends on both $\lambda$ and $x_{\rm cond}$ in the $\lambda/\langle R\rangle \rightarrow 0$ limit, whereas the dielectric loss would be independent of $\lambda$ in the same limit for a bulk dielectric material. The difference in these scalings could come from interference effects. The outgoing wave described by the T-matrix [see Eq.~\eqref{eq:app_T}] contains both the wave reflected from the interface of the powder particle, as well as the wave that transmitted through it. The interference between the reflected and transmitted wave leads to a T-matrix that is strongly wavelength dependent, and hence sensitive to the small imaginary part of the index of refraction, especially at wavelengths where destructive interference occurs. In this case, the main effect of the imaginary part of the index of refraction comes from the Fresnel equations, where a linear $\lambda$ dependence is expected. 
This highlights the importance of future analytical and numerical studies of dielectric loss in dielectric powder media, as well as dedicated experimental measurements to understand and characterize the dielectric loss for both the $\lambda \simeq \langle R\rangle$ and  $\lambda \ll \langle R \rangle$ regimes, and potentially other loss channels due to surface effects \cite{Martin1970}.

\subsection{Disordered Dielectric System in 3D}\label{sec:disorder3D}

In this subsection, we study the conversion of axions and dark photons into photons in a 3D disordered dielectric system, consisting of an ensemble of dielectric spheres. We show that the key features found in Sec.~\ref{sec:iso3D} remain qualitatively the same in 3D. Although multiple scattering of photons between numerous closely spaced dielectric spheres could be solved analytically using a T-matrix formalism, we opted for a fully numerical solution in this work using COMSOL Multiphysics\textsuperscript{\textregistered}\cite{multiphysics3}, following the approach of Ref.~\cite{Jeong:2023bqb}. Maxwell’s equations were solved using the \textit{Electromagnetic Waves, Frequency Domain} module in COMSOL, and the effect of the DM background was incorporated by setting an external current density for each dielectric sphere.

\begin{figure}[t]
    \centering
    \includegraphics[width=0.9\textwidth]{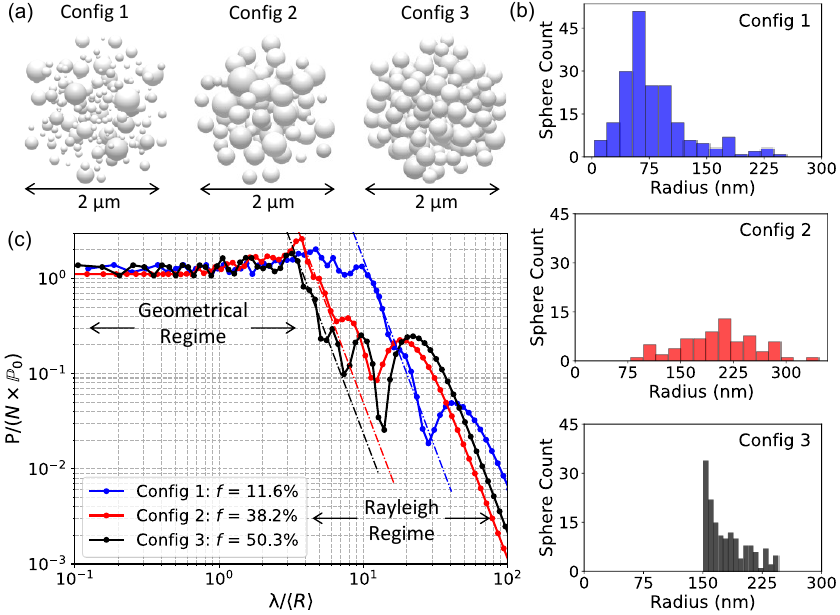}
    \caption{(a) Rendered images of the simulated dielectric powder are shown. Three random configurations are shown, each with varying filling factors ($f$) and distributions of the dielectric spheres' radii. Configurations 1, 2, and 3 have filling factors of 11.6\%, 38.2\%, and 50.3\%, respectively. The size distributions of the spheres are shown in (b). The mean sphere radii, $\langle R \rangle$, of these three configurations are 81 nm, 202 nm, and 180 nm, respectively. (c) The power converted by these three configurations is normalized by the product of Eq.~\eqref{eq:P0_3D} and the number of spheres, and is plotted as a function of the normalized wavelength. The total number of spheres in the three configurations is 189, 73, and 155, respectively. In all three configurations, the total volume, $V_{\rm box}$, of the sphere ensemble is fixed at $2\times2\times2$ $\mathrm{\mu m}^3$, and the refractive index of the dielectric spheres is $n=1.77$ at all wavelengths. We observe that the geometric and Rayleigh regimes are identifiable, with the transition from geometric to Rayleigh following the same behavior as in Fig.~\ref{fig:power_int}(a). The box oscillations are also observed at longer wavelengths.}
    \label{fig:comsol3D}
\end{figure}

Three random configurations of dielectric spheres were simulated, as shown in Fig.\ref{fig:comsol3D}(a). The configurations were generated using a Python script. The total volume of the box, $V_{\rm box}$, was fixed at $8\,{\rm \mu m^3}$ and filled with dielectric spheres to achieve filling factors of 12\%, 38\%, and 50\%. The spheres’ radii were randomly drawn from a truncated Gaussian distribution; however, our sphere placement algorithm did not always produce a Gaussian distribution of sphere radii, due to placement constraints imposed to meet the target filling factors. The sphere distributions are shown in Fig.~\ref{fig:comsol3D}(b), with the mean sphere radii for each configuration being 81 nm, 202 nm, and 180 nm, respectively. The total number of spheres ($N$) in the three configurations is 189, 73, and 155, respectively. Perfectly matched layers were applied as boundary conditions, and the power converted by the spheres ($P$) was integrated over time on all six sides of the simulation box.

In Fig.~\ref{fig:comsol3D}(c), we show the power $P$ normalized by the reference power $\Psingle_{\rm 0}$ (as defined in Eq.~\eqref{eq:P0_3D}) and the number of spheres $N$. Since the interference effects average out in the geometrical and Rayleigh regime, the power in Fig.~\ref{fig:comsol3D}(c) can be approximated by the incoherent sum of the power converted by many spheres. This is given in Eq.~\eqref{eq:power_3D}, which was found by averaging Eq.~\eqref{eq:cyn_approx} over a Gaussian distribution. However, it is not obvious that Eq.~\eqref{eq:power_3D} should continue to hold for a non-Gaussian distribution, as is the case in our simulation. Nonetheless, we compare the power in Fig.~\ref{fig:comsol3D}(c) with Eq.~\eqref{eq:power_3D} when the averages are taken to be over the distributions shown in Fig.~\ref{fig:comsol3D}(b) and find that they do, in fact, agree. This agreement is expected in the Rayleigh regime, since the specific shape of the distribution of the powder radii is not necessary to compute the average power, while for the geometrical regime, the precise form of the distribution is relevant only for corrections of order $\mathcal{O}(\lambda/\langle R\rangle)$.
This allows us to rewrite Eq.~\eqref{eq:power_3D} in terms of the simulation parameters in the geometrical and Rayleigh regime, respectively:
\begin{equation}\label{eq:power_3D_final}
    \begin{split}
    &\langle P_G \rangle = \frac{E_0^2}{2} \left( 1-\frac{1}{n} \right)\left( 1-\frac{1}{n^2} \right) f V_{\rm box} \frac{2\langle R^2 \rangle}{\langle R^3 \rangle}\\
    &\langle P_R \rangle = G(f) \frac{E_0^2}{2} \left( \frac{n^2-1}{n^2+2} \right)^2 f V_{\rm box} \frac{2\langle R^6 \rangle}{\langle R^3 \rangle} \left(\frac{2\pi}{\lambda}\right)^4~,
    \end{split}
\end{equation}
where the averages $\langle \cdot \rangle$ can be taken over any distribution of powder radii, thanks to the agreement between Eq.~\eqref{eq:power_3D} and Fig.~\ref{fig:comsol3D}(c) when the distribution of radii is chosen to be that in Fig.~\ref{fig:comsol3D}(b).

The results presented in Fig.~\ref{fig:comsol3D}(c) are computed only once for a single powder configuration, rather than averaging over many configurations as was done in Sec.~\ref{sec:disorder2D}. Hence the power has more noticeable fluctuations, this is due to a smaller detector size (equivalently, a smaller number of dielectric spheres), and shall average out when the number of dielectric spheres increases. The changing filling factor leads to the same qualitative behavior as observed in Fig.~\ref{fig:power_int}(a). For wavelengths larger than $\sim 100 \langle R \rangle$, the converted photon sees the whole powder medium as an almost uniform dielectric object, and we see the corresponding Rayleigh tail, as well as the preceding oscillations, due to the size of the simulation box. This regime is not relevant experimentally, since the experimental volume is much larger than the wavelength of the signal photon of interest. The agreement between the COMSOL simulation based on finite-element method and the previous semi-analytical analysis serves as a confirmation of our theoretical framework. However, both of these methods become inefficient when considering large volumes of powder; for this reason we only consider small numbers of dielectric particles in Secs.~\ref{sec:disorder2D} and~\ref{sec:disorder3D}. In what follows, we will discuss what changes when considering larger powder volumes.

\subsection{Extending Results to Large Dielectric Powder Volumes}
\label{sec:largenumber}

In Sec.~\ref{sec:disorder2D} and~\ref{sec:disorder3D}, we have demonstrated that the conversion rate of dark photons and axions into photons in a disordered dielectric system scales with the total vacuum-dielectric interface area, and can be expressed simply as a function of the experimental parameters ($V_{\rm box}$, $f$, $\langle R \rangle$, $\sigma_R$). Notably, whereas the conversion power from an individual 2D (3D) powder particle oscillates with the wavelength following Bessel (spherical Bessel) functions (see Figs.~\ref{fig:power_cylinder_no_int} and \ref{fig:power_sphere_no_int}), these oscillations smooth out after averaging over $\mathcal{O}(100)$ powder particles (see Fig.~\ref{fig:electric_field}). Moreover, light emitted from randomly packed dielectric powder exhibits speckle patterns (see Fig.~\ref{fig:electric_field}(b)), which, for any specific realization of small number of particles, leads to oscillations of the conversion rate. These oscillations, again, smooth out after averaging over many configurations, apart from oscillations related to the size of the box, which for realistic experimental systems, will be too low frequency to be relevant.

These results are obtained from both semi-analytical and numerical studies of systems containing a small number of dielectric powder particles. In our proposed experimental setup, the powder volume can be as large as $\SI{e3}{\cm^3}$ and contain as many as $10^{15}$ particles (see Sec.~\ref{sec:exp_setup} for further details). The methods outlined in Secs.~\ref{sec:disorder2D} and~\ref{sec:disorder3D} become computationally impractical for such large systems, due to the matrix inversion required to solve Eq.~\eqref{eq:recursive} and to the extremely fine mesh needed in COMSOL to discretize the wave equation at small wavelengths, which becomes unsustainable for large volumes. Further, the filling-factor-dependent geometrical factor $G(f)$ cannot be precisely predicted based on our semi-analytical and numerical studies in the limit of a large number of particles.
On the other hand, this geometrical factor $G(f)$ characterizes the reduction of the emission power compared to the incoherent sum for $\lambda \gtrsim \langle R\rangle$  due to the interference of {\it photons} in a disordered dielectric medium, and importantly, does not depend on where these photons originate from.
Therefore, as we will elaborate in Sec.~\ref{sec:transport}, it is possible to calibrate the system by shining light through the disordered powder system and measuring the response, both to confirm our understanding of the disordered medium in the geometrical and Rayleigh regime, as well as to precisely determine the sensitivity for $\lambda \simeq \langle R \rangle$.

Our treatment in the previous subsections shows that a broadband sensitivity can be achieved with disordered dielectric media. However, due to the small numbers of dielectric powder particles, we might have  underestimated the significance of two well-known phenomena, i.\,e., photonic bandgaps and  localization, including Anderson localization~\cite{PhysRev.109.1492,M_ximo_2015,scheffold2022transport,Yamilov_2023}. 
Both photonic bandgaps and weak localization might be contributing to the oscillations we see in the previous subsections. Photonic bandgaps can occur in powder systems that we experimentally prepare, for wavelengths that are comparable to the powder size and separation, and would generally depend on the exact arrangement of the powder particles. The local density of optical states (LDOS) of disordered systems might be able to offer more theoretical insights~\cite{Schubert_2010}, which we postpone to dedicated studies in the future.
Luckily, whether a specific disordered system exhibits photonic bandgaps and Anderson localization at specific frequencies can also be calibrated. Since transport properties of disordered dielectric systems do not depend on the source of light, these properties can be characterized by experimentally measuring the transmission coefficient, as explained in more detail in Sec.~\ref{sec:transport}.

The results in Eq.~\eqref{eq:power_2D_final} and Eq.~\eqref{eq:power_3D_final} are interpreted as the total power converted by a small volume of powder, integrated over all directions.
In a realistic experimental setup, the incoherent sum approximation, for both the geometrical and Rayleigh regimes, is expected to improve for larger powder volumes, and the interference effects remain irrelevant.
However, dielectric losses (discussed in Sec.~\ref{sec:disorder2D}) limit the signal photon emission volume, resulting in an effective collection volume $V_{\rm eff}$. The precise formula for $V_{\rm eff}$ as a function of the absorption length for the photon collection chamber we envision is given in Sec.~\ref{sec:transport}.

\subsection{Summary}\label{sec:summary}

In this Section, we demonstrated that a broadband sensitivity to dark photon and axion DM can be achieved with a disordered dielectric medium consisting of dielectric powder with a wide distribution of powder radii. This broadband sensitivity manifests as a smooth signal power from DM to photon conversion as the DM mass (Compton wavelength) is varied, in contrast to a spiky signal power from a resonant experiment (see Sec.~\ref{sec:Section II} for an in-depth explanation), which we elaborate now in the different cases we considered.

Starting from the analysis of DM to photon conversion by a single dielectric object (Secs.~\ref{sec:iso2D} and~\ref{sec:iso3D}), we identified two limiting behaviors of the converted power as a function of wavelength. At wavelengths smaller than the radius of the object ($\lambda < \langle R \rangle$), the converted power is highly oscillatory. These oscillations smooth out after incoherently summing over the power converted by multiple objects of varying radii, as is shown in Figs.~\ref{fig:power_cylinder_no_int} and~\ref{fig:power_sphere_no_int}, with the resulting  signal power being broadband and proportional to the surface area of the dielectric particle. At wavelengths much larger than the radius of the object ($\lambda > \langle R \rangle$), we find a Rayleigh-like scaling, of $1/\lambda^3$ for 2D powder (dielectric fiber) and $1/\lambda^4$ for 3D powder (dielectric spheres). 

In Secs.~\ref{sec:disorder2D} and~\ref{sec:disorder3D}, we used a semi-analytical multiple-scattering method in 2D and COMSOL simulations in 3D to demonstrate that the geometrical and Rayleigh regimes can also be identified when interference effects are taken into account. At small wavelengths ($\lambda \lesssim \langle R \rangle$), a smooth and broadband conversion of DM to photons can be achieved, provided that the radii of and separations between powder particles are sufficiently randomized.
In this regime, the signal power is the same as the results of the incoherent sum in Secs.~\ref{sec:iso2D} and~\ref{sec:iso3D} (see Fig.~\ref{fig:electric_field} and~\ref{fig:comsol3D}). 
However, as the wavelength increases to $\lambda \gtrsim \langle R \rangle$, the behavior differs from the incoherent sum. Whereas the wavelength scaling at very large wavelength ($\lambda \gg \langle R \rangle$) remains the same, the overall amplitude in this Rayleigh regime is suppressed compared to the incoherent sum. This is due to the fact that the transition from the geometrical regime to the Rayleigh regime happens at a lower wavelength. This different behavior at intermediate wavelengths ($\lambda \simeq \langle R \rangle$) leads to a filling factor dependent geometrical factor $G(f)$, which suppresses the signal power in the Rayleigh regime (see Figs.~\ref{fig:power_int}(a) and~\ref{fig:comsol3D}). 
Note that the above-mentioned results are obtained with a small number of powder particles in a small volume, however, as we argued in Sec.~\ref{sec:largenumber}, we expect the smoothness of DM to photon conversion power to improve for larger volumes.

For convenience and easy reference we compile the relevant equations and write the total conversion power in the following compact formula, 
\begin{equation}\label{eq:generic_power}
    \langle P\rangle = G(f) \frac{E_0^2}{2} f V_{\rm eff} \left( \frac{\langle \sigma_{\text{DM}} \rangle}{\langle V_{\rm particle} \rangle} \right)~.
\end{equation}
The DM induced electric field $E_0$ is defined in Eq.~\eqref{eq:E_0} and listed as a function of the local DM density and interactions in Tab.~\ref{tab:E0}. The DM conversion cross section, $\sigma_{\rm DM} \equiv P/(E_0^2/2)$ is defined as the ratio between the power converted from DM into photons and the energy density of the electric field induced by DM. Note that the small coupling of the DM to the photon is contained in $E_0$ in our definition, and the parameter $\sigma_{\rm DM}$ is {independent} of this small coupling.
The ratio $\langle \sigma_{\text{DM}} \rangle/{\langle V_{\rm particle} \rangle}$ in Eq.~\eqref{eq:generic_power}, with values in the different regimes listed in Tab.~\ref{tab:cs}, captures the information about the distribution of powder radii. In the geometrical regime ($\lambda \ll 2\pi\langle R \rangle$), the filling-factor-dependent geometrical factor  $G(f)=1$; 
however, for larger wavelengths ($\lambda \gtrsim 2\pi\langle R \rangle$), 
$G(f)$ cannot be computed analytically, and must be characterized by shining light through the powder medium in a concrete experimental configuration, which we discuss in the next Sections. 

An important qualitative feature in Tab.~\ref{tab:cs} that is worth highlighting is that $\langle \sigma_{\text{DM}} \rangle/\langle V_{\rm particle}\rangle$ increases as powder radius increases in the Rayleigh regime ($\langle R\rangle \ll \lambda$), and decreases as powder radius increases in the Geometrical regime ($\langle R\rangle \gg \lambda$). As a result, for an experimental setup with fixed effective volume $V_{\rm eff}$ and filling factor $f$, there is an optimal mean powder radius $\langle R\rangle\sim \lambda$ that maximizes the sensitivity for any given dark matter wavelength $\lambda$. Such an optimal $\langle R\rangle$ can depend on the filling factor, dielectric loss, and the distribution of the powder radii (see Fig.~\ref{fig:power_int}). In a concrete experimental configuration, the effective volume will generally be wavelength-dependent, however, as we will show in more detail in Sec.~\ref{sec:reach_plot}, this qualitative expectation continues to hold.
In the following, when discussing experimental setups that can scan the axion and dark photon parameter spaces with a finite mass (wavelength) range, the optimal mean powder radius is chosen such that the sensitivity is maximized over the mass (wavelength) range of interest. 

\begin{table}[th!]
    \centering
    \begin{tabular}{|c|c|c|}
    \hline 
        $E_0^2/2$ & Fibers (2D powder) & Spheres (3D powder) \\[1.5mm]
        \hline & \\[-3.5mm]
        axion & $(g_{a\gamma \gamma } B_{\rm ext})^2 \frac{\rho_{\rm DM}}{m^2}$ & $(g_{a\gamma \gamma } B_{\rm ext})^2\frac{\rho_{\rm DM}}{m^2}$ \\[1.5mm]
        \hline & \\[-3.5mm]
        dark photon & $ \frac{1}{3} \varepsilon^2 \rho_{\rm DM}$ & $\frac{2}{3} \varepsilon^2 \rho_{\rm DM}$ \\[1.5mm]
        \hline
    \end{tabular}
\caption{Effective electric field $E_0$ as defined in Eq.~\eqref{eq:E_0} for different DM candidates and experimental setups.  The local DM density $\rho_{\rm DM}$ is given by $\rho_{\text{DM}}=m_a^2 a_0^2/2$ for axions, where $a_0$ is the axion field amplitude [Eq.~\eqref{eq:fourier}], and $\rho_{\text{DM}}= \vec{E}_0^{\prime 2}/2$ for dark photons. 
The component of the electric field parallel to the interface is $1/3$ of the total electric field in the 2D case and $2/3$ in 3D.}
    \label{tab:E0}
\end{table}

\begin{table}[th]
    \centering
    \begin{tabular}{|c|c|c|}
    \hline
        $\langle \sigma_{\text{DM}} \rangle/\langle V_{\rm particle} \rangle$ & Fibers  (2D powder) & Spheres (3D powder)\\
        \hline & \\[-3.5mm]
        Geometrical & $2\left( 1-\frac{1}{n} \right)\left( 1-\frac{1}{n^2} \right) \frac{\langle R \rangle}{\langle R^2 \rangle}$ & $2\left( 1-\frac{1}{n} \right)\left( 1-\frac{1}{n^2} \right)  \frac{\langle R^2 \rangle}{\langle R^3 \rangle}$ \\[1.5mm]
        \hline & \\[-3.5mm]
        Rayleigh & $ \frac{\pi}{4} n^4 \left( 1-\frac{1}{n^2} \right)^2 \frac{\langle R^4 \rangle}{\langle R^2 \rangle} \left( \frac{2\pi}{\lambda} \right)^3$ & $2 \left( \frac{n^2-1}{n^2+2} \right)^2 \frac{\langle R^6 \rangle}{\langle R^3 \rangle} \left(\frac{2\pi}{\lambda} \right)^4$ \\[1.5mm]
        \hline
    \end{tabular}
   \caption{The effective cross section per powder particle volume $\langle \sigma_{\text{DM}} \rangle/\langle V_{\rm particle}\rangle$ as a function of the DM wavelength, the refractive index $n$ and the distribution of the 2D \& 3D powder radii. Note that the small coupling of the DM to the photon is contained in $E_0$ in our definition, and the parameter $\sigma_{\rm DM}$ is \emph{independent} of this small coupling. }
    \label{tab:cs}
\end{table}

Before we close this section, it is worth pointing out that the qualitative behavior of the power in Eq.~\eqref{eq:generic_power} is similar to the scattering of light by dielectric powders. Comparing the cross section $\sigma_{\text{DM}}$ with the standard light scattering cross section, $\sigma_s$, we find that the ratio $\langle \sigma_{\text{DM}} \rangle/\langle \sigma_s \rangle$ is independent of frequency in the geometrical and Rayleigh regimes. In the geometrical regime, $\sigma_{s}$ oscillates as a function of wavelength [see the T-matrix in Eq.~\eqref{eq:app_T}], which, when averaged over the radius of the powder, is proportional to the geometrical cross-section of the powder. While for larger wavelengths, light scatters with a Rayleigh scattering cross-section. These behaviors are exactly what we discovered for DM conversion. This similarity between DM to photon conversion and light scattering suggests that we can use measurements of $\sigma_s$ to calibrate the system, which we elaborate in more detail in Sec.~\ref{sec:transport}.

%%%%%%%%%%%%%%%%%%%%%%%%%%%%%%%%%%%%%%%%%%%%%%%%%%%%%%%%%%%%

\section{Optical Transport and Characterization}\label{sec:transport}

Recall that in the previous section, we found the explicit formulae for the power produced in the geometrical and Rayleigh regimes from a dielectric box of volume $V_\text{box}$. Experimentally, the relevant quantity is the signal power incident on a sensor placed at the boundary of a dielectric medium: $P_\text{det}=\pd V_{\text{eff}}$, where $\pd=P/V_\text{box}$ is the average optical power density injected into the target by DM absorption from the previous section and $V_{\text{eff}}$ is the effective collection volume. In this Section, we use a toy model of diffusion to provide  preliminary estimates of an optimal effective volume  $V_\text{eff}$. We also describe the  macroscopic parameters relevant for DM conversion power, such as  the target geometry and optical transport properties,  and outline schemes to characterize these properties experimentally.

\subsection{Regimes of Optical Transport}

The transport behavior of light in disordered media can be classified into one of several regimes. When scattering is especially weak, ballistic transport occurs. Structures which exhibit ``stealthy" hyperuniformity, such as randomly close-packed spheres of identical size, can host bands of freely propagating optical modes as in a perfect crystal \cite{Leseur2016}. While a weakly-scattering medium is ideal for coupling a large effective volume to the detector, it is also  ineffective for coupling DM and light. 

In strongly-scattering media, interference can cause effects like coherent backscattering (weak localization) and Anderson localization (strong localization). The scattering strength and system size required to observe these effects depends on the dimensionality of the system. While localization is understood to be ubiquitous in 2D systems of sufficient size \cite{Abrahams1979}, 
it has been observed in 3D only for high index contrast media ($n>2$) \cite{Grynko2024,Rezvani2016,Sperling2012}. As Mie resonances can enhance scattering strength, localization is especially likely for $\langle R\rangle\sim\lambda$. While strong scattering can enhance the DM-to-photon conversion in the powder target, localization  substantially limits the effective volume of the detector as observed by a sensor. Strongly scattering disordered media can also support photonic bandgaps, in which the density of optical states is zero and optical transport and DM absorption are suppressed.

Media with intermediate levels of scattering typically support diffusive transport. This is the regime of interest for our experimental setup. Photon diffusion can be modeled as a random walk. The random walk step size is the transport scattering length, $\ell_s^*=\ell_s/(1-g)$, where $g$ is the scattering anisotropy parameter \cite{Lemieux1998} and $\ell_s=\mean{V_\text{particle}}/f\mean{\sigma_s}$ is the photon scattering length in a medium of particles of volume $V_\text{particle}$ and total light scattering cross section $\mean{\sigma_s}$. Dipole scattering, which is characteristic of the Rayleigh regime, is isotropic ($g=0$), while forward-scattering ($0<g<1$) is typical in the geometrical regime. The characteristic length of the total distance of the walk is the absorption length $\bar\ell_a$. We mark this quantity in powder with a bar to distinguish it from the absorption length of the bulk material $\ell_a$. In App.~\ref{app:collector}, we show that, neglecting absorption due to surface states, the absorption length in the powder is $\bar\ell_a\simeq\ell_a/f$. The average net distance traveled by a photon in this regime is the diffusion length, $\ell_d=(\bar{\ell}_a\ell_s^*/3)^{1/2}$ \cite{Lemieux1998} in 3D or $\ell_d=(\bar{\ell}_a\ell_s^*/2)^{1/2}$ in 2D. Infrared-transparent materials like alumina can have $\ell_a \sim \SI{10}{\meter}$ at room temperature \cite{Aleksey2016}. In an alumina powder with $f=0.5$, $\ell_s^*\sim\SI{10}{\mu m}$, and $g\sim 0$, the diffusion length is $\ell_d\sim\SI{1}{\cm}$.

In Section \ref{sec:Characterization} we will describe experimental methods for determining these parameters of diffusive transport in powder and how characterization of the scattering length $\ell_s$ can inform estimates of $\mathcal{P}_V$. In Section \ref{sec:Veff}, we will describe how the effective volume can be increased by using a non-resonant cavity to couple the powder and detector. The required size of this cavity and the achievable $V_\text{eff}$ will depend on $\ell_s^*$ and $\ell_a$.

\subsection{Experimental Methods for Characterizing Diffusive Transport}
\label{sec:Characterization}
Diffuse reflectance and transmittance spectroscopy can be used to determine $\ell_d$, $\bar{\ell}_a$, and $\ell_s$, which in turn characterize both optical transport and DM absorption. Reflectance and transmittance measurements can also help exclude the possibility of non-diffusive transport, which can be recognized in the case of Anderson localization or a photonic band gap by near-zero transmission and near-perfect reflection \cite{scheffold2022transport}.

In the geometrical regime, $\sigma_\text{DM}$ is proportional to the specific surface area. It can be accurately estimated based on the measured density if the size distribution of the powder particles is well-characterized. In the Rayleigh regime, structures at various scales can act as dipole scattering sites, including particles, inter-particle voids, and larger agglomerates or clumps of particles. This leads to a more complicated dependence of $\sigma_\text{DM}$ on $f$ (described by $G(f)$). As shown in the previous section, the DM absorption cross section is identical to the Rayleigh scattering cross section at each site. Therefore the total cross section per volume is $\mean{\sigma_\text{DM}}/\mean{V_\text{particle}}=\mean{\sigma_s}/\mean{V_\text{particle}}=1/\ell_s f$ (note: $\ell_s=\ell_s^*$ in the Rayleigh regime since $g=0$).  

The scattering length can be determined from transmittance measurements using Fourier Transform Infrared Spectroscopy (FTIR) applied to a thin slab of powder. For powders with $\ell_s^*\ll\ell_a$, absorption has a negligible effect on the transmittance, as most non-transmitted power is reflected. A standard treatment of 1D diffusive optical transport in a slab of lossless material with thickness $t$ gives transmission coefficient
\begin{equation}
    \transmittance = e^{-t/\ell_s} + \transmittance_0 \left(\frac{1+z_0}{2z_0+t/\ell_s^*}(1-e^{-t/\ell_s})-\frac{t/\ell_s^*}{2z_0+t/\ell_s^*}e^{-t/\ell_s}\right),
\end{equation}
where the factor $\transmittance_0$ removes photons which are coherently back-reflected \cite{scheffold2022transport} and $z_0$, called the extrapolation length, is the distance away from the sample where the photon concentration is extrapolated to zero from the linear gradient in the slab \cite{Lemieux1998}. The value of $z_0$ is a functional of the angle-dependent reflectance of the slab surface, and can be determined most easily by measuring $\transmittance$ for multiple values of $t$. The first and last terms in this expression account for ballistic and diffusive contributions, repectively, while the middle term models the conversion of incident to diffusive photons. The value of $\ell_s$ can be determined by measuring $\transmittance$ for several values of $t\sim \ell_s$. In the limit $t\gg\ell_s$, we have
\begin{equation}
    \transmittance = \transmittance_0 \frac{1+z_0}{2z_0+t/\ell_s^*},
\end{equation}
and $\ell_s^*$ can be determined by measuring $\transmittance$ for at least two values of $t$. The scattering anisotropy $g$ can be determined by comparing $\ell_s$ and $\ell_s^*$.

The diffuse reflectance $\reflectance_\infty$ is another standard experimental quantity for characterizing optical transport in powder \cite{Johnson1952} and is more suitable for measuring $\ell_a$ when the absorption is small. There are various commonly-used approximations of radiation transport used to determine $\reflectance_\infty$ in terms of the absorption and transport scattering lengths. In App. \ref{app:collector}, we derive
\begin{equation}
    \frac{\ell_s^*}{\bar\ell_a}=\frac{3(1-\reflectance_\infty)^2}{16\reflectance_\infty^2};
    \label{eq:Rinf}
\end{equation}
in the limit $\bar\ell_a\gg\ell_s^*$, $\reflectance_\infty\approx 1-4\ell_d/\bar\ell_a$.\footnote{The value of $\reflectance_\infty$ depends on whether the illumination is collimated or diffuse. A comparison of some alternatives is given in Ref. \cite{kuhn1993infrared}, which favors a `3-flux' model giving $1-\reflectance_\infty\approx 1-5\ell_d/\bar\ell_a$ for collimated illumination in the limit $\ell_a\gg\ell_s^*$.} The dielectric materials of interest for creating a haloscope target have very low loss so that $\ell_s^*/\bar\ell_a\ll 1$ and $\reflectance_\infty$ can be greater than 0.99. Combining the estimate of $\ell_s^*$ from transmittance measurements and $\ell^*_s/\bar\ell_a$ from a reflectance measurement, we can produce estimates of $\bar\ell_a$ and $\ell_d$.

\subsection{Effective Volume and Target Geometry}
\label{sec:Veff}

In addition to the intrinsic scattering and absorbing properties of the dielectric material, the geometry of the absorbing target will have a significant effect on the signal power. Compared to the average steady-state optical energy density $u_\infty$ in the bulk of the powder, the optical energy density $u_0$ near an absorbing boundary (such as a sensor) can be substantially lower. The ratio $u_0/u_\infty$ is related to the effective volume from which converted photons can reach the boundary. In the diffusive transport regime, a sensor with area $A_\text{sensor}\gg\ell_d^2$ placed at the boundary of the powder gathers photons from an effective volume of $V_\text{eff}\approx\ell_d A_\text{sensor}$. A much larger effective volume can be achieved by forming the dielectric medium into a hohlraum - a non-resonant cavity in radiative equilibrium. This geometry, originally used to study blackbody radiation, uniformly couples a large boundary of the dielectric medium to the sensor. In App. \ref{app:collector}, we approximate the optical power gathered by such a cavity using a simple diffusion model. Below is a summary of the results. 

The ratio between the the average intensity entering the cavity and the average intensity leaving the cavity is
\begin{equation}
    \eta_\text{cavity}=\frac{A_\text{cavity}}{A_\text{cavity}+(1-\reflectance_\text{sensor})A_\text{sensor}}\ , 
\end{equation}
where $\reflectance_\text{sensor}$ is the reflectance of the sensor and $A_\text{cavity}$ is the surface area of the cavity. If the surface of the dielectric is charged with isotropic optical energy density $u_0$, then the average intensity entering the cavity is $\bar{c}u_0/4$ and the average intensity leaving the cavity and incident on the sensor is $I_\text{sensor}=\eta_\text{cavity}\bar{c}u_0/4$ where $\bar c$ is the speed of light in the medium. In the bulk of the medium in steady state, $\pd$ is balanced by the optical power density lost to absorption. This balance results in steady-state optical energy density $u_\infty=\pd\bar\ell_a/\bar{c}$. Therefore,
\begin{equation}
    I_\text{sensor}=\frac{1}{4}\eta_\text{cavity}\frac{u_0}{u_\infty}\bar\ell_a\pd
\end{equation}
and the effective volume of the experiment is
\begin{equation}
V_\text{eff}=\frac{1}{4}A_\text{sensor}\eta_\text{cavity}\frac{u_0}{u_\infty}\bar{\ell}_a.
\label{eq:Veff}
\end{equation}

In a large cavity ($A_\text{cavity}\gg A_\text{sensor}$), the circulating intensity is brought into radiative equilibrium with the powder bulk and the effective volume saturates (the ratio $u_0/u_\infty$ and the cavity reflectance will approach 1). The cavity area required to approach saturation will depend on $\ell_d$ via $\reflectance_\infty$. Solving the diffusion equation with appropriate boundary conditions and assuming the cavity interior is lossless, we find that in general,
\begin{equation}
    \frac{u_0}{u_\infty}\approx\frac{(1-\reflectance_\infty)}{(1-\reflectance_\infty)+(1-\eta_\text{cavity})}\ .
    \label{eq:Id}
\end{equation}
Combining this with Eqs. \eqref{eq:Rinf}, \eqref{eq:Veff}, and \eqref{eq:Id}, we find that in the limit $\bar\ell_a\gg\ell_s$,
\begin{equation}
    V_\text{eff} \simeq \frac{1}{4}\frac{\ell_dA_\text{cavity}A_\text{sensor}}{\frac{\ell_d}{\bar\ell_a}A_\text{cavity}+\frac{1}{5}(1-\reflectance_\text{sensor})A_\text{sensor}}.
\end{equation}
When the cavity area is small, the effective volume is proportional to $\ell_dA_\text{cavity}$. When the cavity area is large, 
\begin{equation}
 A_\text{cavity}\gg A_\text{sensor} (1-\reflectance_\text{sensor}) \frac{\bar\ell_a}{5\ell_d},   
\end{equation}
the effective volume saturates to
\begin{equation}
    V_\text{eff}\simeq\frac{1}{4}A_\text{sensor}\bar\ell_a \,,
\end{equation}
enabling the detection of a larger signal than in the absence of a cavity.

\section{Proposed Experimental Setup}\label{sec:exp_setup}

\begin{figure}[t!]
    \centering
    \includegraphics[trim={0, 0, 0, 0}, clip, width=0.85\textwidth]{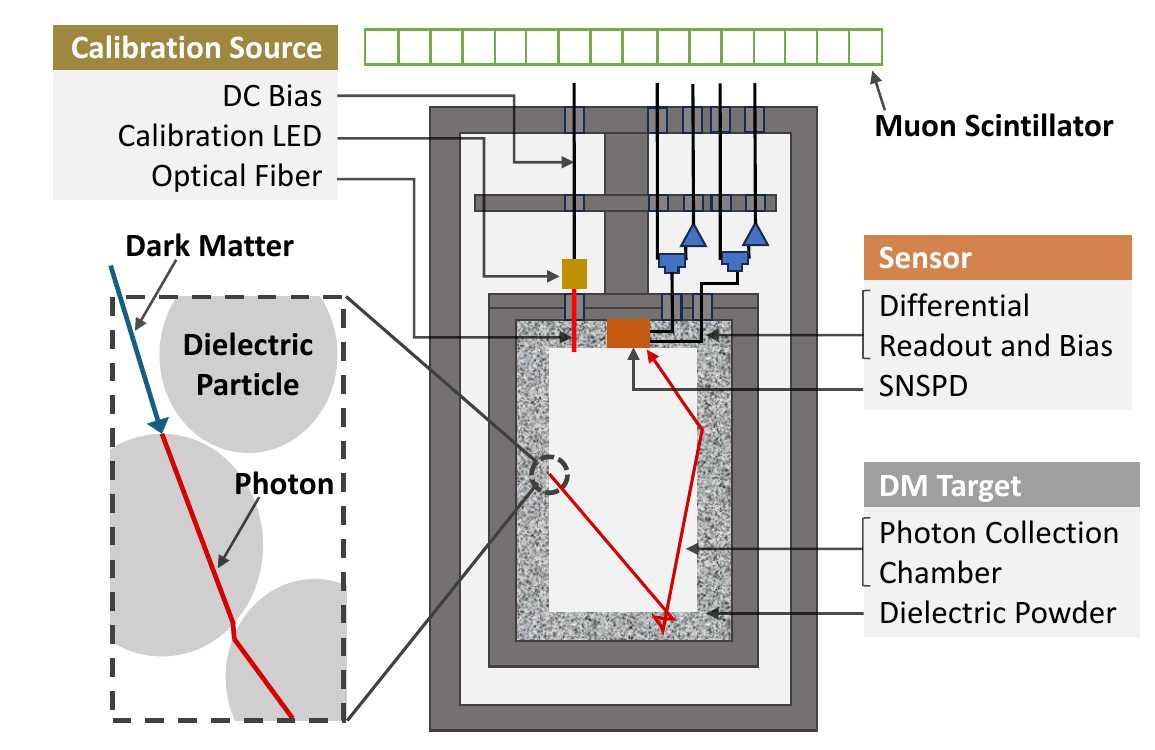}
    \caption{The DPHaSE system schematic is shown. DM is converted to photon at the dielectric-vacuum interfaces in a dielectric powder. The photons diffuse through the powder into a photon collection chamber, where they are detected using an SNSPD. The SNSPD is read out and biased using a differential scheme with two bias tees and two cryogenic amplifiers. By impedance-matching the nanowire to the readout circuit, it is possible to distinguish between single-photon and multi-photon (e.g., cosmic ray) events. An external scintillating detector will complement this internal veto system. \change{The detector, as shown, is configured for a dark photon experiment. For axion detection, an external magnetic field would be required, as well as shielding or field-canceling coils near the sensor.} }
    \label{fig:DPHaSEconcept}
\end{figure}

In this Section, we describe an experimental realization of a dielectric powder haloscope: the Dielectric Powder Haloscope SNSPD Experiment (DPHaSE). In the apparatus shown in Fig.~\ref{fig:DPHaSEconcept}, the absorbing target is a cylindrical shell of dielectric powder which encloses a non-resonant cavity we will call the photon collection chamber. The sensor we propose is a superconducting nanowire single-photon detector (SNSPD) with an impedance-matched differential readout \cite{Colangelo2023}. The optical transport properties of the absorber are calibrated in-situ using a fiber-optic light source. Background events from cosmic rays are vetoed externally by an array of scintillators and internally by detector mechanisms discussed below.

%Cryogenic measurements of $\reflectance_\infty$ for the target can be performed by measuring the sensor count rate from an external optical source coupled to the collection chamber via fiber optic. As the scattering length and anisotropy are not expected to be strongly temperature dependent, changes in $\reflectance_\infty$ after cool-down can be attributed to changes in $\ell_a$.

%2. Estimate $\mathcal{P}_V$ and $\ell_a$ at room temperature using powder reflectance and transmittance spectroscopy as described in above. 

%3. In experiment apparatus, monitor diffuse reflectance during cooldown using an LED source, substituting a photodiode for the SNSPD. The powder should be prepared with the highest possible density (well-packed) to minimize settling. A change in reflectance indicates a change in powder structure (which would affect calibration) or a change in the loss length. If the change is caused by the latter, the change in reflectance should be discontinuous and should exhibit some hysteresis when the temperature is increased again. Otherwise, the change in reflectance can be attributed to the former cause and we can either produce a structure at room temperature with similar reflectance and use that for calibration measurements, or we can lightly sinter the powder to add rigidity to the structure.

\subsection{Dark Matter Absorbing Target}

The absorbing target  consists of a cylindrical shell of dielectric powder enclosing a cavity we will call the photon collection chamber. The thickness of the powder shell should be several times $\ell_d$ so that absorption from the outer detector walls cannot deplete the photon collection chamber. Increasing the thickness much beyond $\ell_d$ will not enhance the signal rate. As shown in the previous section, the effective volume for a cavity-coupled sensor is not limited by diffusive transport and can be as large as $V_\text{max}=\bar\ell_a A_\text{sensor}/4$ when the energy density near the cavity surface is comparable to the energy density in the bulk powder ($u_0/u_\infty\approx 1$).
For the most demanding measurement band, as discussed below, reaching $V_\text{eff}/V_\text{max}=0.8$ for a $\SI{10}{\cm\squared}$ sensor will require a photon collection chamber with surface area of $\SI{2.5}{\meter\squared}$. Reaching $V_\text{eff}/V_\text{max}=0.9$ would require a photon collection chamber of surface area \SI{5.5}{\meter\squared}. If the chamber is formed into a cylinder, the DPHaSE system would be comparable in size to the BREAD experiment~\cite{BREAD:2021tpx}. A more compact experiment can be achieved without diminishing the effective volume by optimizing the surface area per volume of the photon collection chamber. 

The mean particle radius $\langle R \rangle$ is chosen to optimize sensitivity in a particular frequency band. As a heuristic, we choose $\langle R \rangle$ so that the transition between geometrical and Rayleigh scattering occurs approximately at the geometrical mean between the highest and lowest wavelengths of the band. The standard deviation of the particle size $\sigma_R$, must be large enough to dampen interference effects and create a smooth sensitivity curve (see Fig. \ref{fig:power_int}), however choosing too large a standard deviation will result in reduced overall sensitivity. To prepare a target with particles of the desired size distribution, several monodisperse powders can be combined or a single polydisperse powder can be sieved. Care must be taken when depositing the powder to control the packing density and prevent large agglomerates, which would introduce unintended structure and alter the sensitivity curve. A variety of powder dispersion techniques, such as electrostatic spraying, are used in industry for this purpose \cite{kuhn1993infrared}. In order to prevent settling over time, the powder can be sintered to form a rigid shell. The sintering process will generally increase the density of the target and alter its scattering properties. Sections of sintered powder can be analyzed with transmittance and reflectance measurements as described in the previous section to optimize the preparation and verify properties. 

Due to frequency-dependent materials properties, we will divide the experimental bandwidth into 3 regions. In the near- to mid-IR (Compton wavelengths $\lambda=1-\SI{6}{\mu m}$, band {A} in Table \ref{tab: parameters}), alumina is an excellent candidate target material with loss length $\ell_a\gg\SI{1}{\m}$ at low temperature (even in media with high specific surface area \cite{Aleksey2016}) and index of refraction $n=1.7$. Highly pure powders of alumina are commercially available with a variety of particle sizes. To reach $V_\text{eff}/V_\text{max}=0.9$ for a powder with $\reflectance_\infty=0.998$ (from Eq. \eqref{eq:Rinf} with $\bar\ell_a=\SI{20}{\meter}$ and $\ell_s^*=\SI{10}{\mu m}$) and for sensor $\reflectance_\text{sensor}=0$, the required photon collection chamber surface area is $5.5\times 10^3A_\text{sensor}$. 

\change{The sensitivity of the experiment will depend only on the properties of the sensor and the intrinsic properties of the absorber (i.e. $P_\text{det}=\mathcal{P}_VV_\text{eff}\simeq\mathcal{P}_V A_\text{sensor}\bar{\ell}_a/4$). After the absorber material is characterized at room temperature using the transmission and reflection spectroscopy measurements described above, it can be monitored in-situ using a fiber-optic light source to measure $\reflectance_\infty$. As an initial calibration step, the sensor can be replaced with a photodiode to measure $\reflectance_\infty$ during cooldown. In a rigid (sintered) absorber, the scattering length and anisotropy are not expected to be strongly temperature-dependent, and continuous changes in $\reflectance_\infty$ during cool-down can be attributed to changes in $\ell_a$. }

In the mid- to far-IR ($2-\SI{20}{\mu m}$, band B in Table \ref{tab: parameters}), NaCl and KBr are well-known as low-loss powders. Powders of KBr in particular are commonly used to suspend samples for infrared spectroscopy (the mixture is typically pressed into a pellet to minimize scattering) \cite{Olori2021}. In bulk at room temperature, the loss length can be $\ell_a >\SI{100}{\m}$. To our knowledge, there is no published measurement of absorption in NaCl and KBr powder in this bandwidth at cryogenic temperatures. At longer wavelengths, NaCl powder has much higher absorption than the corresponding bulk material due to acoustical modes with wavelengths shorter than $\SI{60}{\mu m}$~\cite{Martin1970}. This surface effect is significantly reduced by cooling to cryogenic temperatures \cite{Hadni1967}. KBr also has an absorption maximum near $\SI{60}{\mu m}$ in both thin film and powder forms \cite{Martin1971} which is substantially reduced at cryogenic temperatures \cite{Hadni1967}. For the purposes of estimating experimental reach in this bandwidth we will suppose $\ell_a=\SI{100}{\meter}$ for a cryogenic powder. If the reflectance of the sensor in band B is $\reflectance_\text{sensor}=0.5$ (as discussed in the next section), then the surface reflectance is $\reflectance_\infty=0.999$ (for $\bar\ell_a=\SI{200}{\meter}$ and $\ell_s^*=\SI{40}{\mu m}$) and the required photon collection chamber surface area for $V_\text{eff}/V_\text{max}=0.9$ is $4.4\times 10^3A_\text{sensor}$.

In the very far IR ($\lambda>\SI{20}{\mu m}$, band C in Table \ref{tab: parameters}), target material choice is very limited. High-resistivity silicon (HR-Si) is one of the few materials with low loss in bulk. The imaginary component of index of refraction is approximately constant for $\lambda>\SI{30}{\mu m}$. At a temperature of \SI{10}{\kelvin}, $\kappa\approx3\times10^{-6}$ (30 times smaller than at room temperature) \cite{Wollack2020}. However the effect of cooling diminishes for $\lambda<\SI{30}{\mu m}$ and $\kappa$ for $\lambda=\SI{20}{\mu m}$ is 30 times larger. Assuming constant $\kappa$ (instead of using Eq. \eqref{eq:complex_index}) leads to a wavelength-dependent loss length
\begin{equation}\label{eq:ell}
  \bar{\ell}_a=\lambda/4\pi \kappa f = \SI{1.6}{\meter} \left(\frac{\lambda}{\SI{30}{\mu m}}\right),
\end{equation}
which we use to generate the values used in Tab.~\ref{tab: parameters}, band C. To achieve $V_\text{eff}/V_\text{max}=0.9$ near the center of the band with $\bar\ell_a=\SI{8}{\meter}$, $\ell_s^*=\SI{40}{\mu m}$, and $\reflectance_\text{sensor}=0.5$ requires a photon collection chamber of area $320\, A_\text{sensor}$.

\subsection{Sensor}
As a sensor technology for detecting the photons converted from the DM interactions with dielectric media, SNSPDs are one of the most promising candidates, offering the sensitivity and low-noise performance required for this proposed DM experiment. Over the past few decades, SNSPDs have been instrumental in various quantum technologies and quantum information applications, and have thus been optimized for near-unity quantum efficiency and extremely low dark count rates. In addition, SNSPDs were key components of the meV-eV DM search experiments and proposals, such as LAMPOST~\cite{Chiles:2021gxk} and BREAD \cite{BREAD:2021tpx}, and have been proposed for GeV-regime direct detection of DM \cite{PhysRevLett.123.151802}, providing compelling motivation to pursue this technology in the present work.

The SNSPD used in the LAMPOST experiment had a relatively large detection area (420\si{\times}420 \si{\mu m^2}) and an ultra-low dark count rate (1 count/day), with saturated detection efficiency at 1550 nm wavelength \cite{Chiles:2021gxk}. Ongoing efforts have been made to extend the sensitivity of SNSPDs to mid- and far-infrared wavelengths, with a recent demonstration of single-photon detection at 29 \si{\mu m} wavelength \cite{Taylor2023}, albeit with much higher dark count rates ($\sim$ 0.1 Hz). Although the current performance metrics of SNSPDs might suffice for conducting some proof-of-concept experiments with proposed DM targets, extending the SNSPD detection area while continuing to optimize mid-infrared sensitivity, magnetic field tolerance, and suppression of background event counts in the SNSPDs needs to be performed to discover new physics beyond the Standard Model.
 
Unlike the LAMPOST experiment, the effective volume of the DPHaSE experiment scales proportionally with the detection area of the SNSPD. Taking full advantage of the $\sim$1000 \si{cm^3} volume available in a standard cryostat will require developing a sufficiently large detection area (ideally $>$1 \si{mm^2}) while maintaining saturated internal detection efficiency. Saturated efficiency means that every photon absorbed by the SNSPD produces a corresponding output pulse. The current state-of-the-art SNSPD has been demonstrated to achieve 1 \si{mm^2} detection area for a single-pixel SNSPD with saturated internal detection efficiency at 1064 nm wavelength \cite{Luskin2023}. However, scaling the area of a single-pixel detector beyond this has been challenging due to non-uniformity and defects in the superconducting thin film, as well as line edge roughness and stitching boundaries introduced during nanofabrication. Multiple sensor chips can be combined to achieve a larger total active area \cite{wollman2019kilopixel}, however a large sensor array will likely require multiplexed readout and biasing given that the number of electrical connections in a cryostat is typically highly constrained. Frequency domain multiplexing has been proposed as particularly well-suited for experiments requiring low bias currents, especially for mid- to far-infrared photon detection \cite{Sypkens2024}.

One promising route to scaling the SNSPD active area without increasing its geometric footprint is coupling impedance-matched nanoantennas or optical metasurfaces to the nanowire~\cite{heath2015nanoantenna}. A properly designed antenna or metasurface will focus incoming radiation on the photon-sensitive nanowire and thus can significantly increase its effective absorption area beyond its geometric area. Since the filling factor of the SNSPD meander in an antenna-coupled configuration is determined by the dimensions and periodicity of the antenna, which scales with the resonant wavelength of the antenna, this antenna-coupled meander geometry will have a significantly reduced meander filling factor compared to conventional SNSPD meander geometry to achieve an equivalent detection area. Thus, in addition to increasing the active area of the sensor, antenna coupling can enhance immunity of the SNSPD to local inhomogeneities and defects in the superconducting film, potentially enabling \si{cm^2}-scale active area. 

While SNSPDs are typically used to detect visible and IR photons, mid-IR sensitivity is required to detect dark photons and axions in the meV mass range. There are two main challenges with extending the sensitivity of SNSPDs to this range. The first is absorption efficiency, which can be improved using nano-antennas \cite{heath2015nanoantenna}. In another project related to mid-infrared SNSPD development, our preliminary simulations showed that an absorption efficiency of 50\% is achievable in SNSPDs with gold bowtie antennas for a plane wave incident at 7.4 \si{\mu m} wavelength. Gold nanostructures have also been shown to dramatically enhance sensor absorption for much longer wavelengths (into the far infrared \cite{Weber2016}). 

The second challenge is to lower the threshold photon energy required to switch a nanowire from a superconducting to normal state. The current approach in the literature is to decrease the critical temperature ($T_c$) of the superconducting films by modifying the film stoichiometry \cite{verma2021single}, which comes at the cost of operating the sensor at milli-Kelvin temperatures. In addition, the superconducting nanowires become narrower and thinner to be sensitive to mid-IR photons \cite{Taylor2023, colangelo2022large}, which reduces photon absorption efficiency while imposing constraints on fabrication and readout performance. The recent state-of-the-art demonstration of 29 \si{\mu m} photon detection in SNSPDs was achieved with WSi film with $T_c$ of 1.3 K, thickness of 3 nm, and 80 nm nanowire width \cite{Taylor2023}. Hence, it is plausible to assume that further extension of SNSPD wavelength sensitivity beyond 29 \si{\mu m} is possible by further decreasing the $T_c$ of the superconducting film; however, the maximum possible wavelength sensitivity of SNSPDs remains an open question in the research community.

For axion DM detection experiments, SNSPDs need to be resilient to external magnetic fields \cite{Baryakhtar:2018doz}. It has been demonstrated in NbN-based SNSPDs that the critical current density ($J_c$) of the SNSPD nanowire follows the relationship $J_c(B) \propto J_c(0) \times B^{-0.02}$, where $B$ is the in-plane magnetic field \cite{polakovic2020superconducting}. This suggests that SNSPDs can withstand higher magnetic fields when the field is applied parallel to the device plane. The reported work showed high photon count rates with low dark counts for SNSPDs operating at 3 K in magnetic fields as high as 5 T (parallel field) and 1 T (perpendicular field). The sputtering conditions of the films were optimized to yield a high density of lattice defects, which was claimed to improve magnetic field tolerance. In addition, further improvement in the magnetic field tolerance of SNSPDs is possible by incorporating nanofabricated pinhole grids into the nanowire structure \cite{zhu2023effective}. 

\begin{table}
\centering
\begin{tabular}{|m{2cm} | m{3cm}| m{3cm}  m{3cm} m{3cm}|}
\cline{2-5}
\multicolumn{1}{c|}{} & \textbf{parameter} & \textbf{Band A} & \textbf{Band B} & \textbf{Band C} \\
\cline{2-5}
\multicolumn{1}{c|}{}& bandwidth & 1-\SI{4}{\mu m}\cite{Oppenheim:62} &  2-\SI{20}{\mu m}\cite{Li1980} & 30-\SI{120}{\mu m} \cite{Wollack2020}\\ 
\multicolumn{1}{c|}{}& material & alumina & KBr/NaCl & HR-Si\\
\multicolumn{1}{c|}{}& $n$ & 1.7 & 1.5 & 3.4  \\
\multicolumn{1}{c|}{}& $\ell_a$ & \SI{10}{\meter}\cite{Aleksey2016} & \SI{100}{\meter}\cite{Li1980} &  1.6-\SI{6.3}{\meter} \cite{Wollack2020} \\
\multicolumn{1}{c|}{}& $\langle R\rangle$ & \SI{0.82}{\mu m} & \SI{2.11}{\mu m} & \SI{20.0}{\mu m}\\
\hline

& $A_\text{sensor}$ & \SI{1}{\mm\squared} &  & \\
\change{\textbf{Pathfinder}} & $V_\text{eff}$ & \SI{5}{\cm^3} & ---  & ---\\
& background rate  & $10^{-5}{\,\rm Hz}$ &  &  \\
\hline

& $A_\text{sensor}$ & \SI{1}{\cm\squared} & \SI{1}{\cm\squared} & \SI{1}{\cm\squared}\\
\textbf{Phase I} & $V_\text{eff}$ & \SI{500}{\cm^3} & \SI{5000}{\cm^3} & 80-\SI{320}{\cm^3} \\
& background rate  & $10^{-3}{\,\rm Hz}$ & $10^{-2}{\,\rm Hz}$ &  $10^{-2}{\,\rm Hz}$\\
\hline
& $A_\text{sensor}$ & \SI{10}{\cm\squared} & \SI{10}{\cm\squared} & \SI{10}{\cm\squared}\\
\textbf{Phase II} & $V_\text{eff}$ & \SI{5000}{\cm^3} & \SI{50000}{\cm^3} & 800-\SI{3200}{\cm^3}\\
& background rate & $10^{-6}{\,\rm Hz}$ & $10^{-6}{\,\rm Hz}$ & $10^{-6}{\,\rm Hz}$\\\hline
\multicolumn{1}{c|}{}& sensor efficiency & 85\% & 50\% & 50\%\\
\cline{2-5}
\end{tabular}
    \caption{Projected parameters for the DPHaSE experiments. To calculate the effective volume, we use the upper limit in Eq. \eqref{eq:Veff}. We also assume a fill factor of $f=0.5$.}\label{tab: parameters}
\end{table}

\subsection{Background Rate and Rejection}  

Background counts in the DPHaSE experiment can arise from intrinsic fluctuation-induced dissipation in the SNSPD (dark counts), as well as from extrinsic sources such as thermal blackbody radiation, luminescence due to the relaxation of mechanical stresses in the powder, energetic cosmic particles and trace radioactive isotopes. Many of these background sources were also present for the LAMPOST experiment, where they added to less than 1 count/day (consistent with a muon background interacting directly with the nanowires)~\cite{Chiles:2021gxk}. While the sensor in the DPHaSE experiment will have to be sensitive to lower energy photons for band B and band C experiments, thermal noise is expected to remain negligible for axion masses $m_a\geq\SI{15}{\milli\eV}$ at liquid helium temperature (\SI{4}{\kelvin}) or below \cite{Baryakhtar:2018doz}. However Cherenkov and bremsstrahlung photons produced within the effective experimental volume may generate background counts at a much higher rate~\cite{Du:2020ldo}. These events must be vetoed with high efficiency. As a primary veto mechanism, we propose using an external tiled array of muon detectors made from scintillator blocks and single-photon avalanche detectors. Such systems can have $>$99.99\% detection efficiency \cite{Artikov2024}.

As a secondary veto mechanism, we will use  the SNSPD with an impedance-matched differential readout and biasing scheme. This gives the SNSPD two new capabilities: spatial resolution from the relative timing of the output pulses on each side of the nanowire \cite{zhao2017single,Colangelo2023} and photon number resolution from the pulse height. 

 The photon number resolution can provide an excellent veto on Cherenkov photon showers. A muon passing through alumina typically produces a photon in the transparency window (\SI{0.2}{\eV} to \SI{3.5}{\eV}) every $\SI{10}{\mu m}$ \cite{Budini1953,Du2024}, causing hundreds of photons to enter the collection chamber where their lifetime is on the order of $\ell_a \sim 10\,\rm{ns}$, which is much smaller than the dead time of the photodetector, on the order of hundreds of ns~\cite{Wollman17}.

The spatial resolution will help identify dark counts arising from material imperfections such as nanowire constrictions and grain boundaries. Since DM-induced detection events in the SNSPD will have a uniform probability of occurring anywhere within the sensor's active area, events occurring at abnormally high frequencies at specific points are likely due to these nanowire constrictions and grain boundaries. SNSPDs with spatial resolution were previously employed to study the geometric origins of dark counts in SNSPDs \cite{zhang2022geometric, andreev2024dark}.

The background count rate of an SNSPD is typically proportional to its area \cite{colangelo2022large}. We can estimate that if the LAMPOST sensor was scaled by a factor of about 600 to an area of \SI{1}{\cm\squared}, the background count rate would increase proportionally to about \SI{1e-3}{\hertz} for the same operating conditions. This is the background count rate we assume for a phase I experiment in Table \ref{tab: parameters}. If the veto systems described in this section are effective, we expect the dark count rate could be substantially reduced. Based on the exponential trend of dark counts with bias current measured on the LAMPOST sensor, we project a background due solely to dark counts could be as low as \SI{1e-6}{\hertz}. This is the background count rate we assume for a phase II experiment.

\section{Expected Reach}\label{sec:reach_plot}

\begin{figure}[t!]
\begin{subfigure}{0.49\textwidth}
    \centering
    \includegraphics[width=\textwidth]{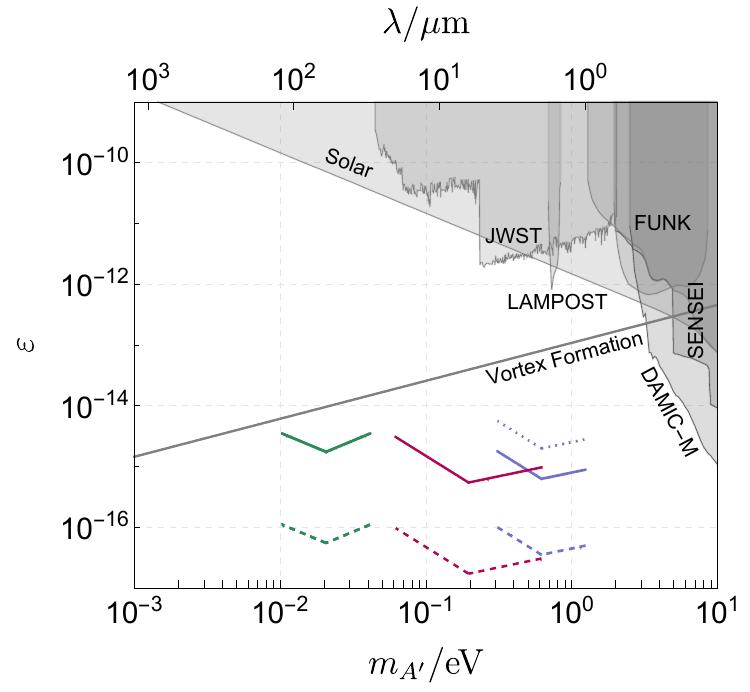}
\end{subfigure}
\begin{subfigure}{0.49\textwidth}
    \centering
    \includegraphics[width=\textwidth]{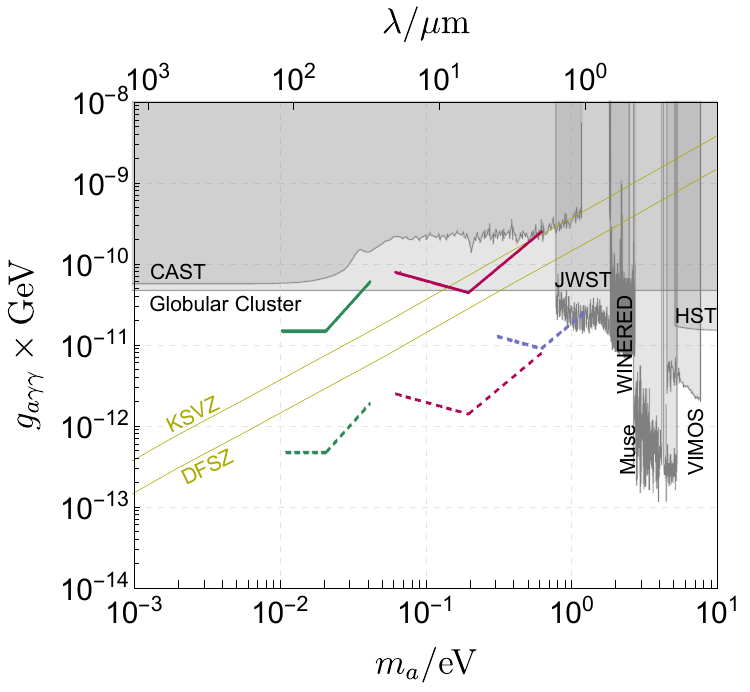}
\end{subfigure}
\caption{Proposed reach for dark photon (\textit{left}) and axion (\textit{right}) DM using 3D disordered dielectric powder. For the dark photon figure we show a vortex formation line (gray), above which additional model-building is required to retain dark photons as a good DM candidate~\cite{Cyncynates:2023zwj, Cyncynates:2024yxm}. The QCD axion KSVZ and DFSZ targets are shown. The values assumed are displayed in Tab.~\ref{tab: parameters}. We also assume a 50\% filling factor and, for the axion, a $10$ T magnetic field. \change{The dotted curve for the dark photon assumes the readily available Pathfinder parameters as given in Tab.~\ref{tab: parameters}, while the solid and dashed curves assume the Phase I and Phase II values, respectively.}
%The solid curves assume Phase I parameter values as given in Tab.~\ref{tab: parameters}, while the dashed curves assume Phase II values.
The transition wavelength is taken to be $3\langle R \rangle$ as can be seen from Fig.~\ref{fig:comsol3D}. The standard deviation is $\sigma_R=0.5 \langle R \rangle$, while the mean radius is a third of the geometric mean of the bandwidth. The blue line corresponds to Band A with mean radius $\langle R \rangle=\SI{0.8}{\mu m}$, the magenta line corresponds to Band B with mean radius $\langle R \rangle=\SI{2}{\mu m}$ and the green line corresponds to Band C with mean radius $\langle R \rangle=\SI{20}{\mu m}$. We assume $\rho_{\rm DM}=0.45~\mathrm{GeV}/\mathrm{cm^3}$. The runtime of the experiment in each band is taken to be 18 months. Existing constraints on dark photons~\cite{Chiles:2021gxk, PhysRevD.102.042001, An:2024kls,DAMIC-M:2025luv,PhysRevLett.134.161002,Li:2023vpv,PhysRevLett.134.011804} and axions~\cite{CAST:2024eil, Dolan:2022kul,Carenza:2023qxh,Todarello:2023hdk,Yin:2024lla,Wang:2023imi} are shown in gray, including the 2022 LAMPOST~\cite{Chiles:2021gxk} result.}
\label{fig:DPHaSE_reach}
\end{figure}

Combining the theoretical computations of the signal power from DM to photon conversion in Sec.~\ref{sec:Section III}, the characterization of the optical transport of the signal photon and the estimate of the effective volume of the experiment in Sec.~\ref{sec:transport}, as well as the experimental design, sensor characterization and background estimate in Sec.~\ref{sec:exp_setup}, we have all the ingredients needed to project the sensitivity of a DM experiment with dielectric powder. Here, we present projections for two benchmark experimental parameter sets across three wavelength bands shown in Tab.~\ref{tab: parameters}, representing near-term and ultimate experimental sensitivities quantitatively.

The projected sensitivity of the experiment, shown in Fig.~\ref{fig:DPHaSE_reach}, is found by using Eq.~\eqref{eq:generic_power} for the 3D case, combined with the experimental setup detailed in Sec.~\ref{sec:exp_setup} and summarized in Tab.~\ref{tab: parameters}. A filling factor of 50\% is assumed. For reference, the maximum theoretical filling factor for spheres of equal radius is around 74\%, while randomly distributed spheres can have a filling factor slightly larger than 60\% \cite{GDScott_1969}. The local DM density is taken to be $\rho_{\rm DM} \simeq \SI{0.45}{\GeV/\cm^3}$ and the runtime of the experiment to be 18 months in each band. The distribution of the powder radii is assumed to be a Gaussian, with standard deviation large enough to ensure a smooth behavior during the transition from the geometrical to Rayleigh regime; specifically we take $\sigma_R = 0.5 \langle R \rangle$, where $\langle R \rangle$ is the mean radius of the powder. 

The geometrical factor $G(f)$ is extracted from the results of the numerical simulations, particularly the 3D data shown in Fig.~\ref{fig:comsol3D}(c), which indicates a transition at around $\lambda_T \simeq 3\langle R \rangle$ for the 50\% filling factor. We choose $G(f)$ such that there is a immediate transition between the Rayleigh regime to the geometric regime, resulting in a continuous sensitivity curve. The $G(f)$ can then be found by the condition that the power in the geometrical and Rayleigh regime are equal at the transition. In a real experiment, however, there can be gaps in the sensitivity, in particular around $\lambda \simeq\langle R \rangle$, due to the presence of photonic bandgaps or the effect of Anderson localization. The frequencies where these gaps appear depend on the exact powder realization and can be identified by transmission experiments, as described previously in Sec.~\ref{sec:transport}. Further, the average values of the powder radii are chosen such that the transition occurs at the geometric mean of each frequency band. The mean radius can be tuned to maximize the sensitivity at a specific frequency in the possibility of a signal. The radii for the blue, magenta, and green lines are, $\SI{0.8}{\mu m}$, $\SI{2}{\mu m}$ and $\SI{20}{\mu m}$, respectively. 

The different colors in Fig.~\ref{fig:DPHaSE_reach} represent the different frequency bands shown in Tab.~\ref{tab: parameters}. Specifically, the blue, magenta and green correspond, respectively, to Band A, B and C. A detailed discussion of the dark count rates and backgrounds can be found in Sec.~\ref{sec:exp_setup}. The solid lines stand for Phase I values and assume a background rate extrapolated from the LAMPOST prototype~\cite{Chiles:2021gxk} dark count rates. The dashed lines stand for the more optimistic Phase II scenario and assume a background rate with an active and detector veto. The material losses of powder at low temperatures have not been fully characterized across all frequencies; in particular, losses at a few THz can be severe and we exclude this range between Bands B and C. At low frequencies (Band C, far-IR), we assume the absorption length increases linearly with the wavelength, as given in Eq.~\eqref{eq:ell}.

Projecting near-term technological improvements (Phase I), we could already improve the reach of dark photon DM in this mass range by three orders of magnitude without a dedicated veto system, reaching below the limit due to studies of vortex formation in the early universe~\cite{East:2022rsi,Cyncynates:2023zwj,Cyncynates:2024yxm}. Further improvements in sensor technology, powder volume, and background suppression in Phase II could extend the reach by an additional two orders of magnitude. For axion DM, a strong magnetic field of about $10\,{\rm T}$ is needed, together with a dedicated veto system to reach new parameter space of an axion like particle. With the Phase II parameters, 
the well-motivated QCD axion can be probed with a broadband experiment, down to masses of $\sim10 \,{\rm meV}$. 
Reaching these very low masses is of particular importance, due to the constraint $m_a \lesssim 60 \,{\rm meV}$ from SN1987A~\cite{Raffelt:1987yt,Chang:2018rso,Carenza:2019pxu,Lucente:2022vuo,Springmann:2024ret,Springmann:2024mjp}.

With the assumption of an immediate transition, the peak sensitivity is achieved at the transition wavelength, $\lambda_T \simeq m^{-1} \simeq 3\langle R \rangle$, where $m$ is the DM mass. If the absorption length $\ell_a$, and therefore the effective volume, is independent of the wavelength (as in the blue and red lines in Fig.~\ref{fig:DPHaSE_reach}), the sensitivity scales as $m^{1/2}$ when $ m>\lambda_T^{-1}$ and as $m^{-3/2}$ when $m < \lambda_T^{-1}$ for dark photons. While for the axion, we have a $m^{3/2}$ scaling when $ m> \lambda_T^{-1} $ and a $m^{-1/2}$ scaling when $m < \lambda_T^{-1}$. For the green line, the absorption length $\ell_a$, and hence the effective volume, scales linearly with the wavelength, see Eq.~\eqref{eq:ell}, therefore all the scalings are multiplied by $m^{1/2}$. At the transition wavelength, we can use Eq.~\eqref{eq:generic_power} to compute the sensitivity of the axion-photon coupling and the dark photon-photon kinetic mixing as a function of the scalable parameters of the experiment, we choose the numerical values to correspond to Band B Phase I (red solid line),
\begin{equation}\label{eq:sensitivity}
    \begin{split}
    g_{a\gamma\gamma} =& \,\SI{4e-11}{\GeV^{-1}}\left( \frac{m_a}{0.2~\text{eV}} \right) \\ 
    \times &\left( \frac{\eta}{0.5} \right)^{-1/2} \left( \frac{\ell_a}{\SI{e4}{\cm}} \right)^{-1/2} \left( \frac{A_{\text{sensor}}}{\SI{}{\cm^2}} \right)^{-1/2} \left( \frac{\Gamma_{\rm bg}}{\SI{e-2}{\hertz}} \right)^{1/4} \left( \frac{T_{\rm exp}}{18~\text{months}} \right)^{-1/4} \left( \frac{B_{\rm ext}}{10~\text{T}} \right)^{-1}, \\ 
    \\
    \varepsilon =& \,\SI{5e-16}{} \left( \frac{\eta}{0.5} \right)^{-1/2} \left( \frac{\ell_a}{\SI{e4}{\cm}} \right)^{-1/2} \left( \frac{A_{\text{sensor}}}{\SI{}{\cm^2}} \right)^{-1/2} \left( \frac{\Gamma_{\rm bg}}{\SI{e-2}{\hertz}} \right)^{1/4} \left( \frac{T_{\rm exp}}{18~\text{months}} \right)^{-1/4}~.\\
    \end{split}
\end{equation}
Equation~\eqref{eq:sensitivity} points to several directions to improve on to achieve a better sensitivity. Apart from increasing the runtime of the experiment ($T_{\rm exp}$), the sensitivity to DM can also be improved by finding materials with a larger absorption length $\ell_a$, and by improving the photon sensor, including increasing the sensor efficiency ($\eta$) and sensor area ($A_{\text{sensor}}$), as well as lowering the intrinsic and environmental background rate ($\Gamma_{\text{bg}}$).

\section{Conclusions}\label{sec:conclusion}
In this paper, we study dark photon and axion DM to photon conversion in a target of {disordered} dielectric material. We also present a novel {broadband} experimental apparatus design combining the dielectric powder conversion target and low-noise SNSPDs. The apparatus can have near-optimal broadband sensitivity to dark photon and axion DM over a wide range of frequencies from $\sim10 \,\rm meV$ to $1\, \rm eV$. Advantages of our proposal include simple target materials and straightforward target preparation, requiring only the purchase of the dielectric powder and its placement in the photon collection chamber, which makes this design convenient to construct and straightforward to scale.

In this novel design, DM to photon conversion occurs on the interfaces between dielectric and vacuum in disordered media. The media can  consist of aligned dielectric fibers (``2D powder'') or dielectric spheres (``3D powder''). DM to photon conversion is broadband and easily calibrated if the powder particles have a wide enough distribution of radii and are randomly packed. At short wavelengths $\lambda \ll 2\pi\langle R \rangle$, the conversion is proportional to the geometric scattering cross-section of light, and is thus determined by the total surface area of the powder packed inside the box. Compared to the area of a  mirror in a similar experimental volume, this leads to an enhancement of $\sim N^{1/3}$, with $N\sim V/\langle R \rangle^3$ the number of 3D powder particles.
 Such an enhancement is comparable to the enhancements achievable in resonant dielectric haloscopes, e.\,g. MADMAX and LAMPOST~\cite{Caldwell:2016dcw,Millar:2016cjp,Baryakhtar:2018doz}, when the radius of the powder is chosen to be comparable with the Compton wavelength of the DM, with much less fabrication/arrangement and no scanning requirements.  At long wavelengths, the conversion is suppressed by a Rayleigh tail; the transition region can be characterized by bulk powder properties such as filling factor and can be calibrated with light transmission measurements.

One disadvantage of the disordered dielectric setup is that the photons produced in the disordered spheres (3D powder) are not collimated, and therefore cannot be focused. To achieve the same sensitivity as a collimated signal setup, photon detectors with larger area, lower dark count rates, and better veto capabilities are needed, suggesting new challenges for detector fabrication. Furthermore, a large volume of dielectric powder, while increasing the signal, will also lead to increased backgrounds due to Cherenkov radiation of cosmic neutrons traveling through the target~\cite{Du:2020ldo}. The properties of these Cherenkov showers are very distinct from the isolated, mono-energetic signal photons and can be vetoed using a combination of a muon veto, multiple photon detectors, and photon number resolution. A full experimental demonstration is of course required, and it is likely that additional backgrounds may arise (e.\,g. from surface modes or relaxation of the dielectric material) as the setup is pushed to lower frequencies than those previously demonstrated.

SNSPDs are continuously in development with improved performance, particularly efforts are ongoing to extend the long-wavelength sensitivity and reduce the sensor dark counts. Combined with internal veto systems including photon number and spatial resolution, the single photon detectors of the near future may achieve effectively background-free operation. 

The current work lays the foundation for several new experimental and theoretical directions. One intriguing option that warrants more careful study is 2D powder. It enjoys the same enhancement of total power over a broad frequency band, while producing signal photons that can be partially focused to increase the signal to background ratio. The signal photons, emitted as cylindrical waves by the 2D powder, can be focused using cylindrical mirrors. Due to the small velocity dispersion of the DM, the momentum of the signal photon along the direction of the fiber is three orders of magnitude smaller than its frequency, and the signal photons can therefore be focused onto a detector that is $10^{-3}$ the area of the cylindrical wavefront.
However, it is unclear how collimated these fibers will need to be in a compact design that optimizes photon focusing. The effect of Anderson localization is also more relevant for 2D powder (as compared to 3D) \cite{Abrahams1979}, and more detailed experimental and theoretical studies are needed to assess the feasibility of using fibers for a DM search. We leave these studies to future work. 

A natural next experimental step is a dedicated experimental study to characterize the properties of dielectric powder across the relevant frequency range to better characterize the properties of the dielectric target. While current literature suggests low absorption in  powders such as alumina, many materials have only been characterized in their bulk form. It is possible that surface modes, e.g. for silica, would give rise to larger absorption in powder than in the bulk; on the other hand, most materials have not been measured at cryogenic temperatures, where losses, especially in so far as they are described the Drude model, would be significantly reduced. Of particular interest is high-resistivity silicon and organic plastic compounds which are transparent in the far-IR and can be used to test the QCD axion at masses below neutron star constraints at $10$s of meV.

Overall, the rapid development of low-noise, single photon detector technology at optical and IR wavelengths motivates new experimental setups to search for axion and dark photon DM. The broadband disordered dielectric approach we present in this paper (see also BREAD~\cite{BREAD:2021tpx}) complements the resonant approaches discussed in~\cite{Caldwell:2016dcw,Baryakhtar:2018doz,MADMAX:2019pub} and could achieve reach to motivated dark photon and axion parameter space in the meV-eV mass range. In the case of a DM signal in a broadband experiment, a resonant experiment can then be designed to precisely characterize the properties of the new particle.

\section{Data Availability}
The data that support the findings of this article are openly available \cite{DPHaSEdata}. 

\begin{acknowledgments}

We thank Baskaran Ganapathy and Robin Kaiser for their insights about Anderson localization, Itay Bloch for providing SENSEI constraint data, and Aaron Chou, Giorgio Gratta, Gray Rybka, Dam Thanh Son, Andrew Sonnenschein,  and Lindley Winslow for helpful discussions. We also thank Phoebe Ayers of the MIT Libraries for assistance with literature review and David Cyncynates and Ella Henry for comments on the manuscript. The COMSOL simulations in this work were performed using the MTL CAD server at MIT. The MIT coauthors acknowledge support for the initial stages of the work from the DOE under the QuantiSED program, Award No. DE-SC0019129. D.\,J.\,P. acknowledges support from the MathWorks Engineering Fellowship. M.\,B. and J.\,H. thank the KITP for hospitality during the completion of this work; this research was supported in part by grant NSF PHY-2309135 to the Kavli Institute for Theoretical Physics (KITP). M.\,B. is supported by the U.\,S. Department of Energy Office of Science under Award Number DE-SC0024375 and the Department of Physics and College of Arts and Science at the University of Washington. M.\,B. is also grateful for the hospitality of Perimeter Institute, where part of this work was carried out. O.~B.~ and J.~H.~ are supported by the Natural Sciences and Engineering Research Council of Canada through a Discovery Grant. Research at Perimeter Institute is supported in part by the Government of Canada through the Department of Innovation, Science and Economic Development and by the Province of Ontario through the Ministry of Colleges and Universities. This work was also supported by a grant from the Simons Foundation (1034867, Dittrich). \change{This material is based upon work supported by the U.S. Department of Energy, Office of Science, Office of High Energy Physics, under Award Number DE-SC0026237}. 

\end{acknowledgments}

\clearpage 
\appendix

\section{Table of Symbols}\label{app:symbols}
\begin{tabular}{p{.1\textwidth}l}
\textbf{symbol} & \textbf{meaning}\\
\hline
$\rho_{\text{DM}}$ & local DM density\\
$a_0$ & axion field amplitude\\
$\vec{E}_0^{\prime}$ & dark photon electric field amplitude\\
$m_a$, $m_{A'}$ & axion mass, dark photon mass\\
$g_{a\gamma\gamma}$ & axion photon coupling \\
$\varepsilon$ & dark photon kinetic mixing parameter \\
$B_{\text{ext}}$ & external magnetic field\\
$E_0$ & axion or dark photon induced electric field, $E_0=g_{a\gamma\gamma}B_{\text{ext}}a_0$ and $E_0=\left|\varepsilon \vec{E}_0^{\prime}\right|$\\
$\vec{S}$ & Poynting vector\\
$\Psingle$ & Signal photon power from a single dielectric particle\\
$P$ & Signal photon power from many dielectric particles\\
$\lambda$ & DM Compton wavelength in vacuum\\
$Q$ & quality factor of a dark matter detector \\
$k$ & wave number of the converted photon $=2\pi/\lambda$ \\
$R$ & dielectric particle radius\\
$\langle R \rangle$ & mean dielectric particle radius\\
$\sigma_R$ & standard deviation of dielectric particle radii\\
$r$ & radial coordinate\\
${\epsilon}$ & (relative) electrical permittivity\\
$\underline{n}$ & dielectric particle complex index of refraction\\
$n$ & real component of index of refraction\\
$\kappa$ & imaginary component of index of refraction\\
$x_{\text{cond}}$ & conductivity in units of $\langle R \rangle$\\
$f$ & filling factor \\ 
$V_\text{box}$ & volume of region containing dielectric podwer particles \\
$V_\text{particle}$ & volume of an individual dielectric particle \\
$V_\text{eff}$ & effective signal-collecting volume \\
$\sigma_s$ & optical scattering cross section \\
$\sigma_\text{DM}$&effective cross section for DM conversion, factoring out DM-photon coupling (see Table.~\eqref{tab: parameters})\\
$G(f)$ & geometric correction factor which adjusts average cross section based on filling factor\\
$\ell_s$ & scattering mean free path\\
$g$ & scattering anisotropy \\
$\ell_s^*$ & transport scattering length \\
$\bar\ell_a$ & effective absorption length of mixed medium \\
$\ell_a$ & absorption length of bulk dielectric material \\
$\ell_d$ & diffusion length\\
$\reflectance_\infty$ & diffuse reflectance of deep bed of dielectric powder\\
$u_\infty$ & optical energy density in powder far from boundary\\
$u_0$ & optical energy density near powder boundary\\
$A_\text{sensor}$ & sensor area \\
$\eta$ & detector efficiency \\
$\Gamma_\text{bg}$ & background count rate\\
\hline
&\qquad\qquad Symbols from Appendix\\
\hline
$\vec{J}$&energy flux\\
$D$&diffusion coefficient\\
$I_0^+$ & average intensity passing into the medium at the boundary\\
$I_0^-$ & average intensity passing out of the medium at the boundary
\end{tabular}

\section{Conversion of Dark Matter into Photons by Dielectric Scatterers}\label{app:computation}

In this Appendix, we present the derivation of the Poynting flux from DM conversion in the presence of a dielectric cylinder and sphere, yielding Eqs. \eqref{eq:poynting_single} and \eqref{eq:poyn_single_sphere} in the main text, and \eqref{eq:recursive}, which are subsequently used to compute the power converted by dielectric structures. In analogy to the dielectric haloscope calculations~\cite{Millar:2016cjp}, we begin by reviewing the scattering of light by a single dielectric fiber, and compute the optical scattering matrix (T-matrix). Building on the machinery of light scattering, we then include a  background DM field and solve for the emission of photons by a single dielectric fiber [Eq.~\eqref{eq:poynting_single}]. Combining the results of optical scattering with a dielectric fiber and of DM to photon conversion at the dielectric fiber, we derive a procedure to compute the DM conversion power from a bundle of fibers using a multiple scattering formalism [Eq.~\eqref{eq:recursive}]. In the last subsection, we compute the power of DM to photon conversion at a dielectric sphere [Eq.~\eqref{eq:poyn_single_sphere}]. The power emitted by a collection of spheres is not discussed here as it is found numerically using COMSOL Multiphysics\textsuperscript{\textregistered} \cite{multiphysics3} (Sec.~\ref{sec:disorder3D}). Throughout this appendix, we restore SI units, including the speed of light $c$, the vacuum electric permittivity $\epsilon_0$ and the vacuum magnetic permeability $\mu_0$.

\subsection{Light scattering by an isolated dielectric fiber}\label{app:onecylinder}

To establish notation and build intuition, we begin by solving the scattering of a plane light wave from a dielectric fiber. The non-magnetic dielectric fiber has radius $R$, length $L\gg R$ (and hence can be treated as infinitely long) and dielectric constant $\epsilon_1$, residing in a infinite medium of dielectric constant $\epsilon_0$. The fiber is aligned along the $z$-axis. Due to cylindrical symmetry, the fields can be decomposed into axial (parallel to the fiber) and transverse (orthogonal to the fiber) directions. With this decomposition, Maxwell's equations are given by \cite{jackson}:
\begin{equation}
\begin{split}
&\vec{B}_t = \frac{1}{k_\alpha^2-k_z^2} \left( \vec{\nabla}_t \left( \pdv{B_z}{z} \right) +i\mu_0\epsilon_\alpha \omega \hat{z} \times \vec{\nabla}_t E_z \right), \\
&\vec{E}_t = \frac{1}{k_\alpha^2-k_z^2} \left( \vec{\nabla}_t \left( \pdv{E_z}{z} \right) -i\omega \hat{z} \times \vec{\nabla}_t B_z \right), \\
&\left( \nabla_t^2  + k_\alpha^2-k_z^2 \right) E_z = 0, \\
&\left( \nabla_t^2 + k_\alpha^2-k_z^2 \right) B_z = 0,
\end{split}
\end{equation}
where $\vec{E}_t$ and $\vec{B}_t$ are the transverse components of the electric and magnetic fields, respectively. The subscript $\alpha=\{0,1\}$ denotes outside and inside the fiber, respectively, with wave vectors $k_0^2=\mu_0 \epsilon_0 \omega^2$ and $k_1=\sqrt{\epsilon_1/\epsilon_0} k_0$. The transverse Laplacian is defined as $\nabla^2_t\equiv\nabla^2-\partial^2/\partial z^2$ \cite{jackson}.

By momentum conservation, in the limit of zero DM velocity, axion and dark photon DM only converts into light with $k_z = 0$. Furthermore, in an external magnetic field that is aligned with the dielectric fiber, axions only convert to photons for $B_z =0$ and $ \pdv{E_z}{z}=0$. Therefore, we also adopt these requirements in this subsection for simplicity. The Maxwell equations then simplify:
\begin{equation}\label{eq:TMnoaxion}
\begin{split}
&\left( \nabla_t^2  + k_\alpha^2 \right) E_z = 0, \\
&\vec{B}_t = \frac{i\mu_0 \epsilon_\alpha \omega }{k_\alpha^2} \left( \hat{z} \times \vec{\nabla}_t E_z \right), \\
&B_z=0~,~~ \vec{E}_t=0~,
\end{split}
\end{equation}
and the problem is reduced to finding the electric field component $E_z$, a scalar function. The solutions of the wave equation in cylindrical coordinates can be decomposed into cylindrical harmonics labeled by index $n$,
\begin{equation}
    \begin{split}
    &E_{z}^{\text{in}} = \sum_n \left[A_n J_n(k_1 \rho) + B_n H^{(1)}_n(k_1 \rho)\right] e^{-in\phi}, \\
    &E_{z}^{\text{out}} = \sum_n \left[ C_n J_n(k_0 \rho) + D_n H^{(1)}_n(k_0 \rho)\right] e^{-in\phi},
    \end{split}
\end{equation}
where $E_{z}^{\text{in}}$ and $E_{z}^{\text{out}}$ are the fields inside and outside the fiber, and $k_1$ and $k_0$ are the momenta inside and outside the fiber, respectively. 
The radial coordinate is denoted by $\rho\in [0,\infty )$, so as to distinguish from $r \gg R$, which appears only in the asymptotic quantities, such as the Poynting vector.
The coefficients $C_n$ of the Bessel functions of the first kind $J_n(x)$ represent the amplitude of the incoming wave on the cylinder, while the coefficients $D_n$ of the Hankel functions of the first kind $H^{(1)}_n(x)$ represent the amplitude of the corresponding outgoing scattered wave that we wish to determine. 

We now determine the coefficients of the expansion. The electric field is non-singular at the origin, which implies that $B_n =0$. Moreover, the fields $E_z$ and $B_t$ are both continuous at the fiber surface ($\rho=R$). Using the orthogonality of Bessel and Hankel functions, these boundary conditions result in the following equalities for each harmonic:
\begin{equation}
    \begin{split}
        &A_n J_n(k_1 R) =  C_n J_n(k_0 R) + D_n H^{(1)}_n(k_0 R), \\
        &\left(\epsilon_1/\epsilon_0\right)^{1/2} A_n  J'_n(k_1 R) = C_n  J'_n(k_0 R) + D_n {H^{(1)}_n}'(k_0 R)~, \\
    \end{split}
\end{equation}
where primes denote a derivative with respect to the argument of the Bessel and Hankel functions.
The scattered wave, determined by the coefficients $D_n$, can then be solved in terms of the incoming wave, determined by the coefficients $C_n$ \cite{Maystre}:
\begin{equation} \label{eq:app_T}
D_n = -\frac{\left(\epsilon_1/\epsilon_0\right)^{1/2} J_n(k_0 R) J'_n(k_1 R) - J_n(k_1 R) J'_n(k_0 R) }{\left(\epsilon_1/\epsilon_0\right)^{1/2}  H_n^{(1)}(k_0 R) J'_n(k_1 R) - J_n(k_1 R){{H^{(1)}_n}}'(k_0 R)} C_n \equiv T_{n,n} C_n.
\end{equation}
The scattering matrix $T_{n,n}$ contains all information about the dielectric fiber. Once the incoming wave is identified, we can use Eq.~\eqref{eq:app_T} to determine the scattered wave from DM conversion to photons by a single dielectric fiber (App.~\ref{app:DMone}) and a bundle of fibers (App.~\ref{app:DMmul}).

\subsection{Dark matter to photon conversion: one fiber}\label{app:DMone}

We now compute the conversion of DM to photons at a single dielectric fiber by modifying Maxwell's equations to include the DM background current. We focus on the case of axion to photon conversion, with the dark photon case being analogous. 
The fiber has permittivity $\epsilon_1$, radius $R$, and is assumed to be infinite and aligned along the $z$-axis. A uniform and constant external magnetic field is applied in the direction of the fiber, $\vec{B}_{\rm ext}=B_{\rm ext}\hat{z}$. 
With the same approximations described around Eq.~\eqref{eq:mod_maxwell}, the modified Maxwell's equations, in the absence of free charges, are \cite{Millar:2016cjp}:
\begin{equation}
\begin{split}
& \vec{\nabla}\cdot\vec{D}=0, \\
& \vec{\nabla}\cdot\vec{B}=0, \\
& \vec{\nabla} \times \vec{E}= - \dot{\vec{B}}, \\
& \vec{\nabla} \times \vec{H} = \dot{\vec{D}} + g_{a\gamma\gamma}\vec{B}_{\rm ext} \dot a, \\
\end{split}
\end{equation}
where $\vec{D}=\epsilon_\alpha \vec{E}$, with $\alpha=0 (1)$  outside  (inside) the fiber. The dielectric fiber is assumed to be non-magnetic so $\vec{H}=\vec{B}/\mu_0$. The wave equation contains an additional axion-induced source term for the electric field:
\begin{equation}\label{eq:DM_wave}
    \begin{split}
       &\nabla^2 \vec{E} - \mu_0 \epsilon_\alpha \ddot{\vec{E}} = \mu_0 g_{a\gamma\gamma}\vec{B}_{\rm ext} \ddot{a}, \\
        &\nabla^2 \vec{B} - \mu_0 \epsilon_\alpha \ddot{\vec{B}} = 0~.\\
    \end{split}
\end{equation}
As in the last subsection, we decompose the fields into transverse and axial components. The axion induced wave satisfies
\begin{equation}\label{eq:app_transversefields}
\begin{split}
&\left( \nabla_t^2  + k_\alpha^2 \right) E_z = -k_\alpha^2 \frac{\epsilon_0}{\epsilon_\alpha} E_0, \\
&\vec{B}_t = \frac{i\mu_0\epsilon_\alpha \omega }{k_\alpha^2} \left( \hat{z} \times \vec{\nabla}_t E_z \right),\\
&B_z=0 ~~,~~ \vec{E}_t=0~,
\end{split}
\end{equation}
where $E_0 \equiv g_{a\gamma\gamma} a_0 |\vec{B}_{\rm ext}|/\epsilon_0$ is the amplitude of the axion-induced electric field, with $a=a_0 e^{-i \omega t}$. We take the limit of zero axion momentum (and therefore $k_z=0$) and all the electromagnetic fields in Eq.~\eqref{eq:app_transversefields} oscillate at the axion frequency $\omega$.

Analogously to the prior analysis, the remaining task is the determination of 
$E_z$. In the DM background, $E_z$ consists of both the homogeneous solution given by the cylindrical harmonics, as well as a particular solution proportional to $E_0$:
\begin{equation}\label{eq:app_hom+part}
    \begin{split}
    &E_{z}^{\text{in}} = \sum_n \left[A_n J_n(k_1 \rho) + B_n H_n^{(1)}(k_1 \rho)\right] e^{-in\phi} - \frac{\epsilon_0}{\epsilon_1}E_0,\\
    &E_{z}^{\text{out}} = \sum_n \left[ C_n J_n(k_0 \rho) + D_n {H^{(1)}_n}'(k_0 \rho)\right] e^{-in\phi}-E_0~.\\
    \end{split}
\end{equation}
The resulting $E_{z}$ has $B_n = 0$ since the electric field cannot be singular at the origin, and $C_n=0$ in the absence of an incoming wave.
The remaining coefficients can then be found by imposing the boundary conditions at the interface $\rho=R$ \cite{Millar:2016cjp}, which leads to the following equations for the cylindrical harmonics:
\begin{equation}
    \begin{split}
        &A_n J_n(k_1 R)  -\delta_{0,n} \frac{E_0}{\left(\epsilon_1/\epsilon_0\right)} =  D_n H_n^{(1)}(k_0 R)-\delta_{0,n}E_0,\\
        &\left(\epsilon_1/\epsilon_0\right)^{1/2} A_n J'_n(k_1 R)  =  D_n {H^{(1)}_n}'(k_0 R).\\
    \end{split}
\end{equation}
Note that the effect of axion to photon conversion only shows up in the zeroeth harmonic in the limit of vanishing axion momentum.

At this point, we can readily see that the $E_0$ term leads to a non-zero $D_n$ coefficient for $n=0$.
This means that the axion sources an outgoing wave in the radial direction, with the electric field pointing in the $z$-direction. This is just an converted wave which flows away from the cylinder, with amplitude determined by the coefficient $D_0$,
\begin{equation} \label{eq:S}
S_{n} \equiv D_n= \delta_{0,n} E_0 \left( 1-\frac{{\epsilon_0}}{{\epsilon_1}} \right) \frac{ J_n'(k_1 R) }{H_n^{(1)}(k_0 R)J'_n(k_1 R) - {({\epsilon_0/\epsilon_1})^{1/2}}J_n(k_1 R) {H^{(1)}_n}'(k_0 R)}.
\end{equation}
 The source coefficient $S_{n}$, as defined here, will be used in the next subsection to denote the wave due to DM to photon conversion, while the coefficient $D_n$ will continue to denote the combined outgoing wave.
With all the coefficients determined, we can find the electric and, correspondingly, the magnetic field using Eq.~\eqref{eq:app_transversefields} and Eq.~\eqref{eq:app_hom+part}, respectively. The Poynting vector is therefore 
\begin{equation}\label{eq:app_poyn_cyl}
    \begin{split}
            \vec{S} &= \frac{E_0^2}{2\mu_0 c} \left(1-\frac{1}{n^2}\right)^2  \\ &\times \frac{|J_0^\prime (k_1 R)|^2}{\Big| H_0^{(1)} (k_0 R) J_0^\prime(k_1 R)  - (1/n) J_0(k_1 R) H_0^{(1)\prime} (k_0 R) \Big|^2} \text{Re}\left[i H_0^{(1)} ( k_0 r) \left(H_0^{(1)\prime} ( k_0 r)\right)^*\right] \hat{r}~,
    \end{split}
\end{equation}
which, in natural units, is Eq.~\eqref{eq:poynting_single}  in the main text. Note that we write this equation in terms of the index of refraction $n^2 \equiv \epsilon_1/\epsilon_0$ to match the main text, not to be confused with the harmonic order $n$, which appears in all equations only as a subscript.

\subsection{Dark matter to photon conversion: multiple fibers}\label{app:DMmul}
In this subsection, we combine the results of Apps.~\ref{app:onecylinder} and~\ref{app:DMone} to compute the electromagnetic field and Poynting vector of a photon generated in a disordered medium consisting of dielectric fibers aligned along the $z$-axis. These fibers are assumed to have a range of radii, as discussed in detail in Sec.~\ref{sec:disorder2D}.

As we found in App.~\ref{app:onecylinder}, the outgoing wave for light scattering off a dielectric fiber is determined by the scattering matrix $(T_{j})_{n,n}$ defined in Eq.~\eqref{eq:app_T}, where the subscript $n$ is the harmonic while $j$ denotes the $j$-th fiber. Further, the effect of DM conversion was in found in App.~\ref{app:DMone} to be the inclusion of an emitted wave sourced by each fiber and given by the coefficients in Eq.~\eqref{eq:S}. These two equations uniquely determine how the outgoing wave at the $j$-th fiber is related to the incoming wave at that fiber. In the case where we have both an incoming wave and DM conversion, the coefficients of the outgoing wave from the $j$-th fiber, $D_{j,n}$, are determined by both the incoming wave at that fiber $C_{j,n}$ and the converted outgoing wave, whose coefficients are denoted through this subsection by $S_{j,n}$ [see Eq.~\eqref{eq:S}]:
\begin{equation}
    D_{j,n} = (T_{j})_{n,n} C_{j,n} + S_{j,n}.
\end{equation}
In what follows we introduce a more compact matrix notation as in \cite{Maystre}, and the previous equation reads,
\begin{equation}\label{eq:D}
    \textbf{D}_{j} = \textbf{T}_{j}\cdot \textbf{C}_{j} + \textbf{S}_{j},
\end{equation}
where $\textbf{D}_{j}$ and $\textbf{C}_{j}$ are column matrices of length $2N+1$ and include effects up to the $N$-th harmonic. $\textbf{T}_{j}$ is a $(2N+1)\times(2N+1)$ diagonal matrix and $\textbf{S}_{j}$ is a column matrix with  only the $n=0$ component nonzero due to the delta function in Eq.~\eqref{eq:S}.
In the simulations of Sec.~\ref{sec:disorder2D}, we truncate the matrix at $N=10$, that is, the harmonics range from $-10$ to $10$. 

A final relation between the coefficients is needed to solve the system: the incoming wave at the $j$-th fiber is related to the outgoing wave from all other fibers. We begin by choosing the $j$-th fiber as the origin of our coordinate system and separating the field outside the $j$-th fiber ($E_j^{\rm out}$) into the incoming and outgoing components. We omit the $z$ index of the electric field in the remainder of the subsection for clarity of notation, and write the electric field at the $j$-th fiber as \cite{Maystre}:
\begin{equation}\label{eq:fields}
\begin{split}
    &E_{j}^{\text{out}} = E_{j}^{\text{outgoing}} + E_{j}^{\text{incoming}},\\
    &E_{j}^{\text{outgoing}}(\rho_j,\phi_j) = \sum_n D_{j,n} H_n(k_0 \rho_j) e^{in\phi_j},\\
    &E_{j}^{\text{incoming}}(\rho_j,\phi_j)  = \sum_n C_{j,n} J_n(k_0 \rho_j) e^{in\phi_j} .
\end{split}
\end{equation}
The subscript index $j$ in the electric field indicates that the field is expressed in the cylindrical coordinate system with the center of the $j$-th fiber as the origin, with $\rho_j$ and $\phi_j$ the respective radial and angular coordinates.  
The incoming wave $E_{j}^{\text{incoming}}$ at the $j$-th fiber is given by the outgoing waves from all other fibers, while $E_{j}^{\text{outgoing}}$ at the $j$-th fiber contains both the photons converted from DM as in Eq.~\eqref{eq:S}, as well as incoming photons $E_{j}^{\text{incoming}}$ which scattered from the $j$-th fiber as in Eq.~\eqref{eq:app_T}. The incoming wave $E_{j}^{\text{incoming}}$ can be expressed as the sum of all outgoing waves in the cylindrical coordinate system of the $j$-th fiber. If we denote the outgoing wave from the $l$-th fiber to the $j$-th fiber by $E_j^{l,\text{outgoing}}$ we have,
\begin{equation}
    E_{j}^{\text{incoming}} = \sum_{l\neq j} E_j^{l,\text{outgoing}}.
\end{equation}
The same quantity $E_j^{l,\text{outgoing}}$ can also be written in the cylindrical coordinate system with the origin at the $l$-th fiber\footnote{The symbols without a superscript $E_{j}^{\text{incoming}}$ and $E_j^{\text{outgoing}}$ should be understood as  $E_{j}^{j,\text{incoming}}$ and $E_j^{j,\text{outgoing}}$, but we omitted the superscript index for simplicity.}. In this coordinate system, the outgoing field is, by the definition above, given in terms of the coefficients $D_{l,m}$ as:
\begin{equation}\label{eq:scat}
E_{l}^{\text{outgoing}} = \sum_m D_{l,m} H_m(k_0 \rho_l) e^{im\phi_l}~.
\end{equation}
The quantities $E_{l}^{\text{outgoing}}$ and $E_j^{l,\text{outgoing}}$ are related by a coordinate transformation, which can be performed using Graf's addition theorem~\cite{NISTBOOK}
for the Hankel function,
\begin{equation}
    H_m(k_0 \rho_l) e^{im\phi_l} = \sum_q e^{i(m-q)\theta_l^j} H_{q-m} (k_0 \rho^l_j) J_q(k_0\rho_j)e^{iq\phi_j},
\end{equation}
where $\rho_j^l\equiv |\vec{\rho_j} - \vec{\rho_l}|$ and $\theta_l^j$ is the angle between $\vec{\rho_j}$ and $\vec{\rho_l}$. 
The terms that depend on $\rho_j^l$ and $\theta_l^j$ are only a function of the relative position of the fibers $j$ and $l$, which is known after we specify the configuration of the fibers, and can be written as a geometrical matrix,
\begin{equation}
    G_{j,l,m,q} =  e^{i(m-q)\theta_l^j} H_{q-m} (k_0 \rho^l_j)~.
\end{equation}
Then the field outgoing from the $l$-th fiber to the $j$-th fiber is (renaming the dummy index):
\begin{equation}
    E_j^{l,\text{outgoing}} = \sum_q D_{l,q} \sum_m G_{j,l,q,m}  J_m(k_0\rho_j)e^{im\phi_j},
\end{equation}
and the total incoming wave on the $j$-th fiber is the sum over $l$ running from all fibers in the system except for the $j$-th fiber,
\begin{equation}\label{eq:incomingalll}
    E_j^{\text{incoming}} = \sum_{l\neq j} E_j^{l,\text{outgoing}} = \sum_{l\neq j} \sum_q D_{l,q} \sum_m G_{j,l,q,m}  J_m(k_0\rho_j)e^{im\phi_j}~.
\end{equation}

Comparing Eq.~\eqref{eq:fields} and Eq.~\eqref{eq:incomingalll}, we can find the coefficients of the incoming wave at the $j$-th fiber in terms of the outgoing waves from all other fibers as,
\begin{equation}
    C_{j,m}=\sum_{l\neq j} \sum_q G_{j,l,q,m}  D_{l,q}~.
\end{equation}
In the matrix notation,
\begin{equation}\label{eq:matrixmaster2}
\textbf{C}_j = \sum_{l\neq j} \textbf{G}_{j,l}\cdot \textbf{D}_l~,
\end{equation}
where $\textbf{C}_{j}$ and $\textbf{D}_l$ are defined as before, while $\textbf{G}_{j,l}$ is a $N\times N$ matrix with components $G_{j,l,q,m}$. This equation together with Eq.~\eqref{eq:D} is enough to determine the outgoing wave at each fiber (Eq.~\eqref{eq:recursive} in the main text):
\begin{equation}
    \begin{split}
        &\textbf{D}_j = \textbf{T}_j \cdot \textbf{C}_j + \textbf{S}_j~,\\
        &\textbf{C}_j = \sum_{l\neq j} \textbf{G}_{j,l} \textbf{D}_l~.\\
    \end{split}
\end{equation}
After we find a solution, we can compute the electric field everywhere, the Poynting vector and total power from dark matter to photon conversion in a 2D dielectric powder consisting of aligned dielectric fiber. 
Note that $\textbf{T}_j$ and $\textbf{S}_j$ depend on the radius of the $j$-th fiber and its index of refraction. While $\textbf{G}_{j,l}$ only depends on the position of the $j$-th fiber with respect to the $l$-th fiber. And all matrices depend on the index of refraction of the medium. Trying to solve for $\textbf{D}$ we get a linear system that can be solved numerically,
\begin{equation}\label{eq:inv}
    \begin{pmatrix} I & -T_1G_{1,2}  & -T_1G_{1,3} &... \\  -T_2G_{2,1}  & I  & -T_2G_{2,3} &... \\   -T_3G_{3,1} & -T_3G_{3,2} & I \\ ... & ... & ... \end{pmatrix} \begin{pmatrix} D_1 \\ D_2 \\ D_3 \\...  \end{pmatrix} = \begin{pmatrix} S_1  \\ S_2 \\ S_3 \\...  \end{pmatrix}.
\end{equation}
We use a python code which is an adaptation of \cite{10.1088/978-1-6817-4301-1} to build the grid of fibers, construct the matrix system and solve it. Once we have the coefficients $\textbf{D}$ for all fibers we can reconstruct the outgoing electric field using Eq.~\eqref{eq:scat} and the total field converted by the system is, $E^{\text{outgoing}} = \sum_l E_j^{l,\text{outgoing}}$. This electric field can then be used to find the transverse magnetic field, the Poynting vector and the integrated emitted power, which we report in the main text.

\subsection{Dark matter to photon conversion: one sphere}\label{app:DMonesphere}
In this subsection we describe the DM conversion by a dielectric sphere. We solve the Maxwell equations for a system consisting of a dielectric sphere of radius $R$ and dielectric constant $\epsilon_1$, residing inside an infinite medium of dielectric constant $\epsilon_0$. The DM acts a source as in Eq.~\eqref{eq:DM_wave},
and the modified wave equations are,
\begin{equation}
    \begin{split}
        &\nabla^2 \vec{E} + k_\alpha^2 \vec{E} = - k_\alpha^2 {\epsilon_0} E_0/\epsilon_\alpha \hat{z}, \\
        &\nabla^2 \vec{B} + k_\alpha^2 \vec{B} = 0 ,\\
    \end{split}
\end{equation}
with $\alpha=0,1$ and $k_1^2 = \epsilon_1/\epsilon_0 \, k_0^2$ as in previous subsections. 
The solution of the homogeneous equation is known from Mie scattering theory~\cite{jackson}, which is given in terms of the vector spherical harmonics $\vec{M}$ and $\vec{N}$, while the particular solution is the same as in App.~\ref{app:DMone}:
\begin{equation} \vec{E} = \sum_m \sum_n \left( A_{emn}\vec{M}_{emn}+A_{omn}\vec{M}_{omn} + B_{emn}\vec{N}_{emn} + B_{omn}\vec{N}_{omn} \right) - {\epsilon_0} E_0/\epsilon_\alpha \hat{z},\end{equation}
where the coefficients $A$ ($B$) are further specified  as $A^{\rm in}$ ($B^{\rm in}$) or $A^{\rm out}$ ($B^{\rm out}$), depending on whether the solution is inside or outside the sphere, respectively. The subscript $e$ denotes the even vector harmonics and $o$ the odd ones. By symmetry, the outgoing field will not depend on $\phi$, so $m=0$ and only the even vector harmonics can have non-zero coefficients,
\begin{equation}\label{eq:electric_field_sphere}
\vec{E} = \sum_n \left( A_{e0n}\vec{M}_{e0n} + B_{e0n}\vec{N}_{e0n} \right) - { \epsilon_0}E_0/\epsilon_\alpha \hat{z}. \end{equation}
The magnetic field can be found with Faraday's law, taking advantage vector spherical harmonics  relations $k_\alpha\vec{M}=\vec{\nabla} \times \vec{N}$ and $k_\alpha\vec{N}=\vec{\nabla} \times \vec{M}$:\begin{equation}\label{eq:magnetic_field_sphere}
\vec{B} = \frac{k_\alpha}{i\omega} \sum_n \left( A_{e0n}\vec{N}_{e0n} + B_{e0n} \vec{M}_{e0n} \right). \end{equation}
Combining the wave equations and the boundary conditions at the interface,
\begin{equation} 
(\vec{E}^{\text{in}} - \vec{E}^{\text{out}}) \cross \hat{r} = (\vec{B}^{\text{in}} - \vec{B}^{\text{out}}) \cross \hat{r} = 0, 
\end{equation}
we find the electric and magnetic field from DM to photon conversion. Since the solutions will only contain the $m=0$ and even vector spherical harmonics, we drop these two subscript indices in the following for notational clarity. After using the spherical harmonics properties to compute the cross products, and using the formulae for the $m=0$ even harmonics, we find: 
\begin{equation}\label{eq:ABharmonics}
    \begin{split}
        &\sum_n A_{n}^{\text{in}} \dv{P^0_n(\cos\theta)}{\theta} j_n (x_{\rm in})  = \sum_n A_n^{\text{out}} \dv{P^0_n(\cos\theta)}{\theta} h^{(1)}_n (x_{\rm out}),  \\
        &\sum_n  B_n^{\text{in}} \dv{P^0_n(\cos\theta)}{\theta} \frac{1}{x_{\rm in}} \dv{(x_{\rm in} j_n(x_{\rm in}))}{x_{\rm in}}  + \frac{\epsilon_0}{\epsilon_1}E_0 \sin\theta = \sum_n B_n^{\text{out}} \dv{P^0_n(\cos\theta)}{\theta} \frac{1}{x_{\text{out}}} \dv{(x_{\text{out}} h^{(1)}_n(x_{\text{out}}))}{x_{\text{out}}} + E_0 \sin\theta, \\
        &\sum_n k_1 B_n^{\text{in}}\dv{P^0_n(\cos\theta)}{\theta} j_n (x_{\rm in}) = \sum_n k_0 B_n^{\text{out}}\dv{P^0_n(\cos\theta)}{\theta} h_n^{(1)} (x_{\rm out}), \\
        &\sum_n k_1 A_n^{\text{in}} \dv{P^0_n(\cos\theta)}{\theta} \frac{1}{x_{\rm in}} \dv{(x_{\rm in} j_n(x_{\rm in}))}{x_{\rm in}} = \sum_n k_0 A_n^{\text{out}} \dv{P^0_n(\cos\theta)}{\theta} \frac{1}{x_{\text{out}}} \dv{(x_{\text{out}} h^{(1)}_n(x_{\text{out}}))}{x_{\text{out}}},
    \end{split}
\end{equation}
where $j_n (x)$ and $h^{(1)}_n(x)$ are spherical Bessel functions and spherical Hankel functions of the first kind, respectively. $P^m_n(\cos\theta)$ are the associated Legendre polynomials, with $P^0_n(\cos\theta) = P_n(\cos\theta)$ the Legendre polynomials. We also temporarily introduce $x_{\rm in}\equiv k_1 R$ and $x_{\rm out} \equiv k_0 R$ for clarity of presentation.  Due to orthogonality, the equations are satisfied for each harmonic independently. Since $\sin\theta = -\dv{P^0_1(\cos\theta)}{\theta}$, the axion sources only the $n=1$ mode. Therefore, Eq.~\eqref{eq:ABharmonics} can be simplified to 
\begin{equation}
    \begin{split}
        &A_1^{\text{in}}  j_1 (x_{\rm in})  = A_1^{\text{out}}  h^{(1)}_1 (x_{\text{out}}),  \\
        &B_1^{\text{in}} \frac{1}{x_{\rm in}} \dv{(x_{\rm in} j_1(x_{\rm in}))}{x_{\rm in} }  +\frac{E_0}{\left(\epsilon_1/\epsilon_0\right)} =  B_1^{\text{out}} \frac{1}{x_{\text{out}}} \dv{(x_{\text{out}} h^{(1)}_1(x_{\text{out}}))}{x_{\text{out}}} + E_0, \\
        &\left(\epsilon_1/\epsilon_0\right)^{1/2} B_1^{\text{in}} j_1(x_{\rm in}) =  B_1^{\text{out}} h^{(1)}_1 (x_{\text{out}}), \\
        &\left(\epsilon_1/\epsilon_0\right)^{1/2} A_1^{\text{in}} \frac{1}{x_{\rm in}} \dv{(x_{\rm in} j_1(x_{\rm in}))}{x_{\rm in}} = A_1^{\text{out}} \frac{1}{x_{\text{out}}} \dv{(x_{\text{out}} h^{(1)}_1(x_{\text{out}}))}{x_{\text{out}}}~. 
    \end{split}
\end{equation}
Due to the orthogonality of the spherical Bessel and spherical Hankel functions, the only solutions are $A_0^{\text{in}}=A_1^{\text{out}}=0$ and
\begin{equation}
    B_n^{\text{out}} = \delta_{n,1} E_0 \left( 1 - \frac{1}{\left(\epsilon_1/\epsilon_0\right)} \right) \frac{\left(\epsilon_1/\epsilon_0\right)^{1/2} k_1 R j_1(k_1 R)}{h^{(1)}_1(k_0 R) [x j_1(x)]^\prime|_{x=k_1 R} - \left(\epsilon_1/\epsilon_0\right) j_1(k_1 R) [x h^{(1)}_1(x)]^\prime|_{x=k_0 R}}~,
\end{equation}
where the primes denote derivatives with respect to $x$. 
Plugging this back into Eqs.~\eqref{eq:electric_field_sphere} and~\eqref{eq:magnetic_field_sphere} and using the explicit formulas for the $m=0$ and $n=1$ even vector spherical harmonics we find: 
\begin{equation}\label{eq:app_sphere_fields}
    \begin{split}
        &\vec{E}^{\rm out} = B_n^{\rm out} \left( 2\cos\theta \frac{h^{(1)}_1(k_0 r)}{k_0 r} \hat{r} - \sin\theta \frac{1}{k_0r} \dv{(x h^{(1)}_1(x))}{x}\Big|_{x=k_0 r} \hat{\theta} \right) - \frac{E_0}{\left(\epsilon_1/\epsilon_0\right)} \hat{z}, \\
        &\vec{B}^{\rm out} = \frac{k_0}{i\omega} B_n^{\rm out} \sin\theta h^{(1)}_1(k_0r) \hat{\phi}~.  \\
    \end{split}
\end{equation}
Combining the electric and magnetic fields, the time averaged Poynting vector can be computed using $\vec{S}=\text{Re}[\vec{E}\times\vec{B}^*]/2\mu_0$. For our application, we are only interested in waves radiating away from the dielectric sphere. Such waves obey the radiation condition which states that for $r\rightarrow \infty$, $\left|\vec{S}\right|\rightarrow 1/r^2$. Keeping only terms that scale as $1/r^2$, the Poynting vector is given by
\begin{equation}\label{eq:app_poyn_sphere}
    \begin{split}
        \vec{S} &= \frac{E_0^2}{2{ \mu_0 c}} \left(1-\frac{1}{n^2}\right)^2 \sin^2\theta
        \\ \times &\Bigg| \frac{n k_1 R j_1(k_1 R)}{h^{(1)}_1(k_0 R) [x j_1(x)]^\prime|_{x=k_1 R} - n^2 j_1(k_1 R) [x h^{(1)}_1(x)]^\prime|_{x=k_0 R}} \Bigg|^2 \text{Re}\left[ i \frac{h_1^{(1)*}( k_0 r) }{k_0 r} [x h_1^{(1)}(x)]^{\prime}\Big|_{x=k_0r} \right] \hat{r}.
    \end{split}
\end{equation}
This corresponds to Eq.~\eqref{eq:poyn_single_sphere} in the main text. Note that we again write the equations in terms of the index of refraction $n^2\equiv{\epsilon_1/\epsilon_0}$ as in Eq.~\eqref{eq:app_poyn_cyl}, not to be confused with the harmonic number $n$, which only appears as a subscript. As expected, the wave converted by a dielectric sphere propagates in the radial direction, and has a dipole-like angular dependence $\sin^2\theta$. 

\section{Radiation gathered by a cavity}
\label{app:collector}

In this section, we analyze a simple diffusion model for radiation transfer in random medium bordering a cavity with a small sensor. First, we model the microscopic dynamics of radiation in the medium of dielectric powder (medium in short hereafter) using a 1D model of a vacuum-dielectric interface. This model motivates definitions for the effective loss length, effective speed of light, and local average energy density. Then we solve the diffusion equation near a vacuum-medium interface where a fraction of the radiation is returned to the interface by a cavity. With this solution, we calculate the average intensity in the cavity, which is also the intensity incident on the sensor.

\subsection{Effective properties of a dielectric mixed medium}

Here, we will use a simple 1D cavity model, shown in the left panel of Fig.~\ref{fig:DMD}, to understand the properties of a mixed medium consisting partly of vacuum and partly of dielectric material. One way such a medium could differ substantially from a bulk dielectric is the presence of surface states, which may increase absorption. The model will not include this or similar effects, but will instead capture the volume-averaged geometric effects of a partial filling factor. Neglecting interference effects,
which we expect will not be important for determining the average properties of a 3D medium, we will use the optical energy $U$ rather than field $E$ as the dynamic variable.  

Consider a 1D cavity of length $L$ bounded by two perfect mirrors. The left portion of the cavity is vacuum with index of refraction $\underline{n}=1$ (vacuum). The right portion of the cavity has a dielectric material with index of refraction $\underline{n}=n+i\kappa$. The fraction of the cavity length filled by the dielectric is $f$. In the dielectric portion, the characteristic loss length is
\begin{equation}
\ell_a=\frac{\lambda}{4\pi\kappa}\ .
\end{equation}
The interface between these regions has reflectivity
\begin{equation}
\reflectance = \frac{(n-1)^2+\kappa^2}{(n+1)^2+\kappa^2}\ .
\end{equation}
Light which passes through the dielectric-vacuum interface and enters the dielectric region returns to the interface after round trip time $\tau_D=2Lfn/c$.
Similarly, the round trip time in the vacuum region is $\tau_V=2L(1-f)/c$. If the energy stored in the vacuum is $U_V(t)$ and the energy stored in the dielectric is $U_D(t)$, then we have
\begin{align}
    \frac{dU_V}{dt}&=-AU_V+BU_D\\
    \frac{dU_D}{dt}&=AU_V-BU_D-CU_D,
\end{align}
where
\begin{align}
A=\frac{(1-\reflectance)}{\tau_V},\quad
B=\frac{(1-\reflectance)}{\tau_D}, \quad
C=\frac{c}{n\ell_a}\ .
\end{align}
Solving this system of equations, we find
\begin{equation}
\begin{bmatrix}
U_V\\
U_D
\end{bmatrix}
=C_+\vec{V}_+e^{-\Gamma_+ t}+C_-\vec{V}_-e^{-\Gamma_- t}\ ,
\end{equation}
where $C_\pm$ are determined by initial conditions, the solution eigenvectors are
\begin{equation}
    \vec{V}_\pm=
    \begin{bmatrix}
        -(1-\Gamma_\mp/A)\\
        1
    \end{bmatrix} ~,
\end{equation}
and the decay rates are
\begin{equation}
\Gamma_\pm=\frac{1}{2}\left(A+B+C\mp\sqrt{(A+B+C)^2-4AC}\right).
\end{equation}
In the limit where $\ell_a\gg L$ (so $A,B\gg C$), we have, to first order in $C/A$,
\begin{align}
\Gamma_+&\approx \frac{1}{2} C\left(1-\frac{B-A}{B+A}\right)=\frac{c}{\ell_a}\frac{f}{1+f(n-1)}\\
\Gamma_-&\approx A+B=\frac{c(1-\reflectance)}{2L(1-f)}\left(1+\frac{(1-f)}{nf}\right)
\end{align}

\begin{align}
    \vec{V}_-&\approx\begin{bmatrix}
        -1-\frac{1}{2}\frac{C}{A}\left(1-\frac{B-A}{B+A}\right)\\1
    \end{bmatrix}\approx\begin{bmatrix}
        -1\\1
    \end{bmatrix}\\
    \vec{V}_+&\approx\begin{bmatrix}
        B/A\\1
    \end{bmatrix}
\end{align}

The eigenvalue $\vec{V}_-$ decays to zero and the system approaches radiative equilibrium at the rate $\Gamma_-\sim c/L$. As the elements of $\vec{V}_-$ sum to zero, this process is not associated with the net dissipation of energy. Energy dissipation takes place at the much slower rate $\Gamma_+\sim c/\ell_a$. From $\vec{V}_+$ we see that for $t\gg1/\Gamma_-$, $U_V/U_D=B/A=(1-f)/fn$. When this ratio is higher, the energy accumulates more in the lossless region of the cavity and so dissipation is slower. We can also write $\Gamma_+=\bar{c}/\bar{\ell_a}$ as the ratio of the effective velocity
\begin{equation}
    \bar{c}=\frac{cU_V+(c/n)U_D}{U_V+U_D}=\frac{c}{1+f(n-1)}\ ,
\end{equation}
and the effective loss length $\bar{\ell}_a=\ell_a/f$. 

When solving the diffusion equation in the next section, we will find the distribution of the local average energy density $u$. Let $u_D=U_D/Lf$ be the energy density in the dielectric portion of the cavity and $u_V=U_V/L(1-f)$ be the energy density in the vacuum portion of the cavity. In radiative equilibrium, the total circulating intensity in the cavity is uniform: $I_\text{total}=I_{\text{forward}}+I_{\text{backward}}=u_Vc=u_Dc/n$. Therefore we will define $u=(U_V+U_D)/L$ so that $I_\text{total}=u\bar{c}$. 
In higher dimensions, the radiation may not be perpendicular to the interface, however in radiative equilibrium the energy densities will still be related by $u_Vc=u_Dc/n=u\bar{c}$ where $u_D=U_D/Vf$ for experiment volume $V$ and likewise for $u_V$.

\subsection{Diffusion equation}

The time-independent diffusion equation comes from Fick's first law 
\begin{equation}
    \vec{J}=-D\vec{\nabla} u \, ,
    \label{Eq:Fick}
\end{equation}
where $D=\bar{c}l_s^*/3$ is the (3D) diffusion constant and $\vec{J}$ is the net intensity (energy flux), plus the steady-state condition
\begin{equation}
    \dot{u}=\mathcal{P}_V-u\bar{c}/\bar{\ell}_a-\vec{\nabla}\cdot\vec{J}=0\ .
\end{equation}
where $\mathcal{P}_V$ is the power density injected by dark matter absorption and $\bar{c}/\bar{\ell}_a$ is the energy dissipation timescale expressed using the effective velocity and effective loss length derived in the last section. Putting these together, we get the diffusion equation
\begin{equation}
    \nabla^2 u = -\frac{3}{\bar{c}\ell_s^*}\mathcal{P}_V+\frac{3}{\bar{\ell}_a\ell_s^*}u \,.
    \label{Eq:Diff}
\end{equation}
From this, we see that the energy density changes (e.g. near boundaries) on a characteristic length scale $\ell_d=(\bar{\ell}_a\ell_s^*/3)^{1/2}$. We also can immediately see that in the bulk, where $\nabla^2u=0$, the steady state energy density is $u_\infty=\mathcal{P}_V \bar{\ell}_a/\bar{c}$. We will assume that the radiation has an isotropic angular distribution and define the circulating intensity $I_\infty$ as the projection of the Poynting vector $|\vec{S}|=u/4\pi$ in arbitrary direction $\hat{z}$ ($\theta=0$) integrated over the surrounding hemisphere,
\begin{equation}
I_\infty=\int_0^{2\pi} d\phi\int_{0}^{\pi/2} d\theta\sin(\theta) \vec{S}(\Omega)\cdot\hat{z}\\ = \bar{c} u/4\,.
\end{equation}

Naively, we may expect that a sensor of area $A_\text{sensor}$ with perfect efficiency would absorb power $A_\text{sensor}I_\infty$. However, the power circulating near an absorbing boundary will be smaller than $I_\infty$.

\subsection{Boundary conditions}

\begin{figure}[t]
    \centering
    \includegraphics[width=0.9\textwidth]{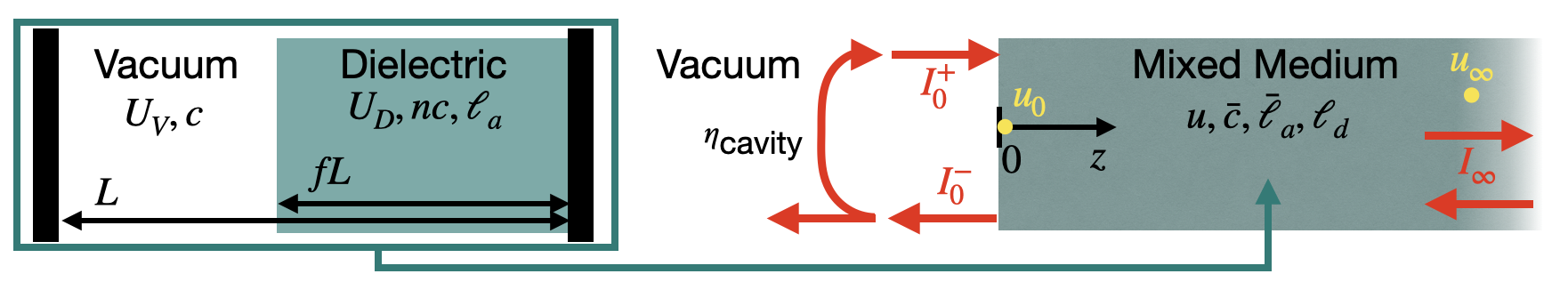}
    \caption{Left: 1D cavity model of the interior of a random mixed medium. A fraction $f$ of the cavity of length $L$ is filled with a dielectric material with index of refraction $n$ and loss length $\ell_a$. We neglect interference effects by tracking energies $U_V$ and $U_D$ stored in the vacuum and dielectric portions, respectively. In radiative equilibrium, we can define an effective speed of light $\bar{c}$, effective loss length $\bar\ell_a$, and local average energy density $u$, which can be used to describe a homogenous mixed medium. Right: Interface between a mixed medium and vacuum. Transport in the mixed medium is modeled as a diffusive random walk with diffusion length $\ell_d$. The subscripts $0$ and $\infty$ indicate quantities which are at or far from the interface, respectively. If the vacuum region is a large cavity surrounded by the mixed medium, the intensity $I_0^+$ incident on the interface is a fraction $\eta_\text{cavity}$ of the intensity $I_0^-$ leaving the interface. }
    \label{fig:DMD}
\end{figure}

Near a boundary, the diffusion model can be inaccurate and the full radiation transport equation has no closed-form solution. Simplified transport models discretize the radiation into $n$ directions ($n$-flux models) \cite{kuhn1993infrared}. Another approach is to add a transition region near the boundary where incoming ballistic photons are ``converted" to diffusive photons \cite{scheffold2022transport}. As we expect the radiation in the cavity to be diffuse, we will treat the entire system using the diffusion equation. 

Consider an interface with vacuum for $z<0$ and a random medium for $z>0$ as shown in Fig.~\ref{fig:DMD}. Let $I^+_0$ and $I^-_0$ be the inflowing and outflowing intensities, respectively: 
\begin{align}
    I^{+}_0&=\int_0^{2\pi} d\phi\int_{0}^{\pi/2} d\theta\sin(\theta) \vec{S}(z=0,\Omega)\cdot\hat{z}\ ,\\
    I^{-}_0&=\int_0^{2\pi} d\phi\int_{\pi/2}^{\pi} d\theta\sin(\theta) \vec{S}(z=0,\Omega)\cdot(-\hat{z})\ .
\end{align}
If the medium near the boundary is charged with energy density $u(0)=u_0$, then, assuming it radiates isotropically, the instantaneous leftward intensity is $I^{-}_0=\bar{c}u_0/4$. In a reflectance or transmittance experiment, a rightward intensity $I^+_0$ is applied to the boundary. From Fick's law [Eq. \eqref{Eq:Fick}] we then get the boundary condition 
\begin{equation}
    D\partial_z u|_{z=0}=I^{-}_0-I^+_0\ .
    \label{BC:R/T}
\end{equation}
Near a partially reflecting boundary, we have $I^+_0=\reflectance I^-_0$ so that $J_z=I^+_0-I^-_0=-(1-\reflectance)I^-_0=-(1-\reflectance)\bar{c}u_0/4$ which gives  boundary condition
\begin{equation}
    D\partial_z u|_{z=0}=(1-\reflectance)\bar{c}u_0/4\ .
    \label{BC:A}
\end{equation}

Combining the diffusion equation [Eq.~\eqref{Eq:Diff}], the boundary condition at the interface [Eq.~\eqref{BC:A}], and the boundary condition at infinity [$u(z\rightarrow\infty)=u_\infty$], the system can be solved. In the next two subsections, we consider two examples with and without dark matter to photon conversion.

\subsection{Diffuse reflectance}

Before proceeding to solve the diffusion equation for a cavity, it will be useful to calculate the diffuse reflectance $\reflectance_\infty$ in the context of this model and obtain Eq. \eqref{eq:Rinf}. This quantity is directly measurable using diffuse reflectance spectroscopy as described in Sec. \ref{sec:transport}, and will be a key parameter for determining the cavity size required for saturating the maximum effective volume of the detector. If intensity $I_0^{+}$ is applied to the boundary from an external source and $I_0^{-}$ is returned, then $\reflectance_\infty=I_0^{+}/I_0^{-}$. In this scenario, we will assume the power density injected by dark matter absorption is negligible ($\mathcal{P}_V=0$) so $u_\infty=0$. 

The ansatz
\begin{equation}
    u(z)=u_0 e^{-z/\ell_d}
\end{equation}
solves the diffusion equation and satisfies the boundary conditions if
\begin{equation}
    u_0\bar{c}/4=\frac{I_0^+}{D/4\ell_d\bar{c}+1}\ .
\end{equation}
The diffuse reflectance is then
\begin{align}\label{eq:Rinfty}
    \reflectance_\infty=\frac{I_0^-}{I_0^+}=\frac{1}{4D/\bar{c}\ell_d+1}\approx 1-4\frac{
    \ell_d}{\bar\ell_a}\ .
\end{align}
Re-arranging, we obtain Eq. \eqref{eq:Rinf}, which is used in the main text to link measurements of $\reflectance_\infty$ to the scattering and absorbing lengths in the powder
\begin{equation}
    \frac{\ell_s^*}{\bar\ell_a}=\frac{3(1-\reflectance_\infty)^2}{16\reflectance_\infty^2}\,.
\end{equation}

\subsection{Energy density in a cavity}

In this subsection, we include the effect of dark matter to photon conversion in the medium. If the cavity is large compared to the diffusion length and effectively establishes an  isotropic and homogeneous radiation environment, the radiation distribution near a flat or gradually curving boundary will be insensitive to the large-scale geometry. Therefore we will again treat this problem in 1D by bending the z-axis along with the surface of the cavity. As shown in the right panel of Fig.~\ref{fig:DMD}, we focus on a small region around the interface at $z=0$ separating the cavity (Left) and the medium (Right). Let $I^{+}_0$ be the isotropic intensity circulating in the cavity (i.e. the intensity measured by a small, one-sided, planar sensor). We can relate $I^{+}_0$ and $I^{-}_0$ using a power balance
\begin{equation}
    A_\text{cavity}(I^{-}_0-I^{+}_0)=(1-\reflectance_\text{sensor})A_\text{sensor}I^{+}_0\ ,
\end{equation}
where the left hand side represents the net power flowing between the powder and the cavity, and the right hand side represents the power absorbed by the sensor. Re-arranging, we have
\begin{equation}
    I^{+}_0=\frac{A_\text{cavity}}{A_\text{cavity}+(1-\reflectance_\text{sensor})A_\text{sensor}}I^{-}_0\equiv \eta_\text{cavity} I^{-}_0\ .
\end{equation}
Solving the diffusion equation (Eq.~\ref{Eq:Diff}) using the boundary condition at the interface
\begin{equation}
    D\partial_z u|_{z=0}=(1-\eta_\text{cavity}) \bar{c}u_0/4\ ,
\end{equation}
and at infinity, $u(z\rightarrow\infty)=u_\infty$, we obtain
\begin{equation}
    u(z)=u_\infty\left(1-(1-u_0/u_\infty)e^{-z/\ell_d}\right)\ ,
\end{equation}
with the energy density at the interface
\begin{equation}
    u_0=u_\infty\left(1+(1-\eta_\text{cavity})\frac{\ell_d\bar{c}}{4D}\right)^{-1}\ .
\end{equation}
Using the expression for $\reflectance_\infty$ from the previous subsection, Eq.~\ref{eq:Rinfty}, we have
\begin{equation}
    u_0\approx u_\infty \frac{1-\reflectance_\infty}{(1-\reflectance_{\infty})+(1-\eta_\text{cavity})}\ .
\end{equation}
Therefore, the intensity in the cavity (also the intensity incident on the detector) is
\begin{align}
    I^{+}_0&=\frac{1}{4}\eta_\text{cavity}u_0\bar{c}\nonumber\\
    &=\frac{1}{4} \frac{\eta_\text{cavity}(1-\reflectance_\infty)}{(1-\reflectance_{\infty})+(1-\eta_\text{cavity})}\bar{c}u_\infty\nonumber\\
    &=\frac{1}{4} \frac{\eta_\text{cavity}(1-\reflectance_\infty)}{(1-\reflectance_{\infty})+(1-\eta_\text{cavity})}\frac{\mathcal{P}_V\ell_a}{f}.
\end{align}
This result can then be used in Sec.~\ref{sec:Veff} to compute effective volume of the dark matter experiment.

\bibliography{Ref}% Produces the bibliography via BibTeX.

\begin{thebibliography}{158}%
\makeatletter
\providecommand \@ifxundefined [1]{%
 \@ifx{#1\undefined}
}%
\providecommand \@ifnum [1]{%
 \ifnum #1\expandafter \@firstoftwo
 \else \expandafter \@secondoftwo
 \fi
}%
\providecommand \@ifx [1]{%
 \ifx #1\expandafter \@firstoftwo
 \else \expandafter \@secondoftwo
 \fi
}%
\providecommand \natexlab [1]{#1}%
\providecommand \enquote  [1]{``#1''}%
\providecommand \bibnamefont  [1]{#1}%
\providecommand \bibfnamefont [1]{#1}%
\providecommand \citenamefont [1]{#1}%
\providecommand \href@noop [0]{\@secondoftwo}%
\providecommand \href [0]{\begingroup \@sanitize@url \@href}%
\providecommand \@href[1]{\@@startlink{#1}\@@href}%
\providecommand \@@href[1]{\endgroup#1\@@endlink}%
\providecommand \@sanitize@url [0]{\catcode `\\12\catcode `\$12\catcode `\&12\catcode `\#12\catcode `\^12\catcode `\_12\catcode `\%12\relax}%
\providecommand \@@startlink[1]{}%
\providecommand \@@endlink[0]{}%
\providecommand \url  [0]{\begingroup\@sanitize@url \@url }%
\providecommand \@url [1]{\endgroup\@href {#1}{\urlprefix }}%
\providecommand \urlprefix  [0]{URL }%
\providecommand \Eprint [0]{\href }%
\providecommand \doibase [0]{https://doi.org/}%
\providecommand \selectlanguage [0]{\@gobble}%
\providecommand \bibinfo  [0]{\@secondoftwo}%
\providecommand \bibfield  [0]{\@secondoftwo}%
\providecommand \translation [1]{[#1]}%
\providecommand \BibitemOpen [0]{}%
\providecommand \bibitemStop [0]{}%
\providecommand \bibitemNoStop [0]{.\EOS\space}%
\providecommand \EOS [0]{\spacefactor3000\relax}%
\providecommand \BibitemShut  [1]{\csname bibitem#1\endcsname}%
\let\auto@bib@innerbib\@empty
%</preamble>
\bibitem [{\citenamefont {Rubin}\ and\ \citenamefont {Ford}(1970)}]{Rubin:1970zza}%
  \BibitemOpen
  \bibfield  {author} {\bibinfo {author} {\bibfnamefont {V.~C.}\ \bibnamefont {Rubin}}\ and\ \bibinfo {author} {\bibfnamefont {W.~K.}\ \bibnamefont {Ford}, \bibfnamefont {Jr.}},\ }\href {https://doi.org/10.1086/150317} {\bibfield  {journal} {\bibinfo  {journal} {Astrophys. J.}\ }\textbf {\bibinfo {volume} {159}},\ \bibinfo {pages} {379} (\bibinfo {year} {1970})}\BibitemShut {NoStop}%
%%CITATION = ASJOA,159,379;%%
\bibitem [{\citenamefont {Spergel}\ \emph {et~al.}(2003)\citenamefont {Spergel} \emph {et~al.}}]{Spergel:2003cb}%
  \BibitemOpen
  \bibfield  {author} {\bibinfo {author} {\bibfnamefont {D.~N.}\ \bibnamefont {Spergel}} \emph {et~al.} (\bibinfo {collaboration} {WMAP}),\ }\href {https://doi.org/10.1086/377226} {\bibfield  {journal} {\bibinfo  {journal} {Astrophys. J. Suppl.}\ }\textbf {\bibinfo {volume} {148}},\ \bibinfo {pages} {175} (\bibinfo {year} {2003})},\ \Eprint {https://arxiv.org/abs/astro-ph/0302209} {arXiv:astro-ph/0302209 [astro-ph]} \BibitemShut {NoStop}%
%%CITATION = ASTRO-PH/0302209;%%
\bibitem [{\citenamefont {Aghanim}\ \emph {et~al.}(2020)\citenamefont {Aghanim} \emph {et~al.}}]{Aghanim:2018eyx}%
  \BibitemOpen
  \bibfield  {author} {\bibinfo {author} {\bibfnamefont {N.}~\bibnamefont {Aghanim}} \emph {et~al.} (\bibinfo {collaboration} {Planck}),\ }\href {https://doi.org/10.1051/0004-6361/201833910} {\bibfield  {journal} {\bibinfo  {journal} {Astron. Astrophys.}\ }\textbf {\bibinfo {volume} {641}},\ \bibinfo {pages} {A6} (\bibinfo {year} {2020})},\ \Eprint {https://arxiv.org/abs/1807.06209} {arXiv:1807.06209 [astro-ph.CO]} \BibitemShut {NoStop}%
\bibitem [{\citenamefont {Peccei}\ and\ \citenamefont {Quinn}(1997)}]{Peccei+1977}%
  \BibitemOpen
  \bibfield  {author} {\bibinfo {author} {\bibfnamefont {R.~D.}\ \bibnamefont {Peccei}}\ and\ \bibinfo {author} {\bibfnamefont {H.~R.}\ \bibnamefont {Quinn}},\ }\href@noop {} {\bibfield  {journal} {\bibinfo  {journal} {Physical Review Letters}\ }\textbf {\bibinfo {volume} {38}},\ \bibinfo {pages} {1440} (\bibinfo {year} {1997})}\BibitemShut {NoStop}%
\bibitem [{\citenamefont {Peccei}\ and\ \citenamefont {Quinn}(1977)}]{Peccei:1977ur}%
  \BibitemOpen
  \bibfield  {author} {\bibinfo {author} {\bibfnamefont {R.~D.}\ \bibnamefont {Peccei}}\ and\ \bibinfo {author} {\bibfnamefont {H.~R.}\ \bibnamefont {Quinn}},\ }\href {https://doi.org/10.1103/PhysRevD.16.1791} {\bibfield  {journal} {\bibinfo  {journal} {Phys. Rev. D}\ }\textbf {\bibinfo {volume} {16}},\ \bibinfo {pages} {1791} (\bibinfo {year} {1977})}\BibitemShut {NoStop}%
\bibitem [{\citenamefont {Weinberg}(1978)}]{Weinberg1978}%
  \BibitemOpen
  \bibfield  {author} {\bibinfo {author} {\bibfnamefont {S.}~\bibnamefont {Weinberg}},\ }\href@noop {} {\bibfield  {journal} {\bibinfo  {journal} {{Physical Review Letters}}\ }\textbf {\bibinfo {volume} {40}},\ \bibinfo {pages} {{223}} (\bibinfo {year} {1978})}\BibitemShut {NoStop}%
\bibitem [{\citenamefont {Wilczek}(1978)}]{Wilczek1978}%
  \BibitemOpen
  \bibfield  {author} {\bibinfo {author} {\bibfnamefont {F.}~\bibnamefont {Wilczek}},\ }\href@noop {} {\bibfield  {journal} {\bibinfo  {journal} {{Physical Review Letters}}\ }\textbf {\bibinfo {volume} {40}},\ \bibinfo {pages} {279} (\bibinfo {year} {1978})}\BibitemShut {NoStop}%
\bibitem [{\citenamefont {Holdom}(1986)}]{Holdom:1985ag}%
  \BibitemOpen
  \bibfield  {author} {\bibinfo {author} {\bibfnamefont {B.}~\bibnamefont {Holdom}},\ }\href {https://doi.org/10.1016/0370-2693(86)91377-8} {\bibfield  {journal} {\bibinfo  {journal} {Phys. Lett.}\ }\textbf {\bibinfo {volume} {B166}},\ \bibinfo {pages} {196} (\bibinfo {year} {1986})}\BibitemShut {NoStop}%
%%CITATION = PHLTA,B166,196;%%
\bibitem [{\citenamefont {Okun}(1982)}]{Okun:1982xi}%
  \BibitemOpen
  \bibfield  {author} {\bibinfo {author} {\bibfnamefont {L.~B.}\ \bibnamefont {Okun}},\ }\href@noop {} {\bibfield  {journal} {\bibinfo  {journal} {Sov. Phys. JETP}\ }\textbf {\bibinfo {volume} {56}},\ \bibinfo {pages} {502} (\bibinfo {year} {1982})}\BibitemShut {NoStop}%
\bibitem [{\citenamefont {Arvanitaki}\ \emph {et~al.}(2010)\citenamefont {Arvanitaki}, \citenamefont {Dimopoulos}, \citenamefont {Dubovsky}, \citenamefont {Kaloper},\ and\ \citenamefont {March-Russell}}]{Arvanitaki:2009fg}%
  \BibitemOpen
  \bibfield  {author} {\bibinfo {author} {\bibfnamefont {A.}~\bibnamefont {Arvanitaki}}, \bibinfo {author} {\bibfnamefont {S.}~\bibnamefont {Dimopoulos}}, \bibinfo {author} {\bibfnamefont {S.}~\bibnamefont {Dubovsky}}, \bibinfo {author} {\bibfnamefont {N.}~\bibnamefont {Kaloper}},\ and\ \bibinfo {author} {\bibfnamefont {J.}~\bibnamefont {March-Russell}},\ }\href {https://doi.org/10.1103/PhysRevD.81.123530} {\bibfield  {journal} {\bibinfo  {journal} {Phys. Rev. D}\ }\textbf {\bibinfo {volume} {81}},\ \bibinfo {pages} {123530} (\bibinfo {year} {2010})},\ \Eprint {https://arxiv.org/abs/0905.4720} {arXiv:0905.4720 [hep-th]} \BibitemShut {NoStop}%
\bibitem [{\citenamefont {Abbott}\ and\ \citenamefont {Sikivie}(1983)}]{Abbott:1982af}%
  \BibitemOpen
  \bibfield  {author} {\bibinfo {author} {\bibfnamefont {L.~F.}\ \bibnamefont {Abbott}}\ and\ \bibinfo {author} {\bibfnamefont {P.}~\bibnamefont {Sikivie}},\ }\href {https://doi.org/10.1016/0370-2693(83)90638-X} {\bibfield  {journal} {\bibinfo  {journal} {Phys. Lett. B}\ }\textbf {\bibinfo {volume} {120}},\ \bibinfo {pages} {133} (\bibinfo {year} {1983})}\BibitemShut {NoStop}%
\bibitem [{\citenamefont {Dine}\ and\ \citenamefont {Fischler}(1983)}]{Dine:1982ah}%
  \BibitemOpen
  \bibfield  {author} {\bibinfo {author} {\bibfnamefont {M.}~\bibnamefont {Dine}}\ and\ \bibinfo {author} {\bibfnamefont {W.}~\bibnamefont {Fischler}},\ }\href {https://doi.org/10.1016/0370-2693(83)90639-1} {\bibfield  {journal} {\bibinfo  {journal} {Phys. Lett. B}\ }\textbf {\bibinfo {volume} {120}},\ \bibinfo {pages} {137} (\bibinfo {year} {1983})}\BibitemShut {NoStop}%
\bibitem [{\citenamefont {Preskill}\ \emph {et~al.}(1983)\citenamefont {Preskill}, \citenamefont {Wise},\ and\ \citenamefont {Wilczek}}]{Preskill:1982cy}%
  \BibitemOpen
  \bibfield  {author} {\bibinfo {author} {\bibfnamefont {J.}~\bibnamefont {Preskill}}, \bibinfo {author} {\bibfnamefont {M.~B.}\ \bibnamefont {Wise}},\ and\ \bibinfo {author} {\bibfnamefont {F.}~\bibnamefont {Wilczek}},\ }\href {https://doi.org/10.1016/0370-2693(83)90637-8} {\bibfield  {journal} {\bibinfo  {journal} {Phys. Lett.}\ }\textbf {\bibinfo {volume} {B120}},\ \bibinfo {pages} {127} (\bibinfo {year} {1983})}\BibitemShut {NoStop}%
\bibitem [{\citenamefont {Sikivie}(1982)}]{Sikivie:1982qv}%
  \BibitemOpen
  \bibfield  {author} {\bibinfo {author} {\bibfnamefont {P.}~\bibnamefont {Sikivie}},\ }\href {https://doi.org/10.1103/PhysRevLett.48.1156} {\bibfield  {journal} {\bibinfo  {journal} {Phys. Rev. Lett.}\ }\textbf {\bibinfo {volume} {48}},\ \bibinfo {pages} {1156} (\bibinfo {year} {1982})}\BibitemShut {NoStop}%
\bibitem [{\citenamefont {Vilenkin}\ and\ \citenamefont {Everett}(1982)}]{Vilenkin:1982ks}%
  \BibitemOpen
  \bibfield  {author} {\bibinfo {author} {\bibfnamefont {A.}~\bibnamefont {Vilenkin}}\ and\ \bibinfo {author} {\bibfnamefont {A.~E.}\ \bibnamefont {Everett}},\ }\href {https://doi.org/10.1103/PhysRevLett.48.1867} {\bibfield  {journal} {\bibinfo  {journal} {Phys. Rev. Lett.}\ }\textbf {\bibinfo {volume} {48}},\ \bibinfo {pages} {1867} (\bibinfo {year} {1982})}\BibitemShut {NoStop}%
\bibitem [{\citenamefont {Vilenkin}(1985)}]{Vilenkin:1984ib}%
  \BibitemOpen
  \bibfield  {author} {\bibinfo {author} {\bibfnamefont {A.}~\bibnamefont {Vilenkin}},\ }\href {https://doi.org/10.1016/0370-1573(85)90033-X} {\bibfield  {journal} {\bibinfo  {journal} {Phys. Rept.}\ }\textbf {\bibinfo {volume} {121}},\ \bibinfo {pages} {263} (\bibinfo {year} {1985})}\BibitemShut {NoStop}%
\bibitem [{\citenamefont {Davis}(1986)}]{Davis:1986xc}%
  \BibitemOpen
  \bibfield  {author} {\bibinfo {author} {\bibfnamefont {R.~L.}\ \bibnamefont {Davis}},\ }\href {https://doi.org/10.1016/0370-2693(86)90300-X} {\bibfield  {journal} {\bibinfo  {journal} {Phys. Lett.}\ }\textbf {\bibinfo {volume} {B180}},\ \bibinfo {pages} {225} (\bibinfo {year} {1986})}\BibitemShut {NoStop}%
%%CITATION = PHLTA,B180,225;%%
\bibitem [{\citenamefont {Borsanyi}\ \emph {et~al.}(2016)\citenamefont {Borsanyi} \emph {et~al.}}]{Borsanyi:2016ksw}%
  \BibitemOpen
  \bibfield  {author} {\bibinfo {author} {\bibfnamefont {S.}~\bibnamefont {Borsanyi}} \emph {et~al.},\ }\href {https://doi.org/10.1038/nature20115} {\bibfield  {journal} {\bibinfo  {journal} {Nature}\ }\textbf {\bibinfo {volume} {539}},\ \bibinfo {pages} {69} (\bibinfo {year} {2016})},\ \Eprint {https://arxiv.org/abs/1606.07494} {arXiv:1606.07494 [hep-lat]} \BibitemShut {NoStop}%
\bibitem [{\citenamefont {Dine}\ \emph {et~al.}(2017)\citenamefont {Dine}, \citenamefont {Draper}, \citenamefont {Stephenson-Haskins},\ and\ \citenamefont {Xu}}]{Dine:2017swf}%
  \BibitemOpen
  \bibfield  {author} {\bibinfo {author} {\bibfnamefont {M.}~\bibnamefont {Dine}}, \bibinfo {author} {\bibfnamefont {P.}~\bibnamefont {Draper}}, \bibinfo {author} {\bibfnamefont {L.}~\bibnamefont {Stephenson-Haskins}},\ and\ \bibinfo {author} {\bibfnamefont {D.}~\bibnamefont {Xu}},\ }\href {https://doi.org/10.1103/PhysRevD.96.095001} {\bibfield  {journal} {\bibinfo  {journal} {Phys. Rev. D}\ }\textbf {\bibinfo {volume} {96}},\ \bibinfo {pages} {095001} (\bibinfo {year} {2017})},\ \Eprint {https://arxiv.org/abs/1705.00676} {arXiv:1705.00676 [hep-ph]} \BibitemShut {NoStop}%
\bibitem [{\citenamefont {Buschmann}\ \emph {et~al.}(2022)\citenamefont {Buschmann}, \citenamefont {Foster}, \citenamefont {Hook}, \citenamefont {Peterson}, \citenamefont {Willcox}, \citenamefont {Zhang},\ and\ \citenamefont {Safdi}}]{Buschmann:2021sdq}%
  \BibitemOpen
  \bibfield  {author} {\bibinfo {author} {\bibfnamefont {M.}~\bibnamefont {Buschmann}}, \bibinfo {author} {\bibfnamefont {J.~W.}\ \bibnamefont {Foster}}, \bibinfo {author} {\bibfnamefont {A.}~\bibnamefont {Hook}}, \bibinfo {author} {\bibfnamefont {A.}~\bibnamefont {Peterson}}, \bibinfo {author} {\bibfnamefont {D.~E.}\ \bibnamefont {Willcox}}, \bibinfo {author} {\bibfnamefont {W.}~\bibnamefont {Zhang}},\ and\ \bibinfo {author} {\bibfnamefont {B.~R.}\ \bibnamefont {Safdi}},\ }\href {https://doi.org/10.1038/s41467-022-28669-y} {\bibfield  {journal} {\bibinfo  {journal} {Nature Commun.}\ }\textbf {\bibinfo {volume} {13}},\ \bibinfo {pages} {1049} (\bibinfo {year} {2022})},\ \Eprint {https://arxiv.org/abs/2108.05368} {arXiv:2108.05368 [hep-ph]} \BibitemShut {NoStop}%
\bibitem [{\citenamefont {Benabou}\ \emph {et~al.}(2024)\citenamefont {Benabou}, \citenamefont {Buschmann}, \citenamefont {Foster},\ and\ \citenamefont {Safdi}}]{Benabou:2024msj}%
  \BibitemOpen
  \bibfield  {author} {\bibinfo {author} {\bibfnamefont {J.~N.}\ \bibnamefont {Benabou}}, \bibinfo {author} {\bibfnamefont {M.}~\bibnamefont {Buschmann}}, \bibinfo {author} {\bibfnamefont {J.~W.}\ \bibnamefont {Foster}},\ and\ \bibinfo {author} {\bibfnamefont {B.~R.}\ \bibnamefont {Safdi}},\ }\href@noop {} {\bibinfo {title} {{Axion mass prediction from adaptive mesh refinement cosmological lattice simulations}}} (\bibinfo {year} {2024}),\ \Eprint {https://arxiv.org/abs/2412.08699} {arXiv:2412.08699 [hep-ph]} \BibitemShut {NoStop}%
\bibitem [{\citenamefont {Arvanitaki}\ \emph {et~al.}(2020)\citenamefont {Arvanitaki}, \citenamefont {Dimopoulos}, \citenamefont {Galanis}, \citenamefont {Lehner}, \citenamefont {Thompson},\ and\ \citenamefont {Van~Tilburg}}]{Arvanitaki:2019rax}%
  \BibitemOpen
  \bibfield  {author} {\bibinfo {author} {\bibfnamefont {A.}~\bibnamefont {Arvanitaki}}, \bibinfo {author} {\bibfnamefont {S.}~\bibnamefont {Dimopoulos}}, \bibinfo {author} {\bibfnamefont {M.}~\bibnamefont {Galanis}}, \bibinfo {author} {\bibfnamefont {L.}~\bibnamefont {Lehner}}, \bibinfo {author} {\bibfnamefont {J.~O.}\ \bibnamefont {Thompson}},\ and\ \bibinfo {author} {\bibfnamefont {K.}~\bibnamefont {Van~Tilburg}},\ }\href {https://doi.org/10.1103/PhysRevD.101.083014} {\bibfield  {journal} {\bibinfo  {journal} {Phys. Rev. D}\ }\textbf {\bibinfo {volume} {101}},\ \bibinfo {pages} {083014} (\bibinfo {year} {2020})},\ \Eprint {https://arxiv.org/abs/1909.11665} {arXiv:1909.11665 [astro-ph.CO]} \BibitemShut {NoStop}%
\bibitem [{\citenamefont {Cyncynates}\ and\ \citenamefont {Thompson}(2023)}]{Cyncynates:2023esj}%
  \BibitemOpen
  \bibfield  {author} {\bibinfo {author} {\bibfnamefont {D.}~\bibnamefont {Cyncynates}}\ and\ \bibinfo {author} {\bibfnamefont {J.~O.}\ \bibnamefont {Thompson}},\ }\href {https://doi.org/10.1103/PhysRevD.108.L091703} {\bibfield  {journal} {\bibinfo  {journal} {Phys. Rev. D}\ }\textbf {\bibinfo {volume} {108}},\ \bibinfo {pages} {L091703} (\bibinfo {year} {2023})},\ \Eprint {https://arxiv.org/abs/2306.04678} {arXiv:2306.04678 [hep-ph]} \BibitemShut {NoStop}%
\bibitem [{\citenamefont {Gorghetto}\ \emph {et~al.}(2021)\citenamefont {Gorghetto}, \citenamefont {Hardy},\ and\ \citenamefont {Villadoro}}]{Gorghetto:2020qws}%
  \BibitemOpen
  \bibfield  {author} {\bibinfo {author} {\bibfnamefont {M.}~\bibnamefont {Gorghetto}}, \bibinfo {author} {\bibfnamefont {E.}~\bibnamefont {Hardy}},\ and\ \bibinfo {author} {\bibfnamefont {G.}~\bibnamefont {Villadoro}},\ }\href {https://doi.org/10.21468/SciPostPhys.10.2.050} {\bibfield  {journal} {\bibinfo  {journal} {SciPost Phys.}\ }\textbf {\bibinfo {volume} {10}},\ \bibinfo {pages} {050} (\bibinfo {year} {2021})},\ \Eprint {https://arxiv.org/abs/2007.04990} {arXiv:2007.04990 [hep-ph]} \BibitemShut {NoStop}%
\bibitem [{\citenamefont {Saikawa}\ \emph {et~al.}(2024)\citenamefont {Saikawa}, \citenamefont {Redondo}, \citenamefont {Vaquero},\ and\ \citenamefont {Kaltschmidt}}]{Saikawa:2024bta}%
  \BibitemOpen
  \bibfield  {author} {\bibinfo {author} {\bibfnamefont {K.}~\bibnamefont {Saikawa}}, \bibinfo {author} {\bibfnamefont {J.}~\bibnamefont {Redondo}}, \bibinfo {author} {\bibfnamefont {A.}~\bibnamefont {Vaquero}},\ and\ \bibinfo {author} {\bibfnamefont {M.}~\bibnamefont {Kaltschmidt}},\ }\href {https://doi.org/10.1088/1475-7516/2024/10/043} {\bibfield  {journal} {\bibinfo  {journal} {JCAP}\ }\textbf {\bibinfo {volume} {{2024}}}\bibfield  {number} {\bibinfo  {number} { ({10})},\ \bibinfo {pages} {043}},\ }\Eprint {https://arxiv.org/abs/2401.17253} {arXiv:2401.17253 [hep-ph]} \BibitemShut {NoStop}%
\bibitem [{\citenamefont {Kim}\ \emph {et~al.}(2024)\citenamefont {Kim}, \citenamefont {Park},\ and\ \citenamefont {Son}}]{Kim:2024wku}%
  \BibitemOpen
  \bibfield  {author} {\bibinfo {author} {\bibfnamefont {H.}~\bibnamefont {Kim}}, \bibinfo {author} {\bibfnamefont {J.}~\bibnamefont {Park}},\ and\ \bibinfo {author} {\bibfnamefont {M.}~\bibnamefont {Son}},\ }\href {https://doi.org/10.1007/JHEP07(2024)150} {\bibfield  {journal} {\bibinfo  {journal} {JHEP}\ }\textbf {\bibinfo {volume} {2024}}\bibfield  {number} {\bibinfo  {number} { (07)},\ \bibinfo {pages} {150}},\ }\Eprint {https://arxiv.org/abs/2402.00741} {arXiv:2402.00741 [hep-ph]} \BibitemShut {NoStop}%
\bibitem [{\citenamefont {Graham}\ \emph {et~al.}(2016)\citenamefont {Graham}, \citenamefont {Mardon},\ and\ \citenamefont {Rajendran}}]{Graham:2015rva}%
  \BibitemOpen
  \bibfield  {author} {\bibinfo {author} {\bibfnamefont {P.~W.}\ \bibnamefont {Graham}}, \bibinfo {author} {\bibfnamefont {J.}~\bibnamefont {Mardon}},\ and\ \bibinfo {author} {\bibfnamefont {S.}~\bibnamefont {Rajendran}},\ }\href {https://doi.org/10.1103/PhysRevD.93.103520} {\bibfield  {journal} {\bibinfo  {journal} {Phys. Rev.}\ }\textbf {\bibinfo {volume} {D93}},\ \bibinfo {pages} {103520} (\bibinfo {year} {2016})},\ \Eprint {https://arxiv.org/abs/1504.02102} {arXiv:1504.02102 [hep-ph]} \BibitemShut {NoStop}%
%%CITATION = ARXIV:1504.02102;%%
\bibitem [{\citenamefont {Agrawal}\ \emph {et~al.}(2020)\citenamefont {Agrawal}, \citenamefont {Kitajima}, \citenamefont {Reece}, \citenamefont {Sekiguchi},\ and\ \citenamefont {Takahashi}}]{Agrawal:2018vin}%
  \BibitemOpen
  \bibfield  {author} {\bibinfo {author} {\bibfnamefont {P.}~\bibnamefont {Agrawal}}, \bibinfo {author} {\bibfnamefont {N.}~\bibnamefont {Kitajima}}, \bibinfo {author} {\bibfnamefont {M.}~\bibnamefont {Reece}}, \bibinfo {author} {\bibfnamefont {T.}~\bibnamefont {Sekiguchi}},\ and\ \bibinfo {author} {\bibfnamefont {F.}~\bibnamefont {Takahashi}},\ }\href {https://doi.org/10.1016/j.physletb.2019.135136} {\bibfield  {journal} {\bibinfo  {journal} {Phys. Lett. B}\ }\textbf {\bibinfo {volume} {801}},\ \bibinfo {pages} {135136} (\bibinfo {year} {2020})},\ \Eprint {https://arxiv.org/abs/1810.07188} {arXiv:1810.07188 [hep-ph]} \BibitemShut {NoStop}%
\bibitem [{\citenamefont {Co}\ \emph {et~al.}(2019)\citenamefont {Co}, \citenamefont {Pierce}, \citenamefont {Zhang},\ and\ \citenamefont {Zhao}}]{Co:2018lka}%
  \BibitemOpen
  \bibfield  {author} {\bibinfo {author} {\bibfnamefont {R.~T.}\ \bibnamefont {Co}}, \bibinfo {author} {\bibfnamefont {A.}~\bibnamefont {Pierce}}, \bibinfo {author} {\bibfnamefont {Z.}~\bibnamefont {Zhang}},\ and\ \bibinfo {author} {\bibfnamefont {Y.}~\bibnamefont {Zhao}},\ }\href {https://doi.org/10.1103/PhysRevD.99.075002} {\bibfield  {journal} {\bibinfo  {journal} {Phys. Rev. D}\ }\textbf {\bibinfo {volume} {99}},\ \bibinfo {pages} {075002} (\bibinfo {year} {2019})},\ \Eprint {https://arxiv.org/abs/1810.07196} {arXiv:1810.07196 [hep-ph]} \BibitemShut {NoStop}%
\bibitem [{\citenamefont {Cyncynates}\ and\ \citenamefont {Weiner}(2025{\natexlab{a}})}]{Cyncynates:2023zwj}%
  \BibitemOpen
  \bibfield  {author} {\bibinfo {author} {\bibfnamefont {D.}~\bibnamefont {Cyncynates}}\ and\ \bibinfo {author} {\bibfnamefont {Z.~J.}\ \bibnamefont {Weiner}},\ }\href {https://doi.org/10.1103/PhysRevLett.134.211002} {\bibfield  {journal} {\bibinfo  {journal} {Phys. Rev. Lett.}\ }\textbf {\bibinfo {volume} {134}},\ \bibinfo {pages} {211002} (\bibinfo {year} {2025}{\natexlab{a}})},\ \Eprint {https://arxiv.org/abs/2310.18397} {arXiv:2310.18397 [hep-ph]} \BibitemShut {NoStop}%
\bibitem [{\citenamefont {Cyncynates}\ and\ \citenamefont {Weiner}(2025{\natexlab{b}})}]{Cyncynates:2024yxm}%
  \BibitemOpen
  \bibfield  {author} {\bibinfo {author} {\bibfnamefont {D.}~\bibnamefont {Cyncynates}}\ and\ \bibinfo {author} {\bibfnamefont {Z.~J.}\ \bibnamefont {Weiner}},\ }\href {https://doi.org/10.1103/PhysRevD.111.103535} {\bibfield  {journal} {\bibinfo  {journal} {Phys. Rev. D}\ }\textbf {\bibinfo {volume} {111}},\ \bibinfo {pages} {103535} (\bibinfo {year} {2025}{\natexlab{b}})},\ \Eprint {https://arxiv.org/abs/2410.14774} {arXiv:2410.14774 [hep-ph]} \BibitemShut {NoStop}%
\bibitem [{\citenamefont {East}\ and\ \citenamefont {Huang}(2022)}]{East:2022rsi}%
  \BibitemOpen
  \bibfield  {author} {\bibinfo {author} {\bibfnamefont {W.~E.}\ \bibnamefont {East}}\ and\ \bibinfo {author} {\bibfnamefont {J.}~\bibnamefont {Huang}},\ }\href {https://doi.org/10.1007/JHEP12(2022)089} {\bibfield  {journal} {\bibinfo  {journal} {JHEP}\ }\textbf {\bibinfo {volume} {2022}}\bibfield  {number} {\bibinfo  {number} { (12)},\ \bibinfo {pages} {089}},\ }\Eprint {https://arxiv.org/abs/2206.12432} {arXiv:2206.12432 [hep-ph]} \BibitemShut {NoStop}%
\bibitem [{\citenamefont {Adams}\ \emph {et~al.}(2022)\citenamefont {Adams} \emph {et~al.}}]{Adams:2022pbo}%
  \BibitemOpen
  \bibfield  {author} {\bibinfo {author} {\bibfnamefont {C.~B.}\ \bibnamefont {Adams}} \emph {et~al.},\ }in\ \href@noop {} {\emph {\bibinfo {booktitle} {{2022 Snowmass Summer Study}}}}\ (\bibinfo {year} {2022})\ \Eprint {https://arxiv.org/abs/2203.14923} {arXiv:2203.14923 [hep-ex]} \BibitemShut {NoStop}%
\bibitem [{\citenamefont {Antypas}\ \emph {et~al.}(2022)\citenamefont {Antypas} \emph {et~al.}}]{Antypas:2022asj}%
  \BibitemOpen
  \bibfield  {author} {\bibinfo {author} {\bibfnamefont {D.}~\bibnamefont {Antypas}} \emph {et~al.},\ }in\ \href@noop {} {\emph {\bibinfo {booktitle} {{2022 Snowmass Summer Study}}}}\ (\bibinfo {year} {2022})\ \Eprint {https://arxiv.org/abs/2203.14915} {arXiv:2203.14915 [hep-ex]} \BibitemShut {NoStop}%
\bibitem [{\citenamefont {Baryakhtar}\ \emph {et~al.}(2025)\citenamefont {Baryakhtar}, \citenamefont {Rosenberg},\ and\ \citenamefont {Rybka}}]{Baryakhtar:2025jwh}%
  \BibitemOpen
  \bibfield  {author} {\bibinfo {author} {\bibfnamefont {M.}~\bibnamefont {Baryakhtar}}, \bibinfo {author} {\bibfnamefont {L.}~\bibnamefont {Rosenberg}},\ and\ \bibinfo {author} {\bibfnamefont {G.}~\bibnamefont {Rybka}},\ }\href@noop {} {\bibinfo {title} {{Searching for the QCD Dark Matter Axion}}} (\bibinfo {year} {2025}),\ \Eprint {https://arxiv.org/abs/2504.10607} {arXiv:2504.10607 [hep-ex]} \BibitemShut {NoStop}%
\bibitem [{\citenamefont {Kim}(1979)}]{Kim:1979if}%
  \BibitemOpen
  \bibfield  {author} {\bibinfo {author} {\bibfnamefont {J.~E.}\ \bibnamefont {Kim}},\ }\href {https://doi.org/10.1103/PhysRevLett.43.103} {\bibfield  {journal} {\bibinfo  {journal} {Phys. Rev. Lett.}\ }\textbf {\bibinfo {volume} {43}},\ \bibinfo {pages} {103} (\bibinfo {year} {1979})}\BibitemShut {NoStop}%
\bibitem [{\citenamefont {Shifman}\ \emph {et~al.}(1980)\citenamefont {Shifman}, \citenamefont {Vainshtein},\ and\ \citenamefont {Zakharov}}]{Shifman:1979if}%
  \BibitemOpen
  \bibfield  {author} {\bibinfo {author} {\bibfnamefont {M.~A.}\ \bibnamefont {Shifman}}, \bibinfo {author} {\bibfnamefont {A.~I.}\ \bibnamefont {Vainshtein}},\ and\ \bibinfo {author} {\bibfnamefont {V.~I.}\ \bibnamefont {Zakharov}},\ }\href {https://doi.org/10.1016/0550-3213(80)90209-6} {\bibfield  {journal} {\bibinfo  {journal} {Nucl. Phys. B}\ }\textbf {\bibinfo {volume} {166}},\ \bibinfo {pages} {493} (\bibinfo {year} {1980})}\BibitemShut {NoStop}%
\bibitem [{\citenamefont {Dine}\ \emph {et~al.}(1981)\citenamefont {Dine}, \citenamefont {Fischler},\ and\ \citenamefont {Srednicki}}]{Dine:1981rt}%
  \BibitemOpen
  \bibfield  {author} {\bibinfo {author} {\bibfnamefont {M.}~\bibnamefont {Dine}}, \bibinfo {author} {\bibfnamefont {W.}~\bibnamefont {Fischler}},\ and\ \bibinfo {author} {\bibfnamefont {M.}~\bibnamefont {Srednicki}},\ }\href {https://doi.org/10.1016/0370-2693(81)90590-6} {\bibfield  {journal} {\bibinfo  {journal} {Phys. Lett. B}\ }\textbf {\bibinfo {volume} {104}},\ \bibinfo {pages} {199} (\bibinfo {year} {1981})}\BibitemShut {NoStop}%
\bibitem [{\citenamefont {Zhitnitsky}(1980)}]{Zhitnitsky:1980tq}%
  \BibitemOpen
  \bibfield  {author} {\bibinfo {author} {\bibfnamefont {A.~R.}\ \bibnamefont {Zhitnitsky}},\ }\href@noop {} {\bibfield  {journal} {\bibinfo  {journal} {Sov. J. Nucl. Phys.}\ }\textbf {\bibinfo {volume} {31}},\ \bibinfo {pages} {260} (\bibinfo {year} {1980})}\BibitemShut {NoStop}%
\bibitem [{\citenamefont {Bartram}\ \emph {et~al.}(2021)\citenamefont {Bartram} \emph {et~al.}}]{ADMX:2021nhd}%
  \BibitemOpen
  \bibfield  {author} {\bibinfo {author} {\bibfnamefont {C.}~\bibnamefont {Bartram}} \emph {et~al.} (\bibinfo {collaboration} {ADMX}),\ }\href {https://doi.org/10.1103/PhysRevLett.127.261803} {\bibfield  {journal} {\bibinfo  {journal} {Phys. Rev. Lett.}\ }\textbf {\bibinfo {volume} {127}},\ \bibinfo {pages} {261803} (\bibinfo {year} {2021})},\ \Eprint {https://arxiv.org/abs/2110.06096} {arXiv:2110.06096 [hep-ex]} \BibitemShut {NoStop}%
\bibitem [{\citenamefont {Zhong}\ \emph {et~al.}(2018)\citenamefont {Zhong} \emph {et~al.}}]{HAYSTAC:2018rwy}%
  \BibitemOpen
  \bibfield  {author} {\bibinfo {author} {\bibfnamefont {L.}~\bibnamefont {Zhong}} \emph {et~al.} (\bibinfo {collaboration} {HAYSTAC}),\ }\href {https://doi.org/10.1103/PhysRevD.97.092001} {\bibfield  {journal} {\bibinfo  {journal} {Phys. Rev. D}\ }\textbf {\bibinfo {volume} {97}},\ \bibinfo {pages} {092001} (\bibinfo {year} {2018})},\ \Eprint {https://arxiv.org/abs/1803.03690} {arXiv:1803.03690 [hep-ex]} \BibitemShut {NoStop}%
\bibitem [{\citenamefont {Backes}\ \emph {et~al.}(2021)\citenamefont {Backes} \emph {et~al.}}]{HAYSTAC:2020kwv}%
  \BibitemOpen
  \bibfield  {author} {\bibinfo {author} {\bibfnamefont {K.~M.}\ \bibnamefont {Backes}} \emph {et~al.} (\bibinfo {collaboration} {HAYSTAC}),\ }\href {https://doi.org/10.1038/s41586-021-03226-7} {\bibfield  {journal} {\bibinfo  {journal} {Nature}\ }\textbf {\bibinfo {volume} {590}},\ \bibinfo {pages} {238} (\bibinfo {year} {2021})},\ \Eprint {https://arxiv.org/abs/2008.01853} {arXiv:2008.01853 [quant-ph]} \BibitemShut {NoStop}%
\bibitem [{\citenamefont {Kwon}\ \emph {et~al.}(2021)\citenamefont {Kwon} \emph {et~al.}}]{CAPP:2020utb}%
  \BibitemOpen
  \bibfield  {author} {\bibinfo {author} {\bibfnamefont {O.}~\bibnamefont {Kwon}} \emph {et~al.} (\bibinfo {collaboration} {CAPP}),\ }\href {https://doi.org/10.1103/PhysRevLett.126.191802} {\bibfield  {journal} {\bibinfo  {journal} {Phys. Rev. Lett.}\ }\textbf {\bibinfo {volume} {126}},\ \bibinfo {pages} {191802} (\bibinfo {year} {2021})},\ \Eprint {https://arxiv.org/abs/2012.10764} {arXiv:2012.10764 [hep-ex]} \BibitemShut {NoStop}%
\bibitem [{\citenamefont {Yi}\ \emph {et~al.}(2023)\citenamefont {Yi} \emph {et~al.}}]{Yi:2022fmn}%
  \BibitemOpen
  \bibfield  {author} {\bibinfo {author} {\bibfnamefont {A.~K.}\ \bibnamefont {Yi}} \emph {et~al.},\ }\href {https://doi.org/10.1103/PhysRevLett.130.071002} {\bibfield  {journal} {\bibinfo  {journal} {Phys. Rev. Lett.}\ }\textbf {\bibinfo {volume} {130}},\ \bibinfo {pages} {071002} (\bibinfo {year} {2023})},\ \Eprint {https://arxiv.org/abs/2210.10961} {arXiv:2210.10961 [hep-ex]} \BibitemShut {NoStop}%
\bibitem [{\citenamefont {Ahn}\ \emph {et~al.}(2024)\citenamefont {Ahn} \emph {et~al.}}]{CAPP:2024dtx}%
  \BibitemOpen
  \bibfield  {author} {\bibinfo {author} {\bibfnamefont {S.}~\bibnamefont {Ahn}} \emph {et~al.} (\bibinfo {collaboration} {CAPP}),\ }\href {https://doi.org/10.1103/PhysRevX.14.031023} {\bibfield  {journal} {\bibinfo  {journal} {Phys. Rev. X}\ }\textbf {\bibinfo {volume} {14}},\ \bibinfo {pages} {031023} (\bibinfo {year} {2024})},\ \Eprint {https://arxiv.org/abs/2402.12892} {arXiv:2402.12892 [hep-ex]} \BibitemShut {NoStop}%
\bibitem [{\citenamefont {McAllister}\ \emph {et~al.}(2017)\citenamefont {McAllister}, \citenamefont {Flower}, \citenamefont {Kruger}, \citenamefont {Ivanov}, \citenamefont {Goryachev}, \citenamefont {Bourhill},\ and\ \citenamefont {Tobar}}]{McAllister:2017lkb}%
  \BibitemOpen
  \bibfield  {author} {\bibinfo {author} {\bibfnamefont {B.~T.}\ \bibnamefont {McAllister}}, \bibinfo {author} {\bibfnamefont {G.}~\bibnamefont {Flower}}, \bibinfo {author} {\bibfnamefont {J.}~\bibnamefont {Kruger}}, \bibinfo {author} {\bibfnamefont {E.~N.}\ \bibnamefont {Ivanov}}, \bibinfo {author} {\bibfnamefont {M.}~\bibnamefont {Goryachev}}, \bibinfo {author} {\bibfnamefont {J.}~\bibnamefont {Bourhill}},\ and\ \bibinfo {author} {\bibfnamefont {M.~E.}\ \bibnamefont {Tobar}},\ }\href {https://doi.org/10.1016/j.dark.2017.09.010} {\bibfield  {journal} {\bibinfo  {journal} {Phys. Dark Univ.}\ }\textbf {\bibinfo {volume} {18}},\ \bibinfo {pages} {67} (\bibinfo {year} {2017})},\ \Eprint {https://arxiv.org/abs/1706.00209} {arXiv:1706.00209 [physics.ins-det]} \BibitemShut {NoStop}%
\bibitem [{\citenamefont {Quiskamp}\ \emph {et~al.}(2022)\citenamefont {Quiskamp}, \citenamefont {McAllister}, \citenamefont {Altin}, \citenamefont {Ivanov}, \citenamefont {Goryachev},\ and\ \citenamefont {Tobar}}]{Quiskamp:2022pks}%
  \BibitemOpen
  \bibfield  {author} {\bibinfo {author} {\bibfnamefont {A.~P.}\ \bibnamefont {Quiskamp}}, \bibinfo {author} {\bibfnamefont {B.~T.}\ \bibnamefont {McAllister}}, \bibinfo {author} {\bibfnamefont {P.}~\bibnamefont {Altin}}, \bibinfo {author} {\bibfnamefont {E.~N.}\ \bibnamefont {Ivanov}}, \bibinfo {author} {\bibfnamefont {M.}~\bibnamefont {Goryachev}},\ and\ \bibinfo {author} {\bibfnamefont {M.~E.}\ \bibnamefont {Tobar}},\ }\href {https://doi.org/10.1126/sciadv.abq3765} {\bibfield  {journal} {\bibinfo  {journal} {Sci. Adv.}\ }\textbf {\bibinfo {volume} {8}},\ \bibinfo {pages} {abq3765} (\bibinfo {year} {2022})},\ \Eprint {https://arxiv.org/abs/2203.12152} {arXiv:2203.12152 [hep-ex]} \BibitemShut {NoStop}%
\bibitem [{\citenamefont {Alesini}\ \emph {et~al.}(2021)\citenamefont {Alesini} \emph {et~al.}}]{Alesini:2020vny}%
  \BibitemOpen
  \bibfield  {author} {\bibinfo {author} {\bibfnamefont {D.}~\bibnamefont {Alesini}} \emph {et~al.},\ }\href {https://doi.org/10.1103/PhysRevD.103.102004} {\bibfield  {journal} {\bibinfo  {journal} {Phys. Rev. D}\ }\textbf {\bibinfo {volume} {103}},\ \bibinfo {pages} {102004} (\bibinfo {year} {2021})},\ \Eprint {https://arxiv.org/abs/2012.09498} {arXiv:2012.09498 [hep-ex]} \BibitemShut {NoStop}%
\bibitem [{\citenamefont {Dixit}\ \emph {et~al.}(2021)\citenamefont {Dixit}, \citenamefont {Chakram}, \citenamefont {He}, \citenamefont {Agrawal}, \citenamefont {Naik}, \citenamefont {Schuster},\ and\ \citenamefont {Chou}}]{Dixit:2020ymh}%
  \BibitemOpen
  \bibfield  {author} {\bibinfo {author} {\bibfnamefont {A.~V.}\ \bibnamefont {Dixit}}, \bibinfo {author} {\bibfnamefont {S.}~\bibnamefont {Chakram}}, \bibinfo {author} {\bibfnamefont {K.}~\bibnamefont {He}}, \bibinfo {author} {\bibfnamefont {A.}~\bibnamefont {Agrawal}}, \bibinfo {author} {\bibfnamefont {R.~K.}\ \bibnamefont {Naik}}, \bibinfo {author} {\bibfnamefont {D.~I.}\ \bibnamefont {Schuster}},\ and\ \bibinfo {author} {\bibfnamefont {A.}~\bibnamefont {Chou}},\ }\href {https://doi.org/10.1103/PhysRevLett.126.141302} {\bibfield  {journal} {\bibinfo  {journal} {Phys. Rev. Lett.}\ }\textbf {\bibinfo {volume} {126}},\ \bibinfo {pages} {141302} (\bibinfo {year} {2021})},\ \Eprint {https://arxiv.org/abs/2008.12231} {arXiv:2008.12231 [hep-ex]} \BibitemShut {NoStop}%
\bibitem [{\citenamefont {Caldwell}\ \emph {et~al.}(2017)\citenamefont {Caldwell}, \citenamefont {Dvali}, \citenamefont {Majorovits}, \citenamefont {Millar}, \citenamefont {Raffelt}, \citenamefont {Redondo}, \citenamefont {Reimann}, \citenamefont {Simon},\ and\ \citenamefont {Steffen}}]{Caldwell:2016dcw}%
  \BibitemOpen
  \bibfield  {author} {\bibinfo {author} {\bibfnamefont {A.}~\bibnamefont {Caldwell}}, \bibinfo {author} {\bibfnamefont {G.}~\bibnamefont {Dvali}}, \bibinfo {author} {\bibfnamefont {B.}~\bibnamefont {Majorovits}}, \bibinfo {author} {\bibfnamefont {A.}~\bibnamefont {Millar}}, \bibinfo {author} {\bibfnamefont {G.}~\bibnamefont {Raffelt}}, \bibinfo {author} {\bibfnamefont {J.}~\bibnamefont {Redondo}}, \bibinfo {author} {\bibfnamefont {O.}~\bibnamefont {Reimann}}, \bibinfo {author} {\bibfnamefont {F.}~\bibnamefont {Simon}},\ and\ \bibinfo {author} {\bibfnamefont {F.}~\bibnamefont {Steffen}} (\bibinfo {collaboration} {MADMAX Working Group}),\ }\href {https://doi.org/10.1103/PhysRevLett.118.091801} {\bibfield  {journal} {\bibinfo  {journal} {Phys. Rev. Lett.}\ }\textbf {\bibinfo {volume} {118}},\ \bibinfo {pages} {091801} (\bibinfo {year} {2017})},\ \Eprint {https://arxiv.org/abs/1611.05865} {arXiv:1611.05865 [physics.ins-det]} \BibitemShut {NoStop}%
\bibitem [{\citenamefont {Brun}\ \emph {et~al.}(2019)\citenamefont {Brun} \emph {et~al.}}]{MADMAX:2019pub}%
  \BibitemOpen
  \bibfield  {author} {\bibinfo {author} {\bibfnamefont {P.}~\bibnamefont {Brun}} \emph {et~al.} (\bibinfo {collaboration} {MADMAX}),\ }\href {https://doi.org/10.1140/epjc/s10052-019-6683-x} {\bibfield  {journal} {\bibinfo  {journal} {Eur. Phys. J. C}\ }\textbf {\bibinfo {volume} {79}},\ \bibinfo {pages} {186} (\bibinfo {year} {2019})},\ \Eprint {https://arxiv.org/abs/1901.07401} {arXiv:1901.07401 [physics.ins-det]} \BibitemShut {NoStop}%
\bibitem [{\citenamefont {Egge}\ \emph {et~al.}(2025)\citenamefont {Egge} \emph {et~al.}}]{MADMAX:2024jnp}%
  \BibitemOpen
  \bibfield  {author} {\bibinfo {author} {\bibfnamefont {J.}~\bibnamefont {Egge}} \emph {et~al.} (\bibinfo {collaboration} {MADMAX}),\ }\href {https://doi.org/10.1103/PhysRevLett.134.151004} {\bibfield  {journal} {\bibinfo  {journal} {Phys. Rev. Lett.}\ }\textbf {\bibinfo {volume} {134}},\ \bibinfo {pages} {151004} (\bibinfo {year} {2025})},\ \Eprint {https://arxiv.org/abs/2408.02368} {arXiv:2408.02368 [hep-ex]} \BibitemShut {NoStop}%
\bibitem [{\citenamefont {Garcia}\ \emph {et~al.}(2024{\natexlab{a}})\citenamefont {Garcia} \emph {et~al.}}]{Garcia:2024xzc}%
  \BibitemOpen
  \bibfield  {author} {\bibinfo {author} {\bibfnamefont {B.~A. d.~S.}\ \bibnamefont {Garcia}} \emph {et~al.},\ }\href@noop {} {\bibinfo {title} {{First search for axion dark matter with a Madmax prototype}}} (\bibinfo {year} {2024}{\natexlab{a}}),\ \Eprint {https://arxiv.org/abs/2409.11777} {arXiv:2409.11777 [hep-ex]} \BibitemShut {NoStop}%
\bibitem [{\citenamefont {Baryakhtar}\ \emph {et~al.}(2018)\citenamefont {Baryakhtar}, \citenamefont {Huang},\ and\ \citenamefont {Lasenby}}]{Baryakhtar:2018doz}%
  \BibitemOpen
  \bibfield  {author} {\bibinfo {author} {\bibfnamefont {M.}~\bibnamefont {Baryakhtar}}, \bibinfo {author} {\bibfnamefont {J.}~\bibnamefont {Huang}},\ and\ \bibinfo {author} {\bibfnamefont {R.}~\bibnamefont {Lasenby}},\ }\href {https://doi.org/10.1103/PhysRevD.98.035006} {\bibfield  {journal} {\bibinfo  {journal} {Phys. Rev.}\ }\textbf {\bibinfo {volume} {D98}},\ \bibinfo {pages} {035006} (\bibinfo {year} {2018})},\ \Eprint {https://arxiv.org/abs/1803.11455} {arXiv:1803.11455 [hep-ph]} \BibitemShut {NoStop}%
%%CITATION = ARXIV:1803.11455;%%
\bibitem [{\citenamefont {Chiles}\ \emph {et~al.}(2022)\citenamefont {Chiles} \emph {et~al.}}]{Chiles:2021gxk}%
  \BibitemOpen
  \bibfield  {author} {\bibinfo {author} {\bibfnamefont {J.}~\bibnamefont {Chiles}} \emph {et~al.},\ }\href {https://doi.org/10.1103/PhysRevLett.128.231802} {\bibfield  {journal} {\bibinfo  {journal} {Phys. Rev. Lett.}\ }\textbf {\bibinfo {volume} {128}},\ \bibinfo {pages} {231802} (\bibinfo {year} {2022})},\ \Eprint {https://arxiv.org/abs/2110.01582} {arXiv:2110.01582 [hep-ex]} \BibitemShut {NoStop}%
\bibitem [{\citenamefont {Carosi}\ \emph {et~al.}(2020)\citenamefont {Carosi}, \citenamefont {Cervantes}, \citenamefont {Kimes}, \citenamefont {Mohapatra}, \citenamefont {Ottens},\ and\ \citenamefont {Rybka}}]{orpheus}%
  \BibitemOpen
  \bibfield  {author} {\bibinfo {author} {\bibfnamefont {G.}~\bibnamefont {Carosi}}, \bibinfo {author} {\bibfnamefont {R.}~\bibnamefont {Cervantes}}, \bibinfo {author} {\bibfnamefont {S.}~\bibnamefont {Kimes}}, \bibinfo {author} {\bibfnamefont {P.}~\bibnamefont {Mohapatra}}, \bibinfo {author} {\bibfnamefont {R.}~\bibnamefont {Ottens}},\ and\ \bibinfo {author} {\bibfnamefont {G.}~\bibnamefont {Rybka}},\ }in\ \href@noop {} {\emph {\bibinfo {booktitle} {Microwave Cavities and Detectors for Axion Research}}},\ \bibinfo {editor} {edited by\ \bibinfo {editor} {\bibfnamefont {G.}~\bibnamefont {Carosi}}\ and\ \bibinfo {editor} {\bibfnamefont {G.}~\bibnamefont {Rybka}}}\ (\bibinfo  {publisher} {Springer International Publishing},\ \bibinfo {address} {Cham},\ \bibinfo {year} {2020})\ pp.\ \bibinfo {pages} {169--175}\BibitemShut {NoStop}%
\bibitem [{\citenamefont {Cervantes}\ \emph {et~al.}(2022{\natexlab{a}})\citenamefont {Cervantes} \emph {et~al.}}]{Cervantes:2022epl}%
  \BibitemOpen
  \bibfield  {author} {\bibinfo {author} {\bibfnamefont {R.}~\bibnamefont {Cervantes}} \emph {et~al.},\ }\href {https://doi.org/10.1103/PhysRevD.106.102002} {\bibfield  {journal} {\bibinfo  {journal} {Phys. Rev. D}\ }\textbf {\bibinfo {volume} {106}},\ \bibinfo {pages} {102002} (\bibinfo {year} {2022}{\natexlab{a}})},\ \Eprint {https://arxiv.org/abs/2204.09475} {arXiv:2204.09475 [hep-ex]} \BibitemShut {NoStop}%
\bibitem [{\citenamefont {Cervantes}\ \emph {et~al.}(2022{\natexlab{b}})\citenamefont {Cervantes} \emph {et~al.}}]{Cervantes:2022yzp}%
  \BibitemOpen
  \bibfield  {author} {\bibinfo {author} {\bibfnamefont {R.}~\bibnamefont {Cervantes}} \emph {et~al.},\ }\href {https://doi.org/10.1103/PhysRevLett.129.201301} {\bibfield  {journal} {\bibinfo  {journal} {Phys. Rev. Lett.}\ }\textbf {\bibinfo {volume} {129}},\ \bibinfo {pages} {201301} (\bibinfo {year} {2022}{\natexlab{b}})},\ \Eprint {https://arxiv.org/abs/2204.03818} {arXiv:2204.03818 [hep-ex]} \BibitemShut {NoStop}%
\bibitem [{\citenamefont {Millar}\ \emph {et~al.}(2023)\citenamefont {Millar} \emph {et~al.}}]{ALPHA:2022rxj}%
  \BibitemOpen
  \bibfield  {author} {\bibinfo {author} {\bibfnamefont {A.~J.}\ \bibnamefont {Millar}} \emph {et~al.} (\bibinfo {collaboration} {ALPHA}),\ }\href {https://doi.org/10.1103/PhysRevD.107.055013} {\bibfield  {journal} {\bibinfo  {journal} {Phys. Rev. D}\ }\textbf {\bibinfo {volume} {107}},\ \bibinfo {pages} {055013} (\bibinfo {year} {2023})},\ \Eprint {https://arxiv.org/abs/2210.00017} {arXiv:2210.00017 [hep-ph]} \BibitemShut {NoStop}%
\bibitem [{\citenamefont {Jaeckel}\ and\ \citenamefont {Redondo}(2013{\natexlab{a}})}]{Jaeckel:2013sqa}%
  \BibitemOpen
  \bibfield  {author} {\bibinfo {author} {\bibfnamefont {J.}~\bibnamefont {Jaeckel}}\ and\ \bibinfo {author} {\bibfnamefont {J.}~\bibnamefont {Redondo}},\ }\href {https://doi.org/10.1088/1475-7516/2013/11/016} {\bibfield  {journal} {\bibinfo  {journal} {JCAP}\ }\textbf {\bibinfo {volume} {2013}}\bibfield  {number} {\bibinfo  {number} { (11)},\ \bibinfo {pages} {016}},\ }\Eprint {https://arxiv.org/abs/1307.7181} {arXiv:1307.7181 [hep-ph]} \BibitemShut {NoStop}%
%%CITATION = ARXIV:1307.7181;%%
\bibitem [{\citenamefont {Jaeckel}\ and\ \citenamefont {Redondo}(2013{\natexlab{b}})}]{Jaeckel:2013eha}%
  \BibitemOpen
  \bibfield  {author} {\bibinfo {author} {\bibfnamefont {J.}~\bibnamefont {Jaeckel}}\ and\ \bibinfo {author} {\bibfnamefont {J.}~\bibnamefont {Redondo}},\ }\href {https://doi.org/10.1103/PhysRevD.88.115002} {\bibfield  {journal} {\bibinfo  {journal} {Phys. Rev.}\ }\textbf {\bibinfo {volume} {D88}},\ \bibinfo {pages} {115002} (\bibinfo {year} {2013}{\natexlab{b}})},\ \Eprint {https://arxiv.org/abs/1308.1103} {arXiv:1308.1103 [hep-ph]} \BibitemShut {NoStop}%
%%CITATION = ARXIV:1308.1103;%%
\bibitem [{\citenamefont {Andrianavalomahefa}\ \emph {et~al.}(2020)\citenamefont {Andrianavalomahefa} \emph {et~al.}}]{PhysRevD.102.042001}%
  \BibitemOpen
  \bibfield  {author} {\bibinfo {author} {\bibfnamefont {A.}~\bibnamefont {Andrianavalomahefa}} \emph {et~al.} (\bibinfo {collaboration} {The FUNK Experiment}),\ }\href {https://doi.org/10.1103/PhysRevD.102.042001} {\bibfield  {journal} {\bibinfo  {journal} {Phys. Rev. D}\ }\textbf {\bibinfo {volume} {102}},\ \bibinfo {pages} {042001} (\bibinfo {year} {2020})}\BibitemShut {NoStop}%
\bibitem [{\citenamefont {Liu}\ \emph {et~al.}(2022)\citenamefont {Liu} \emph {et~al.}}]{BREAD:2021tpx}%
  \BibitemOpen
  \bibfield  {author} {\bibinfo {author} {\bibfnamefont {J.}~\bibnamefont {Liu}} \emph {et~al.} (\bibinfo {collaboration} {BREAD}),\ }\href {https://doi.org/10.1103/PhysRevLett.128.131801} {\bibfield  {journal} {\bibinfo  {journal} {Phys. Rev. Lett.}\ }\textbf {\bibinfo {volume} {128}},\ \bibinfo {pages} {131801} (\bibinfo {year} {2022})},\ \Eprint {https://arxiv.org/abs/2111.12103} {arXiv:2111.12103 [physics.ins-det]} \BibitemShut {NoStop}%
\bibitem [{\citenamefont {Hoshino}\ \emph {et~al.}(2025)\citenamefont {Hoshino}, \citenamefont {Knirck}, \citenamefont {Awida}, \citenamefont {Cancelo}, \citenamefont {Corrodi}, \citenamefont {Di~Federico}, \citenamefont {Knepper}, \citenamefont {Lapuente}, \citenamefont {Littmann}, \citenamefont {Miller}, \citenamefont {Mitchell}, \citenamefont {Rodriguez}, \citenamefont {Ruschman}, \citenamefont {Salemi}, \citenamefont {Sawtell}, \citenamefont {Stefanazzi}, \citenamefont {Sonnenschein}, \citenamefont {Teafoe},\ and\ \citenamefont {Winter}}]{Hoshino:2025fiz}%
  \BibitemOpen
  \bibfield  {author} {\bibinfo {author} {\bibfnamefont {G.}~\bibnamefont {Hoshino}}, \bibinfo {author} {\bibfnamefont {S.}~\bibnamefont {Knirck}}, \bibinfo {author} {\bibfnamefont {M.~H.}\ \bibnamefont {Awida}}, \bibinfo {author} {\bibfnamefont {G.~I.}\ \bibnamefont {Cancelo}}, \bibinfo {author} {\bibfnamefont {S.}~\bibnamefont {Corrodi}}, \bibinfo {author} {\bibfnamefont {M.}~\bibnamefont {Di~Federico}}, \bibinfo {author} {\bibfnamefont {B.}~\bibnamefont {Knepper}}, \bibinfo {author} {\bibfnamefont {A.}~\bibnamefont {Lapuente}}, \bibinfo {author} {\bibfnamefont {M.}~\bibnamefont {Littmann}}, \bibinfo {author} {\bibfnamefont {D.~W.}\ \bibnamefont {Miller}}, \bibinfo {author} {\bibfnamefont {D.~V.}\ \bibnamefont {Mitchell}}, \bibinfo {author} {\bibfnamefont {D.}~\bibnamefont {Rodriguez}}, \bibinfo {author} {\bibfnamefont {M.~K.}\ \bibnamefont {Ruschman}}, \bibinfo {author} {\bibfnamefont {C.~P.}\ \bibnamefont {Salemi}}, \bibinfo {author} {\bibfnamefont {M.~A.}\ \bibnamefont {Sawtell}}, \bibinfo {author}
  {\bibfnamefont {L.}~\bibnamefont {Stefanazzi}}, \bibinfo {author} {\bibfnamefont {A.}~\bibnamefont {Sonnenschein}}, \bibinfo {author} {\bibfnamefont {G.~W.}\ \bibnamefont {Teafoe}},\ and\ \bibinfo {author} {\bibfnamefont {P.}~\bibnamefont {Winter}} (\bibinfo {collaboration} {GigaBREAD Collaboration}),\ }\href {https://doi.org/10.1103/PhysRevLett.134.171002} {\bibfield  {journal} {\bibinfo  {journal} {Phys. Rev. Lett.}\ }\textbf {\bibinfo {volume} {134}},\ \bibinfo {pages} {171002} (\bibinfo {year} {2025})}\BibitemShut {NoStop}%
\bibitem [{\citenamefont {Vinyoles}\ \emph {et~al.}(2015)\citenamefont {Vinyoles}, \citenamefont {Serenelli}, \citenamefont {Villante}, \citenamefont {Basu}, \citenamefont {Redondo},\ and\ \citenamefont {Isern}}]{Vinyoles:2015aba}%
  \BibitemOpen
  \bibfield  {author} {\bibinfo {author} {\bibfnamefont {N.}~\bibnamefont {Vinyoles}}, \bibinfo {author} {\bibfnamefont {A.}~\bibnamefont {Serenelli}}, \bibinfo {author} {\bibfnamefont {F.~L.}\ \bibnamefont {Villante}}, \bibinfo {author} {\bibfnamefont {S.}~\bibnamefont {Basu}}, \bibinfo {author} {\bibfnamefont {J.}~\bibnamefont {Redondo}},\ and\ \bibinfo {author} {\bibfnamefont {J.}~\bibnamefont {Isern}},\ }\href {https://doi.org/10.1088/1475-7516/2015/10/015} {\bibfield  {journal} {\bibinfo  {journal} {JCAP}\ }\textbf {\bibinfo {volume} {1510}}\bibfield  {number} {\bibinfo  {number} { (10)},\ \bibinfo {pages} {015}},\ }\Eprint {https://arxiv.org/abs/1501.01639} {arXiv:1501.01639 [astro-ph.SR]} \BibitemShut {NoStop}%
%%CITATION = ARXIV:1501.01639;%%
\bibitem [{\citenamefont {Ayala}\ \emph {et~al.}(2014)\citenamefont {Ayala}, \citenamefont {Domínguez}, \citenamefont {Giannotti}, \citenamefont {Mirizzi},\ and\ \citenamefont {Straniero}}]{Ayala:2014pea}%
  \BibitemOpen
  \bibfield  {author} {\bibinfo {author} {\bibfnamefont {A.}~\bibnamefont {Ayala}}, \bibinfo {author} {\bibfnamefont {I.}~\bibnamefont {Domínguez}}, \bibinfo {author} {\bibfnamefont {M.}~\bibnamefont {Giannotti}}, \bibinfo {author} {\bibfnamefont {A.}~\bibnamefont {Mirizzi}},\ and\ \bibinfo {author} {\bibfnamefont {O.}~\bibnamefont {Straniero}},\ }\href {https://doi.org/10.1103/PhysRevLett.113.191302} {\bibfield  {journal} {\bibinfo  {journal} {Phys. Rev. Lett.}\ }\textbf {\bibinfo {volume} {113}},\ \bibinfo {pages} {191302} (\bibinfo {year} {2014})},\ \Eprint {https://arxiv.org/abs/1406.6053} {arXiv:1406.6053 [astro-ph.SR]} \BibitemShut {NoStop}%
%%CITATION = ARXIV:1406.6053;%%
\bibitem [{\citenamefont {Dolan}\ \emph {et~al.}(2022)\citenamefont {Dolan}, \citenamefont {Hiskens},\ and\ \citenamefont {Volkas}}]{Dolan:2022kul}%
  \BibitemOpen
  \bibfield  {author} {\bibinfo {author} {\bibfnamefont {M.~J.}\ \bibnamefont {Dolan}}, \bibinfo {author} {\bibfnamefont {F.~J.}\ \bibnamefont {Hiskens}},\ and\ \bibinfo {author} {\bibfnamefont {R.~R.}\ \bibnamefont {Volkas}},\ }\href {https://doi.org/10.1088/1475-7516/2022/10/096} {\bibfield  {journal} {\bibinfo  {journal} {JCAP}\ }\textbf {\bibinfo {volume} {2022}}\bibfield  {number} {\bibinfo  {number} { (10)},\ \bibinfo {pages} {096}},\ }\Eprint {https://arxiv.org/abs/2207.03102} {arXiv:2207.03102 [hep-ph]} \BibitemShut {NoStop}%
\bibitem [{\citenamefont {{\'A}lvarez~Melc{\'o}n}\ \emph {et~al.}(2021)\citenamefont {{\'A}lvarez~Melc{\'o}n}, \citenamefont {Arguedas~Cuendis}, \citenamefont {Baier}, \citenamefont {Barth}, \citenamefont {Br{\"a}uninger}, \citenamefont {Calatroni}, \citenamefont {Cantatore}, \citenamefont {Caspers}, \citenamefont {Castel}, \citenamefont {Cetin}, \citenamefont {Cogollos}, \citenamefont {Dafni}, \citenamefont {Davenport}, \citenamefont {Dermenev}, \citenamefont {Desch}, \citenamefont {D{\'i}az-Morcillo}, \citenamefont {D{\"o}brich}, \citenamefont {Fischer}, \citenamefont {Funk}, \citenamefont {Gallego}, \citenamefont {Garc{\'i}a~Barcel{\'o}}, \citenamefont {Gardikiotis}, \citenamefont {Garza}, \citenamefont {Gimeno}, \citenamefont {Gninenko}, \citenamefont {Golm}, \citenamefont {Hasinoff}, \citenamefont {Hoffmann}, \citenamefont {Irastorza}, \citenamefont {Jakov{\v{c}}i{\'{c}}}, \citenamefont {Kaminski}, \citenamefont {Karuza}, \citenamefont {Laki{\'{c}}}, \citenamefont {Laurent}, \citenamefont
  {Lozano-Guerrero}, \citenamefont {Luz{\'o}n}, \citenamefont {Malbrunot}, \citenamefont {Maroudas}, \citenamefont {Miralda-Escud{\'e}}, \citenamefont {Mirallas}, \citenamefont {Miceli}, \citenamefont {Navarro}, \citenamefont {Ozbey}, \citenamefont {{\"O}zbozduman}, \citenamefont {Pe{\~{n}}a~Garay}, \citenamefont {Pivovaroff}, \citenamefont {Redondo}, \citenamefont {Ruz}, \citenamefont {Ruiz~Ch{\'o}liz}, \citenamefont {Schmidt}, \citenamefont {Schumann}, \citenamefont {Semertzidis}, \citenamefont {Solanki}, \citenamefont {Stewart}, \citenamefont {Tsagris}, \citenamefont {Vafeiadis}, \citenamefont {Vogel}, \citenamefont {Widmann}, \citenamefont {Wuensch},\ and\ \citenamefont {Zioutas}}]{alvarezmeconrades}%
  \BibitemOpen
  \bibfield  {author} {\bibinfo {author} {\bibfnamefont {A.}~\bibnamefont {{\'A}lvarez~Melc{\'o}n}}, \bibinfo {author} {\bibfnamefont {S.}~\bibnamefont {Arguedas~Cuendis}}, \bibinfo {author} {\bibfnamefont {J.}~\bibnamefont {Baier}}, \bibinfo {author} {\bibfnamefont {K.}~\bibnamefont {Barth}}, \bibinfo {author} {\bibfnamefont {H.}~\bibnamefont {Br{\"a}uninger}}, \bibinfo {author} {\bibfnamefont {S.}~\bibnamefont {Calatroni}}, \bibinfo {author} {\bibfnamefont {G.}~\bibnamefont {Cantatore}}, \bibinfo {author} {\bibfnamefont {F.}~\bibnamefont {Caspers}}, \bibinfo {author} {\bibfnamefont {J.~F.}\ \bibnamefont {Castel}}, \bibinfo {author} {\bibfnamefont {S.~A.}\ \bibnamefont {Cetin}}, \bibinfo {author} {\bibfnamefont {C.}~\bibnamefont {Cogollos}}, \bibinfo {author} {\bibfnamefont {T.}~\bibnamefont {Dafni}}, \bibinfo {author} {\bibfnamefont {M.}~\bibnamefont {Davenport}}, \bibinfo {author} {\bibfnamefont {A.}~\bibnamefont {Dermenev}}, \bibinfo {author} {\bibfnamefont {K.}~\bibnamefont {Desch}}, \bibinfo {author}
  {\bibfnamefont {A.}~\bibnamefont {D{\'i}az-Morcillo}}, \bibinfo {author} {\bibfnamefont {B.}~\bibnamefont {D{\"o}brich}}, \bibinfo {author} {\bibfnamefont {H.}~\bibnamefont {Fischer}}, \bibinfo {author} {\bibfnamefont {W.}~\bibnamefont {Funk}}, \bibinfo {author} {\bibfnamefont {J.~D.}\ \bibnamefont {Gallego}}, \bibinfo {author} {\bibfnamefont {J.~M.}\ \bibnamefont {Garc{\'i}a~Barcel{\'o}}}, \bibinfo {author} {\bibfnamefont {A.}~\bibnamefont {Gardikiotis}}, \bibinfo {author} {\bibfnamefont {J.~G.}\ \bibnamefont {Garza}}, \bibinfo {author} {\bibfnamefont {B.}~\bibnamefont {Gimeno}}, \bibinfo {author} {\bibfnamefont {S.}~\bibnamefont {Gninenko}}, \bibinfo {author} {\bibfnamefont {J.}~\bibnamefont {Golm}}, \bibinfo {author} {\bibfnamefont {M.~D.}\ \bibnamefont {Hasinoff}}, \bibinfo {author} {\bibfnamefont {D.~H.~H.}\ \bibnamefont {Hoffmann}}, \bibinfo {author} {\bibfnamefont {I.~G.}\ \bibnamefont {Irastorza}}, \bibinfo {author} {\bibfnamefont {K.}~\bibnamefont {Jakov{\v{c}}i{\'{c}}}}, \bibinfo {author}
  {\bibfnamefont {J.}~\bibnamefont {Kaminski}}, \bibinfo {author} {\bibfnamefont {M.}~\bibnamefont {Karuza}}, \bibinfo {author} {\bibfnamefont {B.}~\bibnamefont {Laki{\'{c}}}}, \bibinfo {author} {\bibfnamefont {J.~M.}\ \bibnamefont {Laurent}}, \bibinfo {author} {\bibfnamefont {A.~J.}\ \bibnamefont {Lozano-Guerrero}}, \bibinfo {author} {\bibfnamefont {G.}~\bibnamefont {Luz{\'o}n}}, \bibinfo {author} {\bibfnamefont {C.}~\bibnamefont {Malbrunot}}, \bibinfo {author} {\bibfnamefont {M.}~\bibnamefont {Maroudas}}, \bibinfo {author} {\bibfnamefont {J.}~\bibnamefont {Miralda-Escud{\'e}}}, \bibinfo {author} {\bibfnamefont {H.}~\bibnamefont {Mirallas}}, \bibinfo {author} {\bibfnamefont {L.}~\bibnamefont {Miceli}}, \bibinfo {author} {\bibfnamefont {P.}~\bibnamefont {Navarro}}, \bibinfo {author} {\bibfnamefont {A.}~\bibnamefont {Ozbey}}, \bibinfo {author} {\bibfnamefont {K.}~\bibnamefont {{\"O}zbozduman}}, \bibinfo {author} {\bibfnamefont {C.}~\bibnamefont {Pe{\~{n}}a~Garay}}, \bibinfo {author} {\bibfnamefont {M.~J.}\
  \bibnamefont {Pivovaroff}}, \bibinfo {author} {\bibfnamefont {J.}~\bibnamefont {Redondo}}, \bibinfo {author} {\bibfnamefont {J.}~\bibnamefont {Ruz}}, \bibinfo {author} {\bibfnamefont {E.}~\bibnamefont {Ruiz~Ch{\'o}liz}}, \bibinfo {author} {\bibfnamefont {S.}~\bibnamefont {Schmidt}}, \bibinfo {author} {\bibfnamefont {M.}~\bibnamefont {Schumann}}, \bibinfo {author} {\bibfnamefont {Y.~K.}\ \bibnamefont {Semertzidis}}, \bibinfo {author} {\bibfnamefont {S.~K.}\ \bibnamefont {Solanki}}, \bibinfo {author} {\bibfnamefont {L.}~\bibnamefont {Stewart}}, \bibinfo {author} {\bibfnamefont {I.}~\bibnamefont {Tsagris}}, \bibinfo {author} {\bibfnamefont {T.}~\bibnamefont {Vafeiadis}}, \bibinfo {author} {\bibfnamefont {J.~K.}\ \bibnamefont {Vogel}}, \bibinfo {author} {\bibfnamefont {E.}~\bibnamefont {Widmann}}, \bibinfo {author} {\bibfnamefont {W.}~\bibnamefont {Wuensch}},\ and\ \bibinfo {author} {\bibfnamefont {K.}~\bibnamefont {Zioutas}},\ }\href {https://doi.org/10.1007/JHEP10(2021)075} {\bibfield  {journal} {\bibinfo
  {journal} {Journal of High Energy Physics}\ }\textbf {\bibinfo {volume} {2021}},\ \bibinfo {pages} {75} (\bibinfo {year} {2021})}\BibitemShut {NoStop}%
\bibitem [{\citenamefont {Janish}\ and\ \citenamefont {Pinetti}(2025)}]{Janish:2023kvi}%
  \BibitemOpen
  \bibfield  {author} {\bibinfo {author} {\bibfnamefont {R.}~\bibnamefont {Janish}}\ and\ \bibinfo {author} {\bibfnamefont {E.}~\bibnamefont {Pinetti}},\ }\href {https://doi.org/10.1103/PhysRevLett.134.071002} {\bibfield  {journal} {\bibinfo  {journal} {Phys. Rev. Lett.}\ }\textbf {\bibinfo {volume} {134}},\ \bibinfo {pages} {071002} (\bibinfo {year} {2025})},\ \Eprint {https://arxiv.org/abs/2310.15395} {arXiv:2310.15395 [hep-ph]} \BibitemShut {NoStop}%
\bibitem [{\citenamefont {Roy}\ \emph {et~al.}(2025)\citenamefont {Roy}, \citenamefont {Blanco}, \citenamefont {Dessert}, \citenamefont {Prabhu},\ and\ \citenamefont {Temim}}]{Roy:2023omw}%
  \BibitemOpen
  \bibfield  {author} {\bibinfo {author} {\bibfnamefont {S.}~\bibnamefont {Roy}}, \bibinfo {author} {\bibfnamefont {C.}~\bibnamefont {Blanco}}, \bibinfo {author} {\bibfnamefont {C.}~\bibnamefont {Dessert}}, \bibinfo {author} {\bibfnamefont {A.}~\bibnamefont {Prabhu}},\ and\ \bibinfo {author} {\bibfnamefont {T.}~\bibnamefont {Temim}},\ }\href {https://doi.org/10.1103/PhysRevLett.134.071003} {\bibfield  {journal} {\bibinfo  {journal} {Phys. Rev. Lett.}\ }\textbf {\bibinfo {volume} {134}},\ \bibinfo {pages} {071003} (\bibinfo {year} {2025})},\ \Eprint {https://arxiv.org/abs/2311.04987} {arXiv:2311.04987 [hep-ph]} \BibitemShut {NoStop}%
\bibitem [{\citenamefont {Pinetti}(2025)}]{Pinetti:2025owq}%
  \BibitemOpen
  \bibfield  {author} {\bibinfo {author} {\bibfnamefont {E.}~\bibnamefont {Pinetti}},\ }\href@noop {} {\bibinfo {title} {{First constraints on QCD axion dark matter using James Webb Space Telescope observations}}} (\bibinfo {year} {2025}),\ \Eprint {https://arxiv.org/abs/2503.11753} {arXiv:2503.11753 [hep-ph]} \BibitemShut {NoStop}%
\bibitem [{\citenamefont {Saha}\ \emph {et~al.}(2025)\citenamefont {Saha}, \citenamefont {Bouri}, \citenamefont {Das}, \citenamefont {Dubey},\ and\ \citenamefont {Laha}}]{Saha:2025any}%
  \BibitemOpen
  \bibfield  {author} {\bibinfo {author} {\bibfnamefont {A.~K.}\ \bibnamefont {Saha}}, \bibinfo {author} {\bibfnamefont {S.}~\bibnamefont {Bouri}}, \bibinfo {author} {\bibfnamefont {A.}~\bibnamefont {Das}}, \bibinfo {author} {\bibfnamefont {A.}~\bibnamefont {Dubey}},\ and\ \bibinfo {author} {\bibfnamefont {R.}~\bibnamefont {Laha}},\ }\href@noop {} {\bibinfo {title} {{Shedding Infrared Light on QCD Axion and ALP Dark Matter with JWST}}} (\bibinfo {year} {2025}),\ \Eprint {https://arxiv.org/abs/2503.14582} {arXiv:2503.14582 [hep-ph]} \BibitemShut {NoStop}%
\bibitem [{\citenamefont {An}\ \emph {et~al.}(2020)\citenamefont {An}, \citenamefont {Pospelov}, \citenamefont {Pradler},\ and\ \citenamefont {Ritz}}]{PhysRevD.102.115022}%
  \BibitemOpen
  \bibfield  {author} {\bibinfo {author} {\bibfnamefont {H.}~\bibnamefont {An}}, \bibinfo {author} {\bibfnamefont {M.}~\bibnamefont {Pospelov}}, \bibinfo {author} {\bibfnamefont {J.}~\bibnamefont {Pradler}},\ and\ \bibinfo {author} {\bibfnamefont {A.}~\bibnamefont {Ritz}},\ }\href {https://doi.org/10.1103/PhysRevD.102.115022} {\bibfield  {journal} {\bibinfo  {journal} {Phys. Rev. D}\ }\textbf {\bibinfo {volume} {102}},\ \bibinfo {pages} {115022} (\bibinfo {year} {2020})}\BibitemShut {NoStop}%
\bibitem [{\citenamefont {Aprile}\ \emph {et~al.}(2022)\citenamefont {Aprile} \emph {et~al.}}]{XENON:2021qze}%
  \BibitemOpen
  \bibfield  {author} {\bibinfo {author} {\bibfnamefont {E.}~\bibnamefont {Aprile}} \emph {et~al.} (\bibinfo {collaboration} {XENON}),\ }\href {https://doi.org/10.1103/PhysRevD.106.022001} {\bibfield  {journal} {\bibinfo  {journal} {Phys. Rev. D}\ }\textbf {\bibinfo {volume} {106}},\ \bibinfo {pages} {022001} (\bibinfo {year} {2022})},\ \bibinfo {note} {[Erratum: Phys.Rev.D 110, 109903 (2024)]},\ \Eprint {https://arxiv.org/abs/2112.12116} {arXiv:2112.12116 [hep-ex]} \BibitemShut {NoStop}%
\bibitem [{\citenamefont {Taylor}\ \emph {et~al.}(2023)\citenamefont {Taylor}, \citenamefont {Walter}, \citenamefont {Korzh}, \citenamefont {Bumble}, \citenamefont {Patel}, \citenamefont {Allmaras}, \citenamefont {Beyer}, \citenamefont {O'Brient}, \citenamefont {Shaw},\ and\ \citenamefont {Wollman}}]{Taylor2023}%
  \BibitemOpen
  \bibfield  {author} {\bibinfo {author} {\bibfnamefont {G.~G.}\ \bibnamefont {Taylor}}, \bibinfo {author} {\bibfnamefont {A.~B.}\ \bibnamefont {Walter}}, \bibinfo {author} {\bibfnamefont {B.}~\bibnamefont {Korzh}}, \bibinfo {author} {\bibfnamefont {B.}~\bibnamefont {Bumble}}, \bibinfo {author} {\bibfnamefont {S.~R.}\ \bibnamefont {Patel}}, \bibinfo {author} {\bibfnamefont {J.~P.}\ \bibnamefont {Allmaras}}, \bibinfo {author} {\bibfnamefont {A.~D.}\ \bibnamefont {Beyer}}, \bibinfo {author} {\bibfnamefont {R.}~\bibnamefont {O'Brient}}, \bibinfo {author} {\bibfnamefont {M.~D.}\ \bibnamefont {Shaw}},\ and\ \bibinfo {author} {\bibfnamefont {E.~E.}\ \bibnamefont {Wollman}},\ }\href {https://doi.org/10.1364/OPTICA.509337} {\bibfield  {journal} {\bibinfo  {journal} {Optica}\ }\textbf {\bibinfo {volume} {10}},\ \bibinfo {pages} {1672} (\bibinfo {year} {2023})}\BibitemShut {NoStop}%
\bibitem [{\citenamefont {Luskin}\ \emph {et~al.}(2023)\citenamefont {Luskin}, \citenamefont {Schmidt}, \citenamefont {Korzh}, \citenamefont {Beyer}, \citenamefont {Bumble}, \citenamefont {Allmaras}, \citenamefont {Walter}, \citenamefont {Wollman}, \citenamefont {Narváez}, \citenamefont {Verma}, \citenamefont {Nam}, \citenamefont {Charaev}, \citenamefont {Colangelo}, \citenamefont {Berggren}, \citenamefont {Peña}, \citenamefont {Spiropulu}, \citenamefont {Garcia-Sciveres}, \citenamefont {Derenzo},\ and\ \citenamefont {Shaw}}]{Luskin2023}%
  \BibitemOpen
  \bibfield  {author} {\bibinfo {author} {\bibfnamefont {J.~S.}\ \bibnamefont {Luskin}}, \bibinfo {author} {\bibfnamefont {E.}~\bibnamefont {Schmidt}}, \bibinfo {author} {\bibfnamefont {B.}~\bibnamefont {Korzh}}, \bibinfo {author} {\bibfnamefont {A.~D.}\ \bibnamefont {Beyer}}, \bibinfo {author} {\bibfnamefont {B.}~\bibnamefont {Bumble}}, \bibinfo {author} {\bibfnamefont {J.~P.}\ \bibnamefont {Allmaras}}, \bibinfo {author} {\bibfnamefont {A.~B.}\ \bibnamefont {Walter}}, \bibinfo {author} {\bibfnamefont {E.~E.}\ \bibnamefont {Wollman}}, \bibinfo {author} {\bibfnamefont {L.}~\bibnamefont {Narváez}}, \bibinfo {author} {\bibfnamefont {V.~B.}\ \bibnamefont {Verma}}, \bibinfo {author} {\bibfnamefont {S.~W.}\ \bibnamefont {Nam}}, \bibinfo {author} {\bibfnamefont {I.}~\bibnamefont {Charaev}}, \bibinfo {author} {\bibfnamefont {M.}~\bibnamefont {Colangelo}}, \bibinfo {author} {\bibfnamefont {K.~K.}\ \bibnamefont {Berggren}}, \bibinfo {author} {\bibfnamefont {C.}~\bibnamefont {Peña}}, \bibinfo {author} {\bibfnamefont
  {M.}~\bibnamefont {Spiropulu}}, \bibinfo {author} {\bibfnamefont {M.}~\bibnamefont {Garcia-Sciveres}}, \bibinfo {author} {\bibfnamefont {S.}~\bibnamefont {Derenzo}},\ and\ \bibinfo {author} {\bibfnamefont {M.~D.}\ \bibnamefont {Shaw}},\ }\href {https://doi.org/10.1063/5.0150282} {\bibfield  {journal} {\bibinfo  {journal} {Applied Physics Letters}\ }\textbf {\bibinfo {volume} {122}},\ \bibinfo {pages} {243506} (\bibinfo {year} {2023})},\ \Eprint {https://arxiv.org/abs/https://pubs.aip.org/aip/apl/article-pdf/doi/10.1063/5.0150282/18004964/243506\_1\_5.0150282.pdf} {https://pubs.aip.org/aip/apl/article-pdf/doi/10.1063/5.0150282/18004964/243506\_1\_5.0150282.pdf} \BibitemShut {NoStop}%
\bibitem [{\citenamefont {Chaudhuri}\ \emph {et~al.}(2018)\citenamefont {Chaudhuri}, \citenamefont {Irwin}, \citenamefont {Graham},\ and\ \citenamefont {Mardon}}]{Chaudhuri:2018rqn}%
  \BibitemOpen
  \bibfield  {author} {\bibinfo {author} {\bibfnamefont {S.}~\bibnamefont {Chaudhuri}}, \bibinfo {author} {\bibfnamefont {K.}~\bibnamefont {Irwin}}, \bibinfo {author} {\bibfnamefont {P.~W.}\ \bibnamefont {Graham}},\ and\ \bibinfo {author} {\bibfnamefont {J.}~\bibnamefont {Mardon}},\ }\href@noop {} {\bibinfo {title} {{Optimal Impedance Matching and Quantum Limits of Electromagnetic Axion and Hidden-Photon Dark Matter Searches}}} (\bibinfo {year} {2018}),\ \Eprint {https://arxiv.org/abs/1803.01627} {arXiv:1803.01627 [hep-ph]} \BibitemShut {NoStop}%
\bibitem [{\citenamefont {Lasenby}(2021)}]{Lasenby:2019hfz}%
  \BibitemOpen
  \bibfield  {author} {\bibinfo {author} {\bibfnamefont {R.}~\bibnamefont {Lasenby}},\ }\href {https://doi.org/10.1103/PhysRevD.103.075007} {\bibfield  {journal} {\bibinfo  {journal} {Phys. Rev. D}\ }\textbf {\bibinfo {volume} {103}},\ \bibinfo {pages} {075007} (\bibinfo {year} {2021})},\ \Eprint {https://arxiv.org/abs/1912.11467} {arXiv:1912.11467 [hep-ph]} \BibitemShut {NoStop}%
\bibitem [{\citenamefont {Korneeva}\ \emph {et~al.}(2018)\citenamefont {Korneeva}, \citenamefont {Vodolazov}, \citenamefont {Semenov}, \citenamefont {Florya}, \citenamefont {Simonov}, \citenamefont {Baeva}, \citenamefont {Korneev}, \citenamefont {Goltsman},\ and\ \citenamefont {Klapwijk}}]{korneeva2018optical}%
  \BibitemOpen
  \bibfield  {author} {\bibinfo {author} {\bibfnamefont {Y.~P.}\ \bibnamefont {Korneeva}}, \bibinfo {author} {\bibfnamefont {D.~Y.}\ \bibnamefont {Vodolazov}}, \bibinfo {author} {\bibfnamefont {A.}~\bibnamefont {Semenov}}, \bibinfo {author} {\bibfnamefont {I.}~\bibnamefont {Florya}}, \bibinfo {author} {\bibfnamefont {N.}~\bibnamefont {Simonov}}, \bibinfo {author} {\bibfnamefont {E.}~\bibnamefont {Baeva}}, \bibinfo {author} {\bibfnamefont {A.}~\bibnamefont {Korneev}}, \bibinfo {author} {\bibfnamefont {G.}~\bibnamefont {Goltsman}},\ and\ \bibinfo {author} {\bibfnamefont {T.}~\bibnamefont {Klapwijk}},\ }\href@noop {} {\bibfield  {journal} {\bibinfo  {journal} {Physical Review Applied}\ }\textbf {\bibinfo {volume} {9}},\ \bibinfo {pages} {064037} (\bibinfo {year} {2018})}\BibitemShut {NoStop}%
\bibitem [{\citenamefont {Bloch}\ \emph {et~al.}(2025{\natexlab{a}})\citenamefont {Bloch}, \citenamefont {Knapen}, \citenamefont {Madden},\ and\ \citenamefont {Marocco}}]{Bloch:2024qqo}%
  \BibitemOpen
  \bibfield  {author} {\bibinfo {author} {\bibfnamefont {I.~M.}\ \bibnamefont {Bloch}}, \bibinfo {author} {\bibfnamefont {S.}~\bibnamefont {Knapen}}, \bibinfo {author} {\bibfnamefont {A.}~\bibnamefont {Madden}},\ and\ \bibinfo {author} {\bibfnamefont {G.}~\bibnamefont {Marocco}},\ }\href {https://doi.org/10.1007/JHEP03(2025)080} {\bibfield  {journal} {\bibinfo  {journal} {Journal of High Energy Physics}\ }\textbf {\bibinfo {volume} {2025}},\ \bibinfo {pages} {80} (\bibinfo {year} {2025}{\natexlab{a}})}\BibitemShut {NoStop}%
\bibitem [{\citenamefont {Wilczek}(1987)}]{PhysRevLett.58.1799}%
  \BibitemOpen
  \bibfield  {author} {\bibinfo {author} {\bibfnamefont {F.}~\bibnamefont {Wilczek}},\ }\href {https://doi.org/10.1103/PhysRevLett.58.1799} {\bibfield  {journal} {\bibinfo  {journal} {Phys. Rev. Lett.}\ }\textbf {\bibinfo {volume} {58}},\ \bibinfo {pages} {1799} (\bibinfo {year} {1987})}\BibitemShut {NoStop}%
%%CITATION = PRLTA,58,1799;%%
\bibitem [{\citenamefont {Sikivie}(1985)}]{PhysRevD.32.2988}%
  \BibitemOpen
  \bibfield  {author} {\bibinfo {author} {\bibfnamefont {P.}~\bibnamefont {Sikivie}},\ }\href {https://doi.org/10.1103/PhysRevD.32.2988} {\bibfield  {journal} {\bibinfo  {journal} {Phys. Rev. D}\ }\textbf {\bibinfo {volume} {32}},\ \bibinfo {pages} {2988} (\bibinfo {year} {1985})},\ \bibinfo {note} {[Erratum: Phys. Rev.D36,974(1987)]}\BibitemShut {NoStop}%
%%CITATION = PHRVA,D32,2988;%%
\bibitem [{\citenamefont {Horns}\ \emph {et~al.}(2013)\citenamefont {Horns}, \citenamefont {Jaeckel}, \citenamefont {Lindner}, \citenamefont {Lobanov}, \citenamefont {Redondo},\ and\ \citenamefont {Ringwald}}]{Horns:2012jf}%
  \BibitemOpen
  \bibfield  {author} {\bibinfo {author} {\bibfnamefont {D.}~\bibnamefont {Horns}}, \bibinfo {author} {\bibfnamefont {J.}~\bibnamefont {Jaeckel}}, \bibinfo {author} {\bibfnamefont {A.}~\bibnamefont {Lindner}}, \bibinfo {author} {\bibfnamefont {A.}~\bibnamefont {Lobanov}}, \bibinfo {author} {\bibfnamefont {J.}~\bibnamefont {Redondo}},\ and\ \bibinfo {author} {\bibfnamefont {A.}~\bibnamefont {Ringwald}},\ }\href {https://doi.org/10.1088/1475-7516/2013/04/016} {\bibfield  {journal} {\bibinfo  {journal} {JCAP}\ }\textbf {\bibinfo {volume} {2013}}\bibfield  {number} {\bibinfo  {number} { (04)},\ \bibinfo {pages} {016}},\ }\Eprint {https://arxiv.org/abs/1212.2970} {arXiv:1212.2970 [hep-ph]} \BibitemShut {NoStop}%
%%CITATION = ARXIV:1212.2970;%%
\bibitem [{\citenamefont {Millar}\ \emph {et~al.}(2017)\citenamefont {Millar}, \citenamefont {Raffelt}, \citenamefont {Redondo},\ and\ \citenamefont {Steffen}}]{Millar:2016cjp}%
  \BibitemOpen
  \bibfield  {author} {\bibinfo {author} {\bibfnamefont {A.~J.}\ \bibnamefont {Millar}}, \bibinfo {author} {\bibfnamefont {G.~G.}\ \bibnamefont {Raffelt}}, \bibinfo {author} {\bibfnamefont {J.}~\bibnamefont {Redondo}},\ and\ \bibinfo {author} {\bibfnamefont {F.~D.}\ \bibnamefont {Steffen}},\ }\href {https://doi.org/10.1088/1475-7516/2017/01/061} {\bibfield  {journal} {\bibinfo  {journal} {JCAP}\ }\textbf {\bibinfo {volume} {01}},\ \bibinfo {pages} {061}},\ \Eprint {https://arxiv.org/abs/1612.07057} {arXiv:1612.07057 [hep-ph]} \BibitemShut {NoStop}%
\bibitem [{\citenamefont {Egge}\ \emph {et~al.}(2020)\citenamefont {Egge}, \citenamefont {Knirck}, \citenamefont {Majorovits}, \citenamefont {Moore},\ and\ \citenamefont {Reimann}}]{Egge:2020hyo}%
  \BibitemOpen
  \bibfield  {author} {\bibinfo {author} {\bibfnamefont {J.}~\bibnamefont {Egge}}, \bibinfo {author} {\bibfnamefont {S.}~\bibnamefont {Knirck}}, \bibinfo {author} {\bibfnamefont {B.}~\bibnamefont {Majorovits}}, \bibinfo {author} {\bibfnamefont {C.}~\bibnamefont {Moore}},\ and\ \bibinfo {author} {\bibfnamefont {O.}~\bibnamefont {Reimann}},\ }\href {https://doi.org/10.1140/epjc/s10052-020-7985-8} {\bibfield  {journal} {\bibinfo  {journal} {Eur. Phys. J. C}\ }\textbf {\bibinfo {volume} {80}},\ \bibinfo {pages} {392} (\bibinfo {year} {2020})},\ \Eprint {https://arxiv.org/abs/2001.04363} {arXiv:2001.04363 [physics.ins-det]} \BibitemShut {NoStop}%
\bibitem [{\citenamefont {Egge}\ \emph {et~al.}(2024)\citenamefont {Egge} \emph {et~al.}}]{Egge:2023cos}%
  \BibitemOpen
  \bibfield  {author} {\bibinfo {author} {\bibfnamefont {J.}~\bibnamefont {Egge}} \emph {et~al.},\ }\href {https://doi.org/10.1088/1475-7516/2024/04/005} {\bibfield  {journal} {\bibinfo  {journal} {JCAP}\ }\textbf {\bibinfo {volume} {04}},\ \bibinfo {pages} {005}},\ \Eprint {https://arxiv.org/abs/2311.13359} {arXiv:2311.13359 [hep-ex]} \BibitemShut {NoStop}%
\bibitem [{\citenamefont {Garcia}\ \emph {et~al.}(2024{\natexlab{b}})\citenamefont {Garcia} \emph {et~al.}}]{MADMAX:2024pil}%
  \BibitemOpen
  \bibfield  {author} {\bibinfo {author} {\bibfnamefont {B.~A. D.~S.}\ \bibnamefont {Garcia}} \emph {et~al.} (\bibinfo {collaboration} {MADMAX}),\ }\href {https://doi.org/10.1088/1748-0221/19/11/T11002} {\bibfield  {journal} {\bibinfo  {journal} {JINST}\ }\textbf {\bibinfo {volume} {19}}\bibfield  {number} {\bibinfo  {number} { (11)},\ \bibinfo {pages} {T11002}},\ }\Eprint {https://arxiv.org/abs/2407.10716} {arXiv:2407.10716 [physics.ins-det]} \BibitemShut {NoStop}%
\bibitem [{\citenamefont {{Masha Baryakhtar, Junwu Huang, Stefan Knirck, David Miller, Andrew Sonnenschein, Joaquin Vieira}}(2024)}]{Baryakhtar_talk}%
  \BibitemOpen
  \bibfield  {author} {\bibinfo {author} {\bibnamefont {{Masha Baryakhtar, Junwu Huang, Stefan Knirck, David Miller, Andrew Sonnenschein, Joaquin Vieira}}},\ }\href@noop {} {\bibinfo {title} {Dielectric enhancements: Terabread and beyond}},\ \bibinfo {howpublished} {\url{https://indico.fnal.gov/event/63051/contributions/287733/}} (\bibinfo {year} {2024}),\ \bibinfo {note} {accessed: 2025-05-08}\BibitemShut {NoStop}%
\bibitem [{\citenamefont {Fan}\ \emph {et~al.}(2025)\citenamefont {Fan}, \citenamefont {Gabrielse}, \citenamefont {Graham}, \citenamefont {Ramani}, \citenamefont {Wong},\ and\ \citenamefont {Xiao}}]{Fan:2024mhm}%
  \BibitemOpen
  \bibfield  {author} {\bibinfo {author} {\bibfnamefont {X.}~\bibnamefont {Fan}}, \bibinfo {author} {\bibfnamefont {G.}~\bibnamefont {Gabrielse}}, \bibinfo {author} {\bibfnamefont {P.~W.}\ \bibnamefont {Graham}}, \bibinfo {author} {\bibfnamefont {H.}~\bibnamefont {Ramani}}, \bibinfo {author} {\bibfnamefont {S.~S.~Y.}\ \bibnamefont {Wong}},\ and\ \bibinfo {author} {\bibfnamefont {Y.}~\bibnamefont {Xiao}},\ }\href {https://doi.org/10.1103/PhysRevD.111.075022} {\bibfield  {journal} {\bibinfo  {journal} {Phys. Rev. D}\ }\textbf {\bibinfo {volume} {111}},\ \bibinfo {pages} {075022} (\bibinfo {year} {2025})},\ \Eprint {https://arxiv.org/abs/2410.05549} {arXiv:2410.05549 [hep-ph]} \BibitemShut {NoStop}%
\bibitem [{\citenamefont {Maystre}\ \emph {et~al.}(2006)\citenamefont {Maystre}, \citenamefont {Enoch},\ and\ \citenamefont {Tayeb}}]{Maystre}%
  \BibitemOpen
  \bibfield  {author} {\bibinfo {author} {\bibfnamefont {D.}~\bibnamefont {Maystre}}, \bibinfo {author} {\bibfnamefont {S.}~\bibnamefont {Enoch}},\ and\ \bibinfo {author} {\bibfnamefont {G.}~\bibnamefont {Tayeb}},\ }\href {https://doi.org/10.1201/9781420026627} {\emph {\bibinfo {title} {Scattering Matrix Method Applied to Photonic Crystals}}}\ (\bibinfo  {publisher} {CRC Press},\ \bibinfo {year} {2006})\ Chap.~\bibinfo {chapter} {1}\BibitemShut {NoStop}%
\bibitem [{\citenamefont {Tayeb}\ and\ \citenamefont {Maystre}(1997)}]{Tayeb:97}%
  \BibitemOpen
  \bibfield  {author} {\bibinfo {author} {\bibfnamefont {G.}~\bibnamefont {Tayeb}}\ and\ \bibinfo {author} {\bibfnamefont {D.}~\bibnamefont {Maystre}},\ }\href {https://doi.org/10.1364/JOSAA.14.003323} {\bibfield  {journal} {\bibinfo  {journal} {J. Opt. Soc. Am. A}\ }\textbf {\bibinfo {volume} {14}},\ \bibinfo {pages} {3323} (\bibinfo {year} {1997})}\BibitemShut {NoStop}%
\bibitem [{\citenamefont {Felbacq}\ \emph {et~al.}(1994)\citenamefont {Felbacq}, \citenamefont {Tayeb},\ and\ \citenamefont {Maystre}}]{Felbacq:94}%
  \BibitemOpen
  \bibfield  {author} {\bibinfo {author} {\bibfnamefont {D.}~\bibnamefont {Felbacq}}, \bibinfo {author} {\bibfnamefont {G.}~\bibnamefont {Tayeb}},\ and\ \bibinfo {author} {\bibfnamefont {D.}~\bibnamefont {Maystre}},\ }\href {https://doi.org/10.1364/JOSAA.11.002526} {\bibfield  {journal} {\bibinfo  {journal} {J. Opt. Soc. Am. A}\ }\textbf {\bibinfo {volume} {11}},\ \bibinfo {pages} {2526} (\bibinfo {year} {1994})}\BibitemShut {NoStop}%
\bibitem [{\citenamefont {Li}\ and\ \citenamefont {Zhang}(1998)}]{PhysRevB.58.9587}%
  \BibitemOpen
  \bibfield  {author} {\bibinfo {author} {\bibfnamefont {L.-M.}\ \bibnamefont {Li}}\ and\ \bibinfo {author} {\bibfnamefont {Z.-Q.}\ \bibnamefont {Zhang}},\ }\href {https://doi.org/10.1103/PhysRevB.58.9587} {\bibfield  {journal} {\bibinfo  {journal} {Phys. Rev. B}\ }\textbf {\bibinfo {volume} {58}},\ \bibinfo {pages} {9587} (\bibinfo {year} {1998})}\BibitemShut {NoStop}%
\bibitem [{\citenamefont {Yasumoto}\ \emph {et~al.}(2004)\citenamefont {Yasumoto}, \citenamefont {Jia},\ and\ \citenamefont {Toyama}}]{yasumoto2004electromagnetic}%
  \BibitemOpen
  \bibfield  {author} {\bibinfo {author} {\bibfnamefont {K.}~\bibnamefont {Yasumoto}}, \bibinfo {author} {\bibfnamefont {H.}~\bibnamefont {Jia}},\ and\ \bibinfo {author} {\bibfnamefont {H.}~\bibnamefont {Toyama}},\ }\href@noop {} {\bibfield  {journal} {\bibinfo  {journal} {IEEJ Transactions on Fundamentals and Materials}\ }\textbf {\bibinfo {volume} {124}},\ \bibinfo {pages} {1141} (\bibinfo {year} {2004})}\BibitemShut {NoStop}%
\bibitem [{\citenamefont {Yasumoto}(2018)}]{yasumoto2018electromagnetic}%
  \BibitemOpen
  \bibfield  {author} {\bibinfo {author} {\bibfnamefont {K.}~\bibnamefont {Yasumoto}},\ }\href@noop {} {\emph {\bibinfo {title} {Electromagnetic theory and applications for photonic crystals}}}\ (\bibinfo  {publisher} {CRC press},\ \bibinfo {year} {2018})\BibitemShut {NoStop}%
\bibitem [{\citenamefont {Jeong}\ \emph {et~al.}(2023)\citenamefont {Jeong}, \citenamefont {Kim}, \citenamefont {Bae},\ and\ \citenamefont {Youn}}]{Jeong:2023bqb}%
  \BibitemOpen
  \bibfield  {author} {\bibinfo {author} {\bibfnamefont {J.}~\bibnamefont {Jeong}}, \bibinfo {author} {\bibfnamefont {Y.}~\bibnamefont {Kim}}, \bibinfo {author} {\bibfnamefont {S.}~\bibnamefont {Bae}},\ and\ \bibinfo {author} {\bibfnamefont {S.}~\bibnamefont {Youn}},\ }\href {https://doi.org/10.1007/s40042-023-00808-8} {\bibfield  {journal} {\bibinfo  {journal} {J. Korean Phys. Soc.}\ }\textbf {\bibinfo {volume} {83}},\ \bibinfo {pages} {161} (\bibinfo {year} {2023})},\ \Eprint {https://arxiv.org/abs/2303.09748} {arXiv:2303.09748 [hep-ph]} \BibitemShut {NoStop}%
\bibitem [{\citenamefont {Jackson}(1999)}]{jackson}%
  \BibitemOpen
  \bibfield  {author} {\bibinfo {author} {\bibfnamefont {J.~D.}\ \bibnamefont {Jackson}},\ }\href {http://cdsweb.cern.ch/record/490457} {\emph {\bibinfo {title} {Classical electrodynamics}}},\ \bibinfo {edition} {3rd}\ ed.\ (\bibinfo  {publisher} {Wiley},\ \bibinfo {address} {New York, {NY}},\ \bibinfo {year} {1999})\BibitemShut {NoStop}%
\bibitem [{\citenamefont {Felbacq}(2016)}]{10.1088/978-1-6817-4301-1}%
  \BibitemOpen
  \bibfield  {author} {\bibinfo {author} {\bibfnamefont {D.}~\bibnamefont {Felbacq}},\ }\href {https://doi.org/10.1088/978-1-6817-4301-1} {\emph {\bibinfo {title} {Advanced Numerical and Theoretical Methods for Photonic Crystals and Metamaterials}}},\ 2053-2571\ (\bibinfo  {publisher} {Morgan \& Claypool Publishers},\ \bibinfo {year} {2016})\BibitemShut {NoStop}%
\bibitem [{\citenamefont {Goodman}(1976)}]{Goodman:76}%
  \BibitemOpen
  \bibfield  {author} {\bibinfo {author} {\bibfnamefont {J.~W.}\ \bibnamefont {Goodman}},\ }\href {https://doi.org/10.1364/JOSA.66.001145} {\bibfield  {journal} {\bibinfo  {journal} {J. Opt. Soc. Am.}\ }\textbf {\bibinfo {volume} {66}},\ \bibinfo {pages} {1145} (\bibinfo {year} {1976})}\BibitemShut {NoStop}%
\bibitem [{\citenamefont {Froufe-P\'erez}\ \emph {et~al.}(2017)\citenamefont {Froufe-P\'erez}, \citenamefont {Engel}, \citenamefont {S\'aenz},\ and\ \citenamefont {Scheffold}}]{FroufePrez2017BandGF}%
  \BibitemOpen
  \bibfield  {author} {\bibinfo {author} {\bibfnamefont {L.~S.}\ \bibnamefont {Froufe-P\'erez}}, \bibinfo {author} {\bibfnamefont {M.}~\bibnamefont {Engel}}, \bibinfo {author} {\bibfnamefont {J.~J.}\ \bibnamefont {S\'aenz}},\ and\ \bibinfo {author} {\bibfnamefont {F.}~\bibnamefont {Scheffold}},\ }\href {https://api.semanticscholar.org/CorpusID:3707170} {\bibfield  {journal} {\bibinfo  {journal} {Proceedings of the National Academy of Sciences}\ }\textbf {\bibinfo {volume} {114}},\ \bibinfo {pages} {9570} (\bibinfo {year} {2017})}\BibitemShut {NoStop}%
\bibitem [{\citenamefont {Scheffold}\ \emph {et~al.}(2022)\citenamefont {Scheffold}, \citenamefont {Haberko}, \citenamefont {Magkiriadou},\ and\ \citenamefont {Froufe-P\'erez}}]{scheffold2022transport}%
  \BibitemOpen
  \bibfield  {author} {\bibinfo {author} {\bibfnamefont {F.}~\bibnamefont {Scheffold}}, \bibinfo {author} {\bibfnamefont {J.}~\bibnamefont {Haberko}}, \bibinfo {author} {\bibfnamefont {S.}~\bibnamefont {Magkiriadou}},\ and\ \bibinfo {author} {\bibfnamefont {L.~S.}\ \bibnamefont {Froufe-P\'erez}},\ }\href {https://doi.org/10.1103/PhysRevLett.129.157402} {\bibfield  {journal} {\bibinfo  {journal} {Phys. Rev. Lett.}\ }\textbf {\bibinfo {volume} {129}},\ \bibinfo {pages} {157402} (\bibinfo {year} {2022})}\BibitemShut {NoStop}%
\bibitem [{\citenamefont {Martin}(1970)}]{Martin1970}%
  \BibitemOpen
  \bibfield  {author} {\bibinfo {author} {\bibfnamefont {T.~P.}\ \bibnamefont {Martin}},\ }\href {https://doi.org/10.1103/PhysRevB.1.3480} {\bibfield  {journal} {\bibinfo  {journal} {Phys. Rev. B}\ }\textbf {\bibinfo {volume} {1}},\ \bibinfo {pages} {3480} (\bibinfo {year} {1970})}\BibitemShut {NoStop}%
\bibitem [{mul()}]{multiphysics3}%
  \BibitemOpen
  \href@noop {} {\bibinfo {title} {{COMSOL Multiphysics\textsuperscript{\textregistered} v. 4.4}}},\ \bibinfo {howpublished} {\url{www.comsol.com}},\ \bibinfo {note} {{COMSOL AB, Stockholm, Sweden}}\BibitemShut {NoStop}%
\bibitem [{\citenamefont {Anderson}(1958)}]{PhysRev.109.1492}%
  \BibitemOpen
  \bibfield  {author} {\bibinfo {author} {\bibfnamefont {P.~W.}\ \bibnamefont {Anderson}},\ }\href {https://doi.org/10.1103/PhysRev.109.1492} {\bibfield  {journal} {\bibinfo  {journal} {Phys. Rev.}\ }\textbf {\bibinfo {volume} {109}},\ \bibinfo {pages} {1492} (\bibinfo {year} {1958})}\BibitemShut {NoStop}%
\bibitem [{\citenamefont {Máximo}\ \emph {et~al.}(2015)\citenamefont {Máximo}, \citenamefont {Piovella}, \citenamefont {Courteille}, \citenamefont {Kaiser},\ and\ \citenamefont {Bachelard}}]{M_ximo_2015}%
  \BibitemOpen
  \bibfield  {author} {\bibinfo {author} {\bibfnamefont {C.~E.}\ \bibnamefont {Máximo}}, \bibinfo {author} {\bibfnamefont {N.}~\bibnamefont {Piovella}}, \bibinfo {author} {\bibfnamefont {P.~W.}\ \bibnamefont {Courteille}}, \bibinfo {author} {\bibfnamefont {R.}~\bibnamefont {Kaiser}},\ and\ \bibinfo {author} {\bibfnamefont {R.}~\bibnamefont {Bachelard}},\ }\bibfield  {journal} {\bibinfo  {journal} {Physical Review A}\ }\textbf {\bibinfo {volume} {92}},\ \href {https://doi.org/10.1103/physreva.92.062702} {10.1103/physreva.92.062702} (\bibinfo {year} {2015})\BibitemShut {NoStop}%
\bibitem [{\citenamefont {Yamilov}\ \emph {et~al.}(2023)\citenamefont {Yamilov}, \citenamefont {Skipetrov}, \citenamefont {Hughes}, \citenamefont {Minkov}, \citenamefont {Yu},\ and\ \citenamefont {Cao}}]{Yamilov_2023}%
  \BibitemOpen
  \bibfield  {author} {\bibinfo {author} {\bibfnamefont {A.}~\bibnamefont {Yamilov}}, \bibinfo {author} {\bibfnamefont {S.~E.}\ \bibnamefont {Skipetrov}}, \bibinfo {author} {\bibfnamefont {T.~W.}\ \bibnamefont {Hughes}}, \bibinfo {author} {\bibfnamefont {M.}~\bibnamefont {Minkov}}, \bibinfo {author} {\bibfnamefont {Z.}~\bibnamefont {Yu}},\ and\ \bibinfo {author} {\bibfnamefont {H.}~\bibnamefont {Cao}},\ }\href {https://doi.org/10.1038/s41567-023-02091-7} {\bibfield  {journal} {\bibinfo  {journal} {Nature Physics}\ }\textbf {\bibinfo {volume} {19}},\ \bibinfo {pages} {1308–1313} (\bibinfo {year} {2023})}\BibitemShut {NoStop}%
\bibitem [{\citenamefont {Schubert}\ \emph {et~al.}(2010)\citenamefont {Schubert}, \citenamefont {Schleede}, \citenamefont {Byczuk}, \citenamefont {Fehske},\ and\ \citenamefont {Vollhardt}}]{Schubert_2010}%
  \BibitemOpen
  \bibfield  {author} {\bibinfo {author} {\bibfnamefont {G.}~\bibnamefont {Schubert}}, \bibinfo {author} {\bibfnamefont {J.}~\bibnamefont {Schleede}}, \bibinfo {author} {\bibfnamefont {K.}~\bibnamefont {Byczuk}}, \bibinfo {author} {\bibfnamefont {H.}~\bibnamefont {Fehske}},\ and\ \bibinfo {author} {\bibfnamefont {D.}~\bibnamefont {Vollhardt}},\ }\bibfield  {journal} {\bibinfo  {journal} {Physical Review B}\ }\textbf {\bibinfo {volume} {81}},\ \href {https://doi.org/10.1103/physrevb.81.155106} {10.1103/physrevb.81.155106} (\bibinfo {year} {2010})\BibitemShut {NoStop}%
\bibitem [{\citenamefont {Leseur}\ \emph {et~al.}(2016)\citenamefont {Leseur}, \citenamefont {Pierrat},\ and\ \citenamefont {Carminati}}]{Leseur2016}%
  \BibitemOpen
  \bibfield  {author} {\bibinfo {author} {\bibfnamefont {O.}~\bibnamefont {Leseur}}, \bibinfo {author} {\bibfnamefont {R.}~\bibnamefont {Pierrat}},\ and\ \bibinfo {author} {\bibfnamefont {R.}~\bibnamefont {Carminati}},\ }\href {https://doi.org/10.1364/OPTICA.3.000763} {\bibfield  {journal} {\bibinfo  {journal} {Optica}\ }\textbf {\bibinfo {volume} {3}},\ \bibinfo {pages} {763} (\bibinfo {year} {2016})}\BibitemShut {NoStop}%
\bibitem [{\citenamefont {Abrahams}\ \emph {et~al.}(1979)\citenamefont {Abrahams}, \citenamefont {Anderson}, \citenamefont {Licciardello},\ and\ \citenamefont {Ramakrishnan}}]{Abrahams1979}%
  \BibitemOpen
  \bibfield  {author} {\bibinfo {author} {\bibfnamefont {E.}~\bibnamefont {Abrahams}}, \bibinfo {author} {\bibfnamefont {P.~W.}\ \bibnamefont {Anderson}}, \bibinfo {author} {\bibfnamefont {D.~C.}\ \bibnamefont {Licciardello}},\ and\ \bibinfo {author} {\bibfnamefont {T.~V.}\ \bibnamefont {Ramakrishnan}},\ }\href {https://doi.org/10.1103/PhysRevLett.42.673} {\bibfield  {journal} {\bibinfo  {journal} {Phys. Rev. Lett.}\ }\textbf {\bibinfo {volume} {42}},\ \bibinfo {pages} {673} (\bibinfo {year} {1979})}\BibitemShut {NoStop}%
\bibitem [{\citenamefont {Grynko}\ \emph {et~al.}(2023)\citenamefont {Grynko}, \citenamefont {Siebert}, \citenamefont {Sperling},\ and\ \citenamefont {F{\"o}rstner}}]{Grynko2024}%
  \BibitemOpen
  \bibfield  {author} {\bibinfo {author} {\bibfnamefont {Y.}~\bibnamefont {Grynko}}, \bibinfo {author} {\bibfnamefont {D.}~\bibnamefont {Siebert}}, \bibinfo {author} {\bibfnamefont {J.}~\bibnamefont {Sperling}},\ and\ \bibinfo {author} {\bibfnamefont {J.}~\bibnamefont {F{\"o}rstner}},\ }\href@noop {} {\bibfield  {journal} {\bibinfo  {journal} {arXiv preprint arXiv:2312.14393}\ } (\bibinfo {year} {2023})}\BibitemShut {NoStop}%
\bibitem [{\citenamefont {Rezvani~Naraghi}\ and\ \citenamefont {Dogariu}(2016)}]{Rezvani2016}%
  \BibitemOpen
  \bibfield  {author} {\bibinfo {author} {\bibfnamefont {R.}~\bibnamefont {Rezvani~Naraghi}}\ and\ \bibinfo {author} {\bibfnamefont {A.}~\bibnamefont {Dogariu}},\ }\href {https://doi.org/10.1103/PhysRevLett.117.263901} {\bibfield  {journal} {\bibinfo  {journal} {Phys. Rev. Lett.}\ }\textbf {\bibinfo {volume} {117}},\ \bibinfo {pages} {263901} (\bibinfo {year} {2016})}\BibitemShut {NoStop}%
\bibitem [{\citenamefont {Sperling}\ \emph {et~al.}(2012)\citenamefont {Sperling}, \citenamefont {B{\"u}hrer}, \citenamefont {Aegerter},\ and\ \citenamefont {Maret}}]{Sperling2012}%
  \BibitemOpen
  \bibfield  {author} {\bibinfo {author} {\bibfnamefont {T.}~\bibnamefont {Sperling}}, \bibinfo {author} {\bibfnamefont {W.}~\bibnamefont {B{\"u}hrer}}, \bibinfo {author} {\bibfnamefont {C.~M.}\ \bibnamefont {Aegerter}},\ and\ \bibinfo {author} {\bibfnamefont {G.}~\bibnamefont {Maret}},\ }\href {https://api.semanticscholar.org/CorpusID:14588037} {\bibfield  {journal} {\bibinfo  {journal} {Nature Photonics}\ }\textbf {\bibinfo {volume} {7}},\ \bibinfo {pages} {48 } (\bibinfo {year} {2012})}\BibitemShut {NoStop}%
\bibitem [{\citenamefont {Lemieux}\ \emph {et~al.}(1998)\citenamefont {Lemieux}, \citenamefont {Vera},\ and\ \citenamefont {Durian}}]{Lemieux1998}%
  \BibitemOpen
  \bibfield  {author} {\bibinfo {author} {\bibfnamefont {P.-A.}\ \bibnamefont {Lemieux}}, \bibinfo {author} {\bibfnamefont {M.~U.}\ \bibnamefont {Vera}},\ and\ \bibinfo {author} {\bibfnamefont {D.~J.}\ \bibnamefont {Durian}},\ }\href {https://doi.org/10.1103/PhysRevE.57.4498} {\bibfield  {journal} {\bibinfo  {journal} {Phys. Rev. E}\ }\textbf {\bibinfo {volume} {57}},\ \bibinfo {pages} {4498} (\bibinfo {year} {1998})}\BibitemShut {NoStop}%
\bibitem [{\citenamefont {Lisitsyn}\ \emph {et~al.}(2016)\citenamefont {Lisitsyn}, \citenamefont {Dombrovsky}, \citenamefont {Mendeleyev}, \citenamefont {Grigorenko}, \citenamefont {Vlaskin},\ and\ \citenamefont {Zhuk}}]{Aleksey2016}%
  \BibitemOpen
  \bibfield  {author} {\bibinfo {author} {\bibfnamefont {A.~V.}\ \bibnamefont {Lisitsyn}}, \bibinfo {author} {\bibfnamefont {L.~A.}\ \bibnamefont {Dombrovsky}}, \bibinfo {author} {\bibfnamefont {V.~Y.}\ \bibnamefont {Mendeleyev}}, \bibinfo {author} {\bibfnamefont {A.~V.}\ \bibnamefont {Grigorenko}}, \bibinfo {author} {\bibfnamefont {M.~S.}\ \bibnamefont {Vlaskin}},\ and\ \bibinfo {author} {\bibfnamefont {A.~Z.}\ \bibnamefont {Zhuk}},\ }\href {https://doi.org/https://doi.org/10.1016/j.infrared.2016.05.028} {\bibfield  {journal} {\bibinfo  {journal} {Infrared Physics \& Technology}\ }\textbf {\bibinfo {volume} {77}},\ \bibinfo {pages} {162} (\bibinfo {year} {2016})}\BibitemShut {NoStop}%
\bibitem [{\citenamefont {Johnson}(1952)}]{Johnson1952}%
  \BibitemOpen
  \bibfield  {author} {\bibinfo {author} {\bibfnamefont {P.~D.}\ \bibnamefont {Johnson}},\ }\href {https://doi.org/10.1364/JOSA.42.000978} {\bibfield  {journal} {\bibinfo  {journal} {J. Opt. Soc. Am.}\ }\textbf {\bibinfo {volume} {42}},\ \bibinfo {pages} {978} (\bibinfo {year} {1952})}\BibitemShut {NoStop}%
\bibitem [{\citenamefont {Kuhn}\ \emph {et~al.}(1993)\citenamefont {Kuhn}, \citenamefont {Korder}, \citenamefont {Arduini-Schuster}, \citenamefont {Caps},\ and\ \citenamefont {Fricke}}]{kuhn1993infrared}%
  \BibitemOpen
  \bibfield  {author} {\bibinfo {author} {\bibfnamefont {J.}~\bibnamefont {Kuhn}}, \bibinfo {author} {\bibfnamefont {S.}~\bibnamefont {Korder}}, \bibinfo {author} {\bibfnamefont {M.}~\bibnamefont {Arduini-Schuster}}, \bibinfo {author} {\bibfnamefont {R.}~\bibnamefont {Caps}},\ and\ \bibinfo {author} {\bibfnamefont {J.}~\bibnamefont {Fricke}},\ }\href {https://doi.org/10.1063/1.1143914} {\bibfield  {journal} {\bibinfo  {journal} {Review of scientific instruments}\ }\textbf {\bibinfo {volume} {64}},\ \bibinfo {pages} {2523} (\bibinfo {year} {1993})},\ \Eprint {https://arxiv.org/abs/https://pubs.aip.org/aip/rsi/article-pdf/64/9/2523/19315455/2523\_1\_online.pdf} {https://pubs.aip.org/aip/rsi/article-pdf/64/9/2523/19315455/2523\_1\_online.pdf} \BibitemShut {NoStop}%
\bibitem [{\citenamefont {Colangelo}\ \emph {et~al.}(2023)\citenamefont {Colangelo}, \citenamefont {Korzh}, \citenamefont {Allmaras}, \citenamefont {Beyer}, \citenamefont {Mueller}, \citenamefont {Briggs}, \citenamefont {Bumble}, \citenamefont {Runyan}, \citenamefont {Stevens}, \citenamefont {McCaughan}, \citenamefont {Zhu}, \citenamefont {Smith}, \citenamefont {Becker}, \citenamefont {Narv\'aez}, \citenamefont {Bienfang}, \citenamefont {Frasca}, \citenamefont {Velasco}, \citenamefont {Ramirez}, \citenamefont {Walter}, \citenamefont {Schmidt}, \citenamefont {Wollman}, \citenamefont {Spiropulu}, \citenamefont {Mirin}, \citenamefont {Nam}, \citenamefont {Berggren},\ and\ \citenamefont {Shaw}}]{Colangelo2023}%
  \BibitemOpen
  \bibfield  {author} {\bibinfo {author} {\bibfnamefont {M.}~\bibnamefont {Colangelo}}, \bibinfo {author} {\bibfnamefont {B.}~\bibnamefont {Korzh}}, \bibinfo {author} {\bibfnamefont {J.~P.}\ \bibnamefont {Allmaras}}, \bibinfo {author} {\bibfnamefont {A.~D.}\ \bibnamefont {Beyer}}, \bibinfo {author} {\bibfnamefont {A.~S.}\ \bibnamefont {Mueller}}, \bibinfo {author} {\bibfnamefont {R.~M.}\ \bibnamefont {Briggs}}, \bibinfo {author} {\bibfnamefont {B.}~\bibnamefont {Bumble}}, \bibinfo {author} {\bibfnamefont {M.}~\bibnamefont {Runyan}}, \bibinfo {author} {\bibfnamefont {M.~J.}\ \bibnamefont {Stevens}}, \bibinfo {author} {\bibfnamefont {A.~N.}\ \bibnamefont {McCaughan}}, \bibinfo {author} {\bibfnamefont {D.}~\bibnamefont {Zhu}}, \bibinfo {author} {\bibfnamefont {S.}~\bibnamefont {Smith}}, \bibinfo {author} {\bibfnamefont {W.}~\bibnamefont {Becker}}, \bibinfo {author} {\bibfnamefont {L.}~\bibnamefont {Narv\'aez}}, \bibinfo {author} {\bibfnamefont {J.~C.}\ \bibnamefont {Bienfang}}, \bibinfo {author} {\bibfnamefont
  {S.}~\bibnamefont {Frasca}}, \bibinfo {author} {\bibfnamefont {A.~E.}\ \bibnamefont {Velasco}}, \bibinfo {author} {\bibfnamefont {E.~E.}\ \bibnamefont {Ramirez}}, \bibinfo {author} {\bibfnamefont {A.~B.}\ \bibnamefont {Walter}}, \bibinfo {author} {\bibfnamefont {E.}~\bibnamefont {Schmidt}}, \bibinfo {author} {\bibfnamefont {E.~E.}\ \bibnamefont {Wollman}}, \bibinfo {author} {\bibfnamefont {M.}~\bibnamefont {Spiropulu}}, \bibinfo {author} {\bibfnamefont {R.}~\bibnamefont {Mirin}}, \bibinfo {author} {\bibfnamefont {S.~W.}\ \bibnamefont {Nam}}, \bibinfo {author} {\bibfnamefont {K.~K.}\ \bibnamefont {Berggren}},\ and\ \bibinfo {author} {\bibfnamefont {M.~D.}\ \bibnamefont {Shaw}},\ }\href {https://doi.org/10.1103/PhysRevApplied.19.044093} {\bibfield  {journal} {\bibinfo  {journal} {Phys. Rev. Appl.}\ }\textbf {\bibinfo {volume} {19}},\ \bibinfo {pages} {044093} (\bibinfo {year} {2023})}\BibitemShut {NoStop}%
\bibitem [{\citenamefont {Olori}\ \emph {et~al.}(2021)\citenamefont {Olori}, \citenamefont {Di~Pietro},\ and\ \citenamefont {Campopiano}}]{Olori2021}%
  \BibitemOpen
  \bibfield  {author} {\bibinfo {author} {\bibfnamefont {A.}~\bibnamefont {Olori}}, \bibinfo {author} {\bibfnamefont {P.}~\bibnamefont {Di~Pietro}},\ and\ \bibinfo {author} {\bibfnamefont {A.}~\bibnamefont {Campopiano}},\ }\href {http://www.scienceijsar.com/sites/default/files/article-pdf/IJSAR-0374.pdf} {\bibfield  {journal} {\bibinfo  {journal} {International Journal of Science Academic Research}\ }\textbf {\bibinfo {volume} {02}},\ \bibinfo {pages} {1015} (\bibinfo {year} {2021})}\BibitemShut {NoStop}%
\bibitem [{\citenamefont {Hadni}\ \emph {et~al.}(1967)\citenamefont {Hadni}, \citenamefont {Claudel}, \citenamefont {Chanal}, \citenamefont {Strimer},\ and\ \citenamefont {Vergnat}}]{Hadni1967}%
  \BibitemOpen
  \bibfield  {author} {\bibinfo {author} {\bibfnamefont {A.}~\bibnamefont {Hadni}}, \bibinfo {author} {\bibfnamefont {J.}~\bibnamefont {Claudel}}, \bibinfo {author} {\bibfnamefont {D.}~\bibnamefont {Chanal}}, \bibinfo {author} {\bibfnamefont {P.}~\bibnamefont {Strimer}},\ and\ \bibinfo {author} {\bibfnamefont {P.}~\bibnamefont {Vergnat}},\ }\href {https://doi.org/10.1103/PhysRev.163.836} {\bibfield  {journal} {\bibinfo  {journal} {Phys. Rev.}\ }\textbf {\bibinfo {volume} {163}},\ \bibinfo {pages} {836} (\bibinfo {year} {1967})}\BibitemShut {NoStop}%
\bibitem [{\citenamefont {Martin}(1971)}]{Martin1971}%
  \BibitemOpen
  \bibfield  {author} {\bibinfo {author} {\bibfnamefont {T.~P.}\ \bibnamefont {Martin}},\ }\href {https://doi.org/https://doi.org/10.1016/0038-1098(71)90231-6} {\bibfield  {journal} {\bibinfo  {journal} {Solid State Communications}\ }\textbf {\bibinfo {volume} {9}},\ \bibinfo {pages} {623} (\bibinfo {year} {1971})}\BibitemShut {NoStop}%
\bibitem [{\citenamefont {Wollack}\ \emph {et~al.}(2020)\citenamefont {Wollack}, \citenamefont {Cataldo}, \citenamefont {Miller},\ and\ \citenamefont {Quijada}}]{Wollack2020}%
  \BibitemOpen
  \bibfield  {author} {\bibinfo {author} {\bibfnamefont {E.~J.}\ \bibnamefont {Wollack}}, \bibinfo {author} {\bibfnamefont {G.}~\bibnamefont {Cataldo}}, \bibinfo {author} {\bibfnamefont {K.~H.}\ \bibnamefont {Miller}},\ and\ \bibinfo {author} {\bibfnamefont {M.~A.}\ \bibnamefont {Quijada}},\ }\href {https://doi.org/10.1364/OL.393847} {\bibfield  {journal} {\bibinfo  {journal} {Opt. Lett.}\ }\textbf {\bibinfo {volume} {45}},\ \bibinfo {pages} {4935} (\bibinfo {year} {2020})}\BibitemShut {NoStop}%
\bibitem [{\citenamefont {Hochberg}\ \emph {et~al.}(2019)\citenamefont {Hochberg}, \citenamefont {Charaev}, \citenamefont {Nam}, \citenamefont {Verma}, \citenamefont {Colangelo},\ and\ \citenamefont {Berggren}}]{PhysRevLett.123.151802}%
  \BibitemOpen
  \bibfield  {author} {\bibinfo {author} {\bibfnamefont {Y.}~\bibnamefont {Hochberg}}, \bibinfo {author} {\bibfnamefont {I.}~\bibnamefont {Charaev}}, \bibinfo {author} {\bibfnamefont {S.-W.}\ \bibnamefont {Nam}}, \bibinfo {author} {\bibfnamefont {V.}~\bibnamefont {Verma}}, \bibinfo {author} {\bibfnamefont {M.}~\bibnamefont {Colangelo}},\ and\ \bibinfo {author} {\bibfnamefont {K.~K.}\ \bibnamefont {Berggren}},\ }\href {https://doi.org/10.1103/PhysRevLett.123.151802} {\bibfield  {journal} {\bibinfo  {journal} {Phys. Rev. Lett.}\ }\textbf {\bibinfo {volume} {123}},\ \bibinfo {pages} {151802} (\bibinfo {year} {2019})}\BibitemShut {NoStop}%
\bibitem [{\citenamefont {Wollman}\ \emph {et~al.}(2019)\citenamefont {Wollman}, \citenamefont {Verma}, \citenamefont {Lita}, \citenamefont {Farr}, \citenamefont {Shaw}, \citenamefont {Mirin},\ and\ \citenamefont {Woo~Nam}}]{wollman2019kilopixel}%
  \BibitemOpen
  \bibfield  {author} {\bibinfo {author} {\bibfnamefont {E.~E.}\ \bibnamefont {Wollman}}, \bibinfo {author} {\bibfnamefont {V.~B.}\ \bibnamefont {Verma}}, \bibinfo {author} {\bibfnamefont {A.~E.}\ \bibnamefont {Lita}}, \bibinfo {author} {\bibfnamefont {W.~H.}\ \bibnamefont {Farr}}, \bibinfo {author} {\bibfnamefont {M.~D.}\ \bibnamefont {Shaw}}, \bibinfo {author} {\bibfnamefont {R.~P.}\ \bibnamefont {Mirin}},\ and\ \bibinfo {author} {\bibfnamefont {S.}~\bibnamefont {Woo~Nam}},\ }\href@noop {} {\bibfield  {journal} {\bibinfo  {journal} {Optics Express}\ }\textbf {\bibinfo {volume} {27}},\ \bibinfo {pages} {35279} (\bibinfo {year} {2019})}\BibitemShut {NoStop}%
\bibitem [{\citenamefont {Sypkens}\ \emph {et~al.}(2024)\citenamefont {Sypkens}, \citenamefont {Minutolo}, \citenamefont {Patel}, \citenamefont {Knehr}, \citenamefont {Walter}, \citenamefont {Leduc}, \citenamefont {Narváez}, \citenamefont {Chamberlin}, \citenamefont {Jamison-Hooks}, \citenamefont {Shaw}, \citenamefont {Day},\ and\ \citenamefont {Korzh}}]{Sypkens2024}%
  \BibitemOpen
  \bibfield  {author} {\bibinfo {author} {\bibfnamefont {S.}~\bibnamefont {Sypkens}}, \bibinfo {author} {\bibfnamefont {L.}~\bibnamefont {Minutolo}}, \bibinfo {author} {\bibfnamefont {S.}~\bibnamefont {Patel}}, \bibinfo {author} {\bibfnamefont {E.}~\bibnamefont {Knehr}}, \bibinfo {author} {\bibfnamefont {A.~B.}\ \bibnamefont {Walter}}, \bibinfo {author} {\bibfnamefont {H.~G.}\ \bibnamefont {Leduc}}, \bibinfo {author} {\bibfnamefont {L.}~\bibnamefont {Narváez}}, \bibinfo {author} {\bibfnamefont {R.}~\bibnamefont {Chamberlin}}, \bibinfo {author} {\bibfnamefont {T.}~\bibnamefont {Jamison-Hooks}}, \bibinfo {author} {\bibfnamefont {M.~D.}\ \bibnamefont {Shaw}}, \bibinfo {author} {\bibfnamefont {P.~K.}\ \bibnamefont {Day}},\ and\ \bibinfo {author} {\bibfnamefont {B.}~\bibnamefont {Korzh}},\ }\href {https://doi.org/10.1063/5.0220090} {\bibfield  {journal} {\bibinfo  {journal} {Applied Physics Letters}\ }\textbf {\bibinfo {volume} {124}},\ \bibinfo {pages} {262602} (\bibinfo {year} {2024})},\ \Eprint
  {https://arxiv.org/abs/https://pubs.aip.org/aip/apl/article-pdf/doi/10.1063/5.0220090/20011699/262602\_1\_5.0220090.pdf} {https://pubs.aip.org/aip/apl/article-pdf/doi/10.1063/5.0220090/20011699/262602\_1\_5.0220090.pdf} \BibitemShut {NoStop}%
\bibitem [{\citenamefont {Heath}\ \emph {et~al.}(2015)\citenamefont {Heath}, \citenamefont {Tanner}, \citenamefont {Drysdale}, \citenamefont {Miki}, \citenamefont {Giannini}, \citenamefont {Maier},\ and\ \citenamefont {Hadfield}}]{heath2015nanoantenna}%
  \BibitemOpen
  \bibfield  {author} {\bibinfo {author} {\bibfnamefont {R.~M.}\ \bibnamefont {Heath}}, \bibinfo {author} {\bibfnamefont {M.~G.}\ \bibnamefont {Tanner}}, \bibinfo {author} {\bibfnamefont {T.~D.}\ \bibnamefont {Drysdale}}, \bibinfo {author} {\bibfnamefont {S.}~\bibnamefont {Miki}}, \bibinfo {author} {\bibfnamefont {V.}~\bibnamefont {Giannini}}, \bibinfo {author} {\bibfnamefont {S.~A.}\ \bibnamefont {Maier}},\ and\ \bibinfo {author} {\bibfnamefont {R.~H.}\ \bibnamefont {Hadfield}},\ }\href@noop {} {\bibfield  {journal} {\bibinfo  {journal} {Nano Letters}\ }\textbf {\bibinfo {volume} {15}},\ \bibinfo {pages} {819} (\bibinfo {year} {2015})}\BibitemShut {NoStop}%
\bibitem [{\citenamefont {Weber}\ \emph {et~al.}(2016)\citenamefont {Weber}, \citenamefont {Nesterov}, \citenamefont {Weiss}, \citenamefont {Scherer}, \citenamefont {Hentschel}, \citenamefont {Vogt}, \citenamefont {Huck}, \citenamefont {Li}, \citenamefont {Dressel}, \citenamefont {Giessen},\ and\ \citenamefont {Neubrech}}]{Weber2016}%
  \BibitemOpen
  \bibfield  {author} {\bibinfo {author} {\bibfnamefont {K.}~\bibnamefont {Weber}}, \bibinfo {author} {\bibfnamefont {M.}~\bibnamefont {Nesterov}}, \bibinfo {author} {\bibfnamefont {T.}~\bibnamefont {Weiss}}, \bibinfo {author} {\bibfnamefont {M.}~\bibnamefont {Scherer}}, \bibinfo {author} {\bibfnamefont {M.}~\bibnamefont {Hentschel}}, \bibinfo {author} {\bibfnamefont {J.}~\bibnamefont {Vogt}}, \bibinfo {author} {\bibfnamefont {C.}~\bibnamefont {Huck}}, \bibinfo {author} {\bibfnamefont {W.}~\bibnamefont {Li}}, \bibinfo {author} {\bibfnamefont {M.}~\bibnamefont {Dressel}}, \bibinfo {author} {\bibfnamefont {H.}~\bibnamefont {Giessen}},\ and\ \bibinfo {author} {\bibfnamefont {F.}~\bibnamefont {Neubrech}},\ }\href {https://doi.org/10.1021/acsphotonics.6b00534} {\bibfield  {journal} {\bibinfo  {journal} {ACS Photonics}\ }\textbf {\bibinfo {volume} {4}} (\bibinfo {year} {2016})}\BibitemShut {NoStop}%
\bibitem [{\citenamefont {Verma}\ \emph {et~al.}(2021)\citenamefont {Verma}, \citenamefont {Korzh}, \citenamefont {Walter}, \citenamefont {Lita}, \citenamefont {Briggs}, \citenamefont {Colangelo}, \citenamefont {Zhai}, \citenamefont {Wollman}, \citenamefont {Beyer}, \citenamefont {Allmaras} \emph {et~al.}}]{verma2021single}%
  \BibitemOpen
  \bibfield  {author} {\bibinfo {author} {\bibfnamefont {V.}~\bibnamefont {Verma}}, \bibinfo {author} {\bibfnamefont {B.}~\bibnamefont {Korzh}}, \bibinfo {author} {\bibfnamefont {A.~B.}\ \bibnamefont {Walter}}, \bibinfo {author} {\bibfnamefont {A.~E.}\ \bibnamefont {Lita}}, \bibinfo {author} {\bibfnamefont {R.~M.}\ \bibnamefont {Briggs}}, \bibinfo {author} {\bibfnamefont {M.}~\bibnamefont {Colangelo}}, \bibinfo {author} {\bibfnamefont {Y.}~\bibnamefont {Zhai}}, \bibinfo {author} {\bibfnamefont {E.~E.}\ \bibnamefont {Wollman}}, \bibinfo {author} {\bibfnamefont {A.~D.}\ \bibnamefont {Beyer}}, \bibinfo {author} {\bibfnamefont {J.~P.}\ \bibnamefont {Allmaras}}, \emph {et~al.},\ }\href@noop {} {\bibfield  {journal} {\bibinfo  {journal} {APL Photonics}\ }\textbf {\bibinfo {volume} {6}} (\bibinfo {year} {2021})}\BibitemShut {NoStop}%
\bibitem [{\citenamefont {Colangelo}\ \emph {et~al.}(2022)\citenamefont {Colangelo}, \citenamefont {Walter}, \citenamefont {Korzh}, \citenamefont {Schmidt}, \citenamefont {Bumble}, \citenamefont {Lita}, \citenamefont {Beyer}, \citenamefont {Allmaras}, \citenamefont {Briggs}, \citenamefont {Kozorezov} \emph {et~al.}}]{colangelo2022large}%
  \BibitemOpen
  \bibfield  {author} {\bibinfo {author} {\bibfnamefont {M.}~\bibnamefont {Colangelo}}, \bibinfo {author} {\bibfnamefont {A.~B.}\ \bibnamefont {Walter}}, \bibinfo {author} {\bibfnamefont {B.~A.}\ \bibnamefont {Korzh}}, \bibinfo {author} {\bibfnamefont {E.}~\bibnamefont {Schmidt}}, \bibinfo {author} {\bibfnamefont {B.}~\bibnamefont {Bumble}}, \bibinfo {author} {\bibfnamefont {A.~E.}\ \bibnamefont {Lita}}, \bibinfo {author} {\bibfnamefont {A.~D.}\ \bibnamefont {Beyer}}, \bibinfo {author} {\bibfnamefont {J.~P.}\ \bibnamefont {Allmaras}}, \bibinfo {author} {\bibfnamefont {R.~M.}\ \bibnamefont {Briggs}}, \bibinfo {author} {\bibfnamefont {A.~G.}\ \bibnamefont {Kozorezov}}, \emph {et~al.},\ }\href@noop {} {\bibfield  {journal} {\bibinfo  {journal} {Nano Letters}\ }\textbf {\bibinfo {volume} {22}},\ \bibinfo {pages} {5667} (\bibinfo {year} {2022})}\BibitemShut {NoStop}%
\bibitem [{\citenamefont {Polakovic}\ \emph {et~al.}(2020)\citenamefont {Polakovic}, \citenamefont {Armstrong}, \citenamefont {Yefremenko}, \citenamefont {Pearson}, \citenamefont {Hafidi}, \citenamefont {Karapetrov}, \citenamefont {Meziani},\ and\ \citenamefont {Novosad}}]{polakovic2020superconducting}%
  \BibitemOpen
  \bibfield  {author} {\bibinfo {author} {\bibfnamefont {T.}~\bibnamefont {Polakovic}}, \bibinfo {author} {\bibfnamefont {W.}~\bibnamefont {Armstrong}}, \bibinfo {author} {\bibfnamefont {V.}~\bibnamefont {Yefremenko}}, \bibinfo {author} {\bibfnamefont {J.~E.}\ \bibnamefont {Pearson}}, \bibinfo {author} {\bibfnamefont {K.}~\bibnamefont {Hafidi}}, \bibinfo {author} {\bibfnamefont {G.}~\bibnamefont {Karapetrov}}, \bibinfo {author} {\bibfnamefont {Z.-E.}\ \bibnamefont {Meziani}},\ and\ \bibinfo {author} {\bibfnamefont {V.}~\bibnamefont {Novosad}},\ }\href {https://doi.org/https://doi.org/10.1016/j.nima.2020.163543} {\bibfield  {journal} {\bibinfo  {journal} {Nuclear Instruments and Methods in Physics Research Section A: Accelerators, Spectrometers, Detectors and Associated Equipment}\ }\textbf {\bibinfo {volume} {959}},\ \bibinfo {pages} {163543} (\bibinfo {year} {2020})}\BibitemShut {NoStop}%
\bibitem [{\citenamefont {Zhu}\ \emph {et~al.}(2023)\citenamefont {Zhu}, \citenamefont {Charaev},\ and\ \citenamefont {Schilling}}]{zhu2023effective}%
  \BibitemOpen
  \bibfield  {author} {\bibinfo {author} {\bibfnamefont {D.}~\bibnamefont {Zhu}}, \bibinfo {author} {\bibfnamefont {I.}~\bibnamefont {Charaev}},\ and\ \bibinfo {author} {\bibfnamefont {A.}~\bibnamefont {Schilling}},\ }\href@noop {} {\bibfield  {journal} {\bibinfo  {journal} {Superconductor Science and Technology}\ }\textbf {\bibinfo {volume} {36}},\ \bibinfo {pages} {105012} (\bibinfo {year} {2023})}\BibitemShut {NoStop}%
\bibitem [{\citenamefont {Oppenheim}\ and\ \citenamefont {Even}(1962)}]{Oppenheim:62}%
  \BibitemOpen
  \bibfield  {author} {\bibinfo {author} {\bibfnamefont {U.~P.}\ \bibnamefont {Oppenheim}}\ and\ \bibinfo {author} {\bibfnamefont {U.}~\bibnamefont {Even}},\ }\href {https://doi.org/10.1364/JOSA.52.1078_1} {\bibfield  {journal} {\bibinfo  {journal} {J. Opt. Soc. Am.}\ }\textbf {\bibinfo {volume} {52}},\ \bibinfo {pages} {1078} (\bibinfo {year} {1962})}\BibitemShut {NoStop}%
\bibitem [{\citenamefont {Li}(1980)}]{Li1980}%
  \BibitemOpen
  \bibfield  {author} {\bibinfo {author} {\bibfnamefont {H.}~\bibnamefont {Li}},\ }\href {https://doi.org/10.1007/BF00506275} {\bibfield  {journal} {\bibinfo  {journal} {Int J Thermophys}\ }\textbf {\bibinfo {volume} {1980}},\ \bibinfo {pages} {97–134} (\bibinfo {year} {1980})}\BibitemShut {NoStop}%
\bibitem [{\citenamefont {Du}\ \emph {et~al.}(2022)\citenamefont {Du}, \citenamefont {Egana-Ugrinovic}, \citenamefont {Essig},\ and\ \citenamefont {Sholapurkar}}]{Du:2020ldo}%
  \BibitemOpen
  \bibfield  {author} {\bibinfo {author} {\bibfnamefont {P.}~\bibnamefont {Du}}, \bibinfo {author} {\bibfnamefont {D.}~\bibnamefont {Egana-Ugrinovic}}, \bibinfo {author} {\bibfnamefont {R.}~\bibnamefont {Essig}},\ and\ \bibinfo {author} {\bibfnamefont {M.}~\bibnamefont {Sholapurkar}},\ }\href {https://doi.org/10.1103/PhysRevX.12.011009} {\bibfield  {journal} {\bibinfo  {journal} {Phys. Rev. X}\ }\textbf {\bibinfo {volume} {12}},\ \bibinfo {pages} {011009} (\bibinfo {year} {2022})},\ \Eprint {https://arxiv.org/abs/2011.13939} {arXiv:2011.13939 [hep-ph]} \BibitemShut {NoStop}%
\bibitem [{\citenamefont {Artikov}\ \emph {et~al.}(2024)\citenamefont {Artikov}, \citenamefont {Baranov}, \citenamefont {Boikov}, \citenamefont {Chokheli}, \citenamefont {Davydov}, \citenamefont {Glagolev}, \citenamefont {Simonenko}, \citenamefont {Tsamalaidze}, \citenamefont {Vasilyev},\ and\ \citenamefont {Zimin}}]{Artikov2024}%
  \BibitemOpen
  \bibfield  {author} {\bibinfo {author} {\bibfnamefont {A.}~\bibnamefont {Artikov}}, \bibinfo {author} {\bibfnamefont {V.}~\bibnamefont {Baranov}}, \bibinfo {author} {\bibfnamefont {A.}~\bibnamefont {Boikov}}, \bibinfo {author} {\bibfnamefont {D.}~\bibnamefont {Chokheli}}, \bibinfo {author} {\bibfnamefont {Y.}~\bibnamefont {Davydov}}, \bibinfo {author} {\bibfnamefont {V.}~\bibnamefont {Glagolev}}, \bibinfo {author} {\bibfnamefont {A.}~\bibnamefont {Simonenko}}, \bibinfo {author} {\bibfnamefont {Z.}~\bibnamefont {Tsamalaidze}}, \bibinfo {author} {\bibfnamefont {I.}~\bibnamefont {Vasilyev}},\ and\ \bibinfo {author} {\bibfnamefont {I.}~\bibnamefont {Zimin}},\ }\href {https://doi.org/https://doi.org/10.1016/j.nima.2024.169436} {\bibfield  {journal} {\bibinfo  {journal} {Nuclear Instruments and Methods in Physics Research Section A: Accelerators, Spectrometers, Detectors and Associated Equipment}\ }\textbf {\bibinfo {volume} {1064}},\ \bibinfo {pages} {169436} (\bibinfo {year} {2024})}\BibitemShut {NoStop}%
\bibitem [{\citenamefont {Zhao}\ \emph {et~al.}(2017)\citenamefont {Zhao}, \citenamefont {Zhu}, \citenamefont {Calandri}, \citenamefont {Dane}, \citenamefont {McCaughan}, \citenamefont {Bellei}, \citenamefont {Wang}, \citenamefont {Santavicca},\ and\ \citenamefont {Berggren}}]{zhao2017single}%
  \BibitemOpen
  \bibfield  {author} {\bibinfo {author} {\bibfnamefont {Q.-Y.}\ \bibnamefont {Zhao}}, \bibinfo {author} {\bibfnamefont {D.}~\bibnamefont {Zhu}}, \bibinfo {author} {\bibfnamefont {N.}~\bibnamefont {Calandri}}, \bibinfo {author} {\bibfnamefont {A.~E.}\ \bibnamefont {Dane}}, \bibinfo {author} {\bibfnamefont {A.~N.}\ \bibnamefont {McCaughan}}, \bibinfo {author} {\bibfnamefont {F.}~\bibnamefont {Bellei}}, \bibinfo {author} {\bibfnamefont {H.-Z.}\ \bibnamefont {Wang}}, \bibinfo {author} {\bibfnamefont {D.~F.}\ \bibnamefont {Santavicca}},\ and\ \bibinfo {author} {\bibfnamefont {K.~K.}\ \bibnamefont {Berggren}},\ }\href@noop {} {\bibfield  {journal} {\bibinfo  {journal} {Nature Photonics}\ }\textbf {\bibinfo {volume} {11}},\ \bibinfo {pages} {247} (\bibinfo {year} {2017})}\BibitemShut {NoStop}%
\bibitem [{\citenamefont {Budini}(1953)}]{Budini1953}%
  \BibitemOpen
  \bibfield  {author} {\bibinfo {author} {\bibfnamefont {P.}~\bibnamefont {Budini}},\ }\href {https://doi.org/10.1103/PhysRev.89.1147} {\bibfield  {journal} {\bibinfo  {journal} {Phys. Rev.}\ }\textbf {\bibinfo {volume} {89}},\ \bibinfo {pages} {1147} (\bibinfo {year} {1953})}\BibitemShut {NoStop}%
\bibitem [{\citenamefont {Du}\ \emph {et~al.}(2024)\citenamefont {Du}, \citenamefont {Egaña-Ugrinovic}, \citenamefont {Essig},\ and\ \citenamefont {Sholapurkar}}]{Du2024}%
  \BibitemOpen
  \bibfield  {author} {\bibinfo {author} {\bibfnamefont {P.}~\bibnamefont {Du}}, \bibinfo {author} {\bibfnamefont {D.}~\bibnamefont {Egaña-Ugrinovic}}, \bibinfo {author} {\bibfnamefont {R.}~\bibnamefont {Essig}},\ and\ \bibinfo {author} {\bibfnamefont {M.}~\bibnamefont {Sholapurkar}},\ }\href {https://doi.org/10.1007/JHEP01(2024)164} {\bibfield  {journal} {\bibinfo  {journal} {Journal of High Energy Physics}\ }\textbf {\bibinfo {volume} {2024}} (\bibinfo {year} {2024})}\BibitemShut {NoStop}%
\bibitem [{\citenamefont {Wollman}\ \emph {et~al.}(2017)\citenamefont {Wollman} \emph {et~al.}}]{Wollman17}%
  \BibitemOpen
  \bibfield  {author} {\bibinfo {author} {\bibfnamefont {E.~E.}\ \bibnamefont {Wollman}} \emph {et~al.},\ }\href {https://doi.org/10.1364/OE.25.026792} {\bibfield  {journal} {\bibinfo  {journal} {Opt. Express}\ }\textbf {\bibinfo {volume} {25}},\ \bibinfo {pages} {26792} (\bibinfo {year} {2017})}\BibitemShut {NoStop}%
\bibitem [{\citenamefont {Zhang}\ \emph {et~al.}(2022)\citenamefont {Zhang}, \citenamefont {Zhang}, \citenamefont {Huang}, \citenamefont {Yang}, \citenamefont {You}, \citenamefont {Liu}, \citenamefont {Hu}, \citenamefont {Xiao}, \citenamefont {Zhang}, \citenamefont {Wang}, \citenamefont {Li}, \citenamefont {Wang},\ and\ \citenamefont {Li}}]{zhang2022geometric}%
  \BibitemOpen
  \bibfield  {author} {\bibinfo {author} {\bibfnamefont {X.}~\bibnamefont {Zhang}}, \bibinfo {author} {\bibfnamefont {X.}~\bibnamefont {Zhang}}, \bibinfo {author} {\bibfnamefont {J.}~\bibnamefont {Huang}}, \bibinfo {author} {\bibfnamefont {C.}~\bibnamefont {Yang}}, \bibinfo {author} {\bibfnamefont {L.}~\bibnamefont {You}}, \bibinfo {author} {\bibfnamefont {X.}~\bibnamefont {Liu}}, \bibinfo {author} {\bibfnamefont {P.}~\bibnamefont {Hu}}, \bibinfo {author} {\bibfnamefont {Y.}~\bibnamefont {Xiao}}, \bibinfo {author} {\bibfnamefont {W.}~\bibnamefont {Zhang}}, \bibinfo {author} {\bibfnamefont {Y.}~\bibnamefont {Wang}}, \bibinfo {author} {\bibfnamefont {L.}~\bibnamefont {Li}}, \bibinfo {author} {\bibfnamefont {Z.}~\bibnamefont {Wang}},\ and\ \bibinfo {author} {\bibfnamefont {H.}~\bibnamefont {Li}},\ }\href@noop {} {\bibfield  {journal} {\bibinfo  {journal} {Superconductivity}\ }\textbf {\bibinfo {volume} {1}},\ \bibinfo {pages} {100006} (\bibinfo {year} {2022})}\BibitemShut {NoStop}%
\bibitem [{\citenamefont {Andreev}\ \emph {et~al.}(2024)\citenamefont {Andreev}, \citenamefont {Semenov}, \citenamefont {Manova}, \citenamefont {Seleznev}, \citenamefont {Svyatodukh}, \citenamefont {Divochiy}, \citenamefont {Morozov},\ and\ \citenamefont {Goltsman}}]{andreev2024dark}%
  \BibitemOpen
  \bibfield  {author} {\bibinfo {author} {\bibfnamefont {V.}~\bibnamefont {Andreev}}, \bibinfo {author} {\bibfnamefont {A.}~\bibnamefont {Semenov}}, \bibinfo {author} {\bibfnamefont {N.}~\bibnamefont {Manova}}, \bibinfo {author} {\bibfnamefont {V.}~\bibnamefont {Seleznev}}, \bibinfo {author} {\bibfnamefont {S.}~\bibnamefont {Svyatodukh}}, \bibinfo {author} {\bibfnamefont {A.}~\bibnamefont {Divochiy}}, \bibinfo {author} {\bibfnamefont {P.}~\bibnamefont {Morozov}},\ and\ \bibinfo {author} {\bibfnamefont {G.}~\bibnamefont {Goltsman}},\ }\href@noop {} {\bibfield  {journal} {\bibinfo  {journal} {IEEE Transactions on Applied Superconductivity}\ } (\bibinfo {year} {2024})}\BibitemShut {NoStop}%
\bibitem [{\citenamefont {An}\ \emph {et~al.}(2024)\citenamefont {An}, \citenamefont {Ge}, \citenamefont {Liu},\ and\ \citenamefont {Lu}}]{An:2024kls}%
  \BibitemOpen
  \bibfield  {author} {\bibinfo {author} {\bibfnamefont {H.}~\bibnamefont {An}}, \bibinfo {author} {\bibfnamefont {S.}~\bibnamefont {Ge}}, \bibinfo {author} {\bibfnamefont {J.}~\bibnamefont {Liu}},\ and\ \bibinfo {author} {\bibfnamefont {Z.}~\bibnamefont {Lu}},\ }\href@noop {} {\bibfield  {journal} {\bibinfo  {journal} {arXiv preprint}\ } (\bibinfo {year} {2024})},\ \Eprint {https://arxiv.org/abs/2402.17140} {arXiv:2402.17140 [hep-ph]} \BibitemShut {NoStop}%
\bibitem [{\citenamefont {Aggarwal}\ \emph {et~al.}(2025)\citenamefont {Aggarwal} \emph {et~al.}}]{DAMIC-M:2025luv}%
  \BibitemOpen
  \bibfield  {author} {\bibinfo {author} {\bibfnamefont {K.}~\bibnamefont {Aggarwal}} \emph {et~al.} (\bibinfo {collaboration} {DAMIC-M}),\ }\href@noop {} {\bibfield  {journal} {\bibinfo  {journal} {arXiv preprint}\ } (\bibinfo {year} {2025})},\ \Eprint {https://arxiv.org/abs/2503.14617} {arXiv:2503.14617 [hep-ex]} \BibitemShut {NoStop}%
\bibitem [{\citenamefont {Bloch}\ \emph {et~al.}(2025{\natexlab{b}})\citenamefont {Bloch}, \citenamefont {Botti}, \citenamefont {Cababie}, \citenamefont {Cancelo}, \citenamefont {Cervantes-Vergara}, \citenamefont {Daal}, \citenamefont {Desai}, \citenamefont {Drlica-Wagner}, \citenamefont {Essig}, \citenamefont {Estrada}, \citenamefont {Etzion}, \citenamefont {Moroni}, \citenamefont {Holland}, \citenamefont {Kehat}, \citenamefont {Lawson}, \citenamefont {Luoma}, \citenamefont {Orly}, \citenamefont {Perez}, \citenamefont {Rodrigues}, \citenamefont {Saffold}, \citenamefont {Scorza}, \citenamefont {Sofo-Haro}, \citenamefont {Stifter}, \citenamefont {Tiffenberg}, \citenamefont {Uemura}, \citenamefont {Villalpando}, \citenamefont {Volansky}, \citenamefont {Winkel}, \citenamefont {Wu},\ and\ \citenamefont {Yu}}]{PhysRevLett.134.161002}%
  \BibitemOpen
  \bibfield  {author} {\bibinfo {author} {\bibfnamefont {I.~M.}\ \bibnamefont {Bloch}}, \bibinfo {author} {\bibfnamefont {A.~M.}\ \bibnamefont {Botti}}, \bibinfo {author} {\bibfnamefont {M.}~\bibnamefont {Cababie}}, \bibinfo {author} {\bibfnamefont {G.}~\bibnamefont {Cancelo}}, \bibinfo {author} {\bibfnamefont {B.~A.}\ \bibnamefont {Cervantes-Vergara}}, \bibinfo {author} {\bibfnamefont {M.}~\bibnamefont {Daal}}, \bibinfo {author} {\bibfnamefont {A.}~\bibnamefont {Desai}}, \bibinfo {author} {\bibfnamefont {A.}~\bibnamefont {Drlica-Wagner}}, \bibinfo {author} {\bibfnamefont {R.}~\bibnamefont {Essig}}, \bibinfo {author} {\bibfnamefont {J.}~\bibnamefont {Estrada}}, \bibinfo {author} {\bibfnamefont {E.}~\bibnamefont {Etzion}}, \bibinfo {author} {\bibfnamefont {G.~F.}\ \bibnamefont {Moroni}}, \bibinfo {author} {\bibfnamefont {S.~E.}\ \bibnamefont {Holland}}, \bibinfo {author} {\bibfnamefont {J.}~\bibnamefont {Kehat}}, \bibinfo {author} {\bibfnamefont {I.}~\bibnamefont {Lawson}}, \bibinfo {author} {\bibfnamefont
  {S.}~\bibnamefont {Luoma}}, \bibinfo {author} {\bibfnamefont {A.}~\bibnamefont {Orly}}, \bibinfo {author} {\bibfnamefont {S.~E.}\ \bibnamefont {Perez}}, \bibinfo {author} {\bibfnamefont {D.}~\bibnamefont {Rodrigues}}, \bibinfo {author} {\bibfnamefont {N.~A.}\ \bibnamefont {Saffold}}, \bibinfo {author} {\bibfnamefont {S.}~\bibnamefont {Scorza}}, \bibinfo {author} {\bibfnamefont {M.}~\bibnamefont {Sofo-Haro}}, \bibinfo {author} {\bibfnamefont {K.}~\bibnamefont {Stifter}}, \bibinfo {author} {\bibfnamefont {J.}~\bibnamefont {Tiffenberg}}, \bibinfo {author} {\bibfnamefont {S.}~\bibnamefont {Uemura}}, \bibinfo {author} {\bibfnamefont {E.~M.}\ \bibnamefont {Villalpando}}, \bibinfo {author} {\bibfnamefont {T.}~\bibnamefont {Volansky}}, \bibinfo {author} {\bibfnamefont {F.}~\bibnamefont {Winkel}}, \bibinfo {author} {\bibfnamefont {Y.}~\bibnamefont {Wu}},\ and\ \bibinfo {author} {\bibfnamefont {T.-T.}\ \bibnamefont {Yu}} (\bibinfo {collaboration} {SENSEI Collaboration}),\ }\href
  {https://doi.org/10.1103/PhysRevLett.134.161002} {\bibfield  {journal} {\bibinfo  {journal} {Phys. Rev. Lett.}\ }\textbf {\bibinfo {volume} {134}},\ \bibinfo {pages} {161002} (\bibinfo {year} {2025}{\natexlab{b}})}\BibitemShut {NoStop}%
\bibitem [{\citenamefont {Li}\ and\ \citenamefont {Xu}(2023)}]{Li:2023vpv}%
  \BibitemOpen
  \bibfield  {author} {\bibinfo {author} {\bibfnamefont {S.-P.}\ \bibnamefont {Li}}\ and\ \bibinfo {author} {\bibfnamefont {X.-J.}\ \bibnamefont {Xu}},\ }\href {https://doi.org/10.1088/1475-7516/2023/09/009} {\bibfield  {journal} {\bibinfo  {journal} {JCAP}\ }\textbf {\bibinfo {volume} {{2023}}}\bibfield  {number} {\bibinfo  {number} { (09)},\ \bibinfo {pages} {009}},\ }\Eprint {https://arxiv.org/abs/2304.12907} {arXiv:2304.12907 [hep-ph]} \BibitemShut {NoStop}%
\bibitem [{\citenamefont {Adari}\ \emph {et~al.}(2025)\citenamefont {Adari}, \citenamefont {Bloch}, \citenamefont {Botti}, \citenamefont {Cababie}, \citenamefont {Cancelo}, \citenamefont {Cervantes-Vergara}, \citenamefont {Crisler}, \citenamefont {Daal}, \citenamefont {Desai}, \citenamefont {Drlica-Wagner}, \citenamefont {Essig}, \citenamefont {Estrada}, \citenamefont {Etzion}, \citenamefont {Moroni}, \citenamefont {Holland}, \citenamefont {Kehat}, \citenamefont {Korn}, \citenamefont {Lawson}, \citenamefont {Luoma}, \citenamefont {Orly}, \citenamefont {Perez}, \citenamefont {Rodrigues}, \citenamefont {Saffold}, \citenamefont {Scorza}, \citenamefont {Singal}, \citenamefont {Sofo-Haro}, \citenamefont {Stefanazzi}, \citenamefont {Stifter}, \citenamefont {Tiffenberg}, \citenamefont {Uemura}, \citenamefont {Villalpando}, \citenamefont {Volansky}, \citenamefont {Wu}, \citenamefont {Yu}, \citenamefont {Emken},\ and\ \citenamefont {Xu}}]{PhysRevLett.134.011804}%
  \BibitemOpen
  \bibfield  {author} {\bibinfo {author} {\bibfnamefont {P.}~\bibnamefont {Adari}}, \bibinfo {author} {\bibfnamefont {I.~M.}\ \bibnamefont {Bloch}}, \bibinfo {author} {\bibfnamefont {A.~M.}\ \bibnamefont {Botti}}, \bibinfo {author} {\bibfnamefont {M.}~\bibnamefont {Cababie}}, \bibinfo {author} {\bibfnamefont {G.}~\bibnamefont {Cancelo}}, \bibinfo {author} {\bibfnamefont {B.~A.}\ \bibnamefont {Cervantes-Vergara}}, \bibinfo {author} {\bibfnamefont {M.}~\bibnamefont {Crisler}}, \bibinfo {author} {\bibfnamefont {M.}~\bibnamefont {Daal}}, \bibinfo {author} {\bibfnamefont {A.}~\bibnamefont {Desai}}, \bibinfo {author} {\bibfnamefont {A.}~\bibnamefont {Drlica-Wagner}}, \bibinfo {author} {\bibfnamefont {R.}~\bibnamefont {Essig}}, \bibinfo {author} {\bibfnamefont {J.}~\bibnamefont {Estrada}}, \bibinfo {author} {\bibfnamefont {E.}~\bibnamefont {Etzion}}, \bibinfo {author} {\bibfnamefont {G.~F.}\ \bibnamefont {Moroni}}, \bibinfo {author} {\bibfnamefont {S.~E.}\ \bibnamefont {Holland}}, \bibinfo {author} {\bibfnamefont
  {J.}~\bibnamefont {Kehat}}, \bibinfo {author} {\bibfnamefont {Y.}~\bibnamefont {Korn}}, \bibinfo {author} {\bibfnamefont {I.}~\bibnamefont {Lawson}}, \bibinfo {author} {\bibfnamefont {S.}~\bibnamefont {Luoma}}, \bibinfo {author} {\bibfnamefont {A.}~\bibnamefont {Orly}}, \bibinfo {author} {\bibfnamefont {S.~E.}\ \bibnamefont {Perez}}, \bibinfo {author} {\bibfnamefont {D.}~\bibnamefont {Rodrigues}}, \bibinfo {author} {\bibfnamefont {N.~A.}\ \bibnamefont {Saffold}}, \bibinfo {author} {\bibfnamefont {S.}~\bibnamefont {Scorza}}, \bibinfo {author} {\bibfnamefont {A.}~\bibnamefont {Singal}}, \bibinfo {author} {\bibfnamefont {M.}~\bibnamefont {Sofo-Haro}}, \bibinfo {author} {\bibfnamefont {L.}~\bibnamefont {Stefanazzi}}, \bibinfo {author} {\bibfnamefont {K.}~\bibnamefont {Stifter}}, \bibinfo {author} {\bibfnamefont {J.}~\bibnamefont {Tiffenberg}}, \bibinfo {author} {\bibfnamefont {S.}~\bibnamefont {Uemura}}, \bibinfo {author} {\bibfnamefont {E.~M.}\ \bibnamefont {Villalpando}}, \bibinfo {author} {\bibfnamefont
  {T.}~\bibnamefont {Volansky}}, \bibinfo {author} {\bibfnamefont {Y.}~\bibnamefont {Wu}}, \bibinfo {author} {\bibfnamefont {T.-T.}\ \bibnamefont {Yu}}, \bibinfo {author} {\bibfnamefont {T.}~\bibnamefont {Emken}},\ and\ \bibinfo {author} {\bibfnamefont {H.}~\bibnamefont {Xu}} (\bibinfo {collaboration} {SENSEI Collaboration}),\ }\href {https://doi.org/10.1103/PhysRevLett.134.011804} {\bibfield  {journal} {\bibinfo  {journal} {Phys. Rev. Lett.}\ }\textbf {\bibinfo {volume} {134}},\ \bibinfo {pages} {011804} (\bibinfo {year} {2025})}\BibitemShut {NoStop}%
\bibitem [{\citenamefont {Altenm\"uller}\ \emph {et~al.}(2024)\citenamefont {Altenm\"uller} \emph {et~al.}}]{CAST:2024eil}%
  \BibitemOpen
  \bibfield  {author} {\bibinfo {author} {\bibfnamefont {K.}~\bibnamefont {Altenm\"uller}} \emph {et~al.} (\bibinfo {collaboration} {CAST}),\ }\href {https://doi.org/10.1103/PhysRevLett.133.221005} {\bibfield  {journal} {\bibinfo  {journal} {Phys. Rev. Lett.}\ }\textbf {\bibinfo {volume} {133}},\ \bibinfo {pages} {221005} (\bibinfo {year} {2024})},\ \Eprint {https://arxiv.org/abs/2406.16840} {arXiv:2406.16840 [hep-ex]} \BibitemShut {NoStop}%
\bibitem [{\citenamefont {Carenza}\ \emph {et~al.}(2023)\citenamefont {Carenza}, \citenamefont {Lucente},\ and\ \citenamefont {Vitagliano}}]{Carenza:2023qxh}%
  \BibitemOpen
  \bibfield  {author} {\bibinfo {author} {\bibfnamefont {P.}~\bibnamefont {Carenza}}, \bibinfo {author} {\bibfnamefont {G.}~\bibnamefont {Lucente}},\ and\ \bibinfo {author} {\bibfnamefont {E.}~\bibnamefont {Vitagliano}},\ }\href {https://doi.org/10.1103/PhysRevD.107.083032} {\bibfield  {journal} {\bibinfo  {journal} {Phys. Rev. D}\ }\textbf {\bibinfo {volume} {107}},\ \bibinfo {pages} {083032} (\bibinfo {year} {2023})},\ \Eprint {https://arxiv.org/abs/2301.06560} {arXiv:2301.06560 [hep-ph]} \BibitemShut {NoStop}%
\bibitem [{\citenamefont {Todarello}\ \emph {et~al.}(2024)\citenamefont {Todarello}, \citenamefont {Regis}, \citenamefont {Reynoso-Cordova}, \citenamefont {Taoso}, \citenamefont {Vaz}, \citenamefont {Brinchmann}, \citenamefont {Steinmetz},\ and\ \citenamefont {Zoutendijke}}]{Todarello:2023hdk}%
  \BibitemOpen
  \bibfield  {author} {\bibinfo {author} {\bibfnamefont {E.}~\bibnamefont {Todarello}}, \bibinfo {author} {\bibfnamefont {M.}~\bibnamefont {Regis}}, \bibinfo {author} {\bibfnamefont {J.}~\bibnamefont {Reynoso-Cordova}}, \bibinfo {author} {\bibfnamefont {M.}~\bibnamefont {Taoso}}, \bibinfo {author} {\bibfnamefont {D.}~\bibnamefont {Vaz}}, \bibinfo {author} {\bibfnamefont {J.}~\bibnamefont {Brinchmann}}, \bibinfo {author} {\bibfnamefont {M.}~\bibnamefont {Steinmetz}},\ and\ \bibinfo {author} {\bibfnamefont {S.~L.}\ \bibnamefont {Zoutendijke}},\ }\href {https://doi.org/10.1088/1475-7516/2024/05/043} {\bibfield  {journal} {\bibinfo  {journal} {JCAP}\ }\textbf {\bibinfo {volume} {2024}}\bibfield  {number} {\bibinfo  {number} { (05)},\ \bibinfo {pages} {043}},\ }\Eprint {https://arxiv.org/abs/2307.07403} {arXiv:2307.07403 [astro-ph.CO]} \BibitemShut {NoStop}%
\bibitem [{\citenamefont {Yin}\ \emph {et~al.}(2025)\citenamefont {Yin} \emph {et~al.}}]{Yin:2024lla}%
  \BibitemOpen
  \bibfield  {author} {\bibinfo {author} {\bibfnamefont {W.}~\bibnamefont {Yin}} \emph {et~al.},\ }\href {https://doi.org/10.1103/PhysRevLett.134.051004} {\bibfield  {journal} {\bibinfo  {journal} {Phys. Rev. Lett.}\ }\textbf {\bibinfo {volume} {134}},\ \bibinfo {pages} {051004} (\bibinfo {year} {2025})},\ \Eprint {https://arxiv.org/abs/2402.07976} {arXiv:2402.07976 [astro-ph.CO]} \BibitemShut {NoStop}%
\bibitem [{\citenamefont {Wang}\ \emph {et~al.}(2024)\citenamefont {Wang} \emph {et~al.}}]{Wang:2023imi}%
  \BibitemOpen
  \bibfield  {author} {\bibinfo {author} {\bibfnamefont {H.}~\bibnamefont {Wang}} \emph {et~al.},\ }\href {https://doi.org/10.1103/PhysRevD.110.103007} {\bibfield  {journal} {\bibinfo  {journal} {Phys. Rev. D}\ }\textbf {\bibinfo {volume} {110}},\ \bibinfo {pages} {103007} (\bibinfo {year} {2024})},\ \Eprint {https://arxiv.org/abs/2311.05476} {arXiv:2311.05476 [astro-ph.CO]} \BibitemShut {NoStop}%
\bibitem [{\citenamefont {Scott}\ and\ \citenamefont {Kilgour}(1969)}]{GDScott_1969}%
  \BibitemOpen
  \bibfield  {author} {\bibinfo {author} {\bibfnamefont {G.~D.}\ \bibnamefont {Scott}}\ and\ \bibinfo {author} {\bibfnamefont {D.~M.}\ \bibnamefont {Kilgour}},\ }\href {https://doi.org/10.1088/0022-3727/2/6/311} {\bibfield  {journal} {\bibinfo  {journal} {Journal of Physics D: Applied Physics}\ }\textbf {\bibinfo {volume} {2}},\ \bibinfo {pages} {863} (\bibinfo {year} {1969})}\BibitemShut {NoStop}%
\bibitem [{\citenamefont {Raffelt}\ and\ \citenamefont {Seckel}(1988)}]{Raffelt:1987yt}%
  \BibitemOpen
  \bibfield  {author} {\bibinfo {author} {\bibfnamefont {G.}~\bibnamefont {Raffelt}}\ and\ \bibinfo {author} {\bibfnamefont {D.}~\bibnamefont {Seckel}},\ }\href {https://doi.org/10.1103/PhysRevLett.60.1793} {\bibfield  {journal} {\bibinfo  {journal} {Phys. Rev. Lett.}\ }\textbf {\bibinfo {volume} {60}},\ \bibinfo {pages} {1793} (\bibinfo {year} {1988})}\BibitemShut {NoStop}%
\bibitem [{\citenamefont {Chang}\ \emph {et~al.}(2018)\citenamefont {Chang}, \citenamefont {Essig},\ and\ \citenamefont {McDermott}}]{Chang:2018rso}%
  \BibitemOpen
  \bibfield  {author} {\bibinfo {author} {\bibfnamefont {J.~H.}\ \bibnamefont {Chang}}, \bibinfo {author} {\bibfnamefont {R.}~\bibnamefont {Essig}},\ and\ \bibinfo {author} {\bibfnamefont {S.~D.}\ \bibnamefont {McDermott}},\ }\href {https://doi.org/10.1007/JHEP09(2018)051} {\bibfield  {journal} {\bibinfo  {journal} {JHEP}\ }\textbf {\bibinfo {volume} {{2018}}}\bibfield  {number} {\bibinfo  {number} { ({9})},\ \bibinfo {pages} {051}},\ }\Eprint {https://arxiv.org/abs/1803.00993} {arXiv:1803.00993 [hep-ph]} \BibitemShut {NoStop}%
\bibitem [{\citenamefont {Carenza}\ \emph {et~al.}(2019)\citenamefont {Carenza}, \citenamefont {Fischer}, \citenamefont {Giannotti}, \citenamefont {Guo}, \citenamefont {Mart\'\i{}nez-Pinedo},\ and\ \citenamefont {Mirizzi}}]{Carenza:2019pxu}%
  \BibitemOpen
  \bibfield  {author} {\bibinfo {author} {\bibfnamefont {P.}~\bibnamefont {Carenza}}, \bibinfo {author} {\bibfnamefont {T.}~\bibnamefont {Fischer}}, \bibinfo {author} {\bibfnamefont {M.}~\bibnamefont {Giannotti}}, \bibinfo {author} {\bibfnamefont {G.}~\bibnamefont {Guo}}, \bibinfo {author} {\bibfnamefont {G.}~\bibnamefont {Mart\'\i{}nez-Pinedo}},\ and\ \bibinfo {author} {\bibfnamefont {A.}~\bibnamefont {Mirizzi}},\ }\href {https://doi.org/10.1088/1475-7516/2019/10/016} {\bibfield  {journal} {\bibinfo  {journal} {JCAP}\ }\textbf {\bibinfo {volume} {10}}\bibfield  {number} {\bibinfo  {number} { (10)},\ \bibinfo {pages} {016}},\ }\bibinfo {note} {[Erratum: JCAP 05, E01 (2020)]},\ \Eprint {https://arxiv.org/abs/1906.11844} {arXiv:1906.11844 [hep-ph]} \BibitemShut {NoStop}%
\bibitem [{\citenamefont {Lucente}\ \emph {et~al.}(2022)\citenamefont {Lucente}, \citenamefont {Mastrototaro}, \citenamefont {Carenza}, \citenamefont {Di~Luzio}, \citenamefont {Giannotti},\ and\ \citenamefont {Mirizzi}}]{Lucente:2022vuo}%
  \BibitemOpen
  \bibfield  {author} {\bibinfo {author} {\bibfnamefont {G.}~\bibnamefont {Lucente}}, \bibinfo {author} {\bibfnamefont {L.}~\bibnamefont {Mastrototaro}}, \bibinfo {author} {\bibfnamefont {P.}~\bibnamefont {Carenza}}, \bibinfo {author} {\bibfnamefont {L.}~\bibnamefont {Di~Luzio}}, \bibinfo {author} {\bibfnamefont {M.}~\bibnamefont {Giannotti}},\ and\ \bibinfo {author} {\bibfnamefont {A.}~\bibnamefont {Mirizzi}},\ }\href {https://doi.org/10.1103/PhysRevD.105.123020} {\bibfield  {journal} {\bibinfo  {journal} {Phys. Rev. D}\ }\textbf {\bibinfo {volume} {105}},\ \bibinfo {pages} {123020} (\bibinfo {year} {2022})},\ \Eprint {https://arxiv.org/abs/2203.15812} {arXiv:2203.15812 [hep-ph]} \BibitemShut {NoStop}%
\bibitem [{\citenamefont {Springmann}\ \emph {et~al.}(2024)\citenamefont {Springmann}, \citenamefont {Stadlbauer}, \citenamefont {Stelzl},\ and\ \citenamefont {Weiler}}]{Springmann:2024ret}%
  \BibitemOpen
  \bibfield  {author} {\bibinfo {author} {\bibfnamefont {K.}~\bibnamefont {Springmann}}, \bibinfo {author} {\bibfnamefont {M.}~\bibnamefont {Stadlbauer}}, \bibinfo {author} {\bibfnamefont {S.}~\bibnamefont {Stelzl}},\ and\ \bibinfo {author} {\bibfnamefont {A.}~\bibnamefont {Weiler}},\ }\href@noop {} {\bibinfo {title} {{A Universal Bound on QCD Axions from Supernovae}}} (\bibinfo {year} {2024}),\ \Eprint {https://arxiv.org/abs/2410.19902} {arXiv:2410.19902 [hep-ph]} \BibitemShut {NoStop}%
\bibitem [{\citenamefont {Springmann}\ \emph {et~al.}(2025)\citenamefont {Springmann}, \citenamefont {Stadlbauer}, \citenamefont {Stelzl},\ and\ \citenamefont {Weiler}}]{Springmann:2024mjp}%
  \BibitemOpen
  \bibfield  {author} {\bibinfo {author} {\bibfnamefont {K.}~\bibnamefont {Springmann}}, \bibinfo {author} {\bibfnamefont {M.}~\bibnamefont {Stadlbauer}}, \bibinfo {author} {\bibfnamefont {S.}~\bibnamefont {Stelzl}},\ and\ \bibinfo {author} {\bibfnamefont {A.}~\bibnamefont {Weiler}},\ }\href {https://doi.org/10.1007/JHEP02(2025)138} {\bibfield  {journal} {\bibinfo  {journal} {JHEP}\ }\textbf {\bibinfo {volume} {{2025}}}\bibfield  {number} {\bibinfo  {number} { ({02})},\ \bibinfo {pages} {138}},\ }\Eprint {https://arxiv.org/abs/2410.10945} {arXiv:2410.10945 [hep-ph]} \BibitemShut {NoStop}%
\bibitem [{\citenamefont {Olver}\ \emph {et~al.}(2010)\citenamefont {Olver}, \citenamefont {Lozier}, \citenamefont {Boisvert},\ and\ \citenamefont {Clark}}]{NISTBOOK}%
  \BibitemOpen
  \bibfield  {author} {\bibinfo {author} {\bibfnamefont {F.}~\bibnamefont {Olver}}, \bibinfo {author} {\bibfnamefont {D.}~\bibnamefont {Lozier}}, \bibinfo {author} {\bibfnamefont {R.}~\bibnamefont {Boisvert}},\ and\ \bibinfo {author} {\bibfnamefont {C.}~\bibnamefont {Clark}},\ }\href@noop {} {\emph {\bibinfo {title} {The NIST Handbook of Mathematical Functions}}}\ (\bibinfo  {publisher} {Cambridge University Press, New York, NY},\ \bibinfo {year} {2010})\BibitemShut {NoStop}%
\end{thebibliography}%

\end{document}